\documentclass[twocolumn, tighten]{aastex62}

\usepackage{amsmath}
\usepackage{multirow}
\usepackage{url}
\usepackage{booktabs}

\newcommand{\sniip}{SN~2016gfy}
\newcommand{\hostglx}{NGC~2276}
\newcommand{\maghundred}{mag (100 d)$^{-1}$}
\newcommand{\magfifty}{mag (50 d)$^{-1}$}
\newcommand{\magday}{mag (d)$^{-1}$}
\newcommand{\ebv}{$E(B-V)$}
\newcommand{\kms}{$\rm km\ s^{-1}$}

\received{April 4, 2019}
\revised{\today}
\accepted{--}
\submitjournal{ApJ}

\shorttitle{Type II SN~2016gfy}
\shortauthors{Singh et al. 2019a}

\begin{document}

\title{Observational signature of circumstellar interaction and $\rm ^{56}Ni$-mixing in the Type II Supernova 2016gfy}

\correspondingauthor{Avinash Singh}
\email{avinash21292@gmail.com, avinash.singh@iiap.res.in}

\author[0000-0003-2091-622X]{Avinash Singh}
\affiliation{Indian Institute of Astrophysics, Koramangala 2nd Block, Bengaluru 560034, India}
\affiliation{Joint Astronomy Programme, Department of Physics, Indian Institute of Science, Bengaluru 560012, India}

\author[0000-0001-7225-2475]{Brajesh Kumar}
\affiliation{Indian Institute of Astrophysics, Koramangala 2nd Block, Bengaluru 560034, India}

\author[0000-0003-1169-1954]{Takashi J. Moriya}
\affiliation{Division of Science, National Astronomical Observatory of Japan, National Institutes of Natural Sciences, 2-21-1 Osawa, \\
Mitaka, Tokyo 181-8588, Japan}

\author{G.C. Anupama}
\affiliation{Indian Institute of Astrophysics, Koramangala 2nd Block, Bengaluru 560034, India}

\author{D.K. Sahu}
\affiliation{Indian Institute of Astrophysics, Koramangala 2nd Block, Bengaluru 560034, India}

\author[0000-0001-6272-5507]{Peter J. Brown}
\affiliation{Department of Physics and Astronomy, George P. and Cynthia Woods Mitchell Institute for Fundamental Physics \& Astronomy,\\
Texas A\&M University, 4242 TAMU, College Station, TX 77843, USA}

\author[0000-0003-0123-0062]{Jennifer E. Andrews}
\affiliation{Steward Observatory, University of Arizona, 933 North Cherry Avenue, Tucson, AZ 85721, USA}

\author{Nathan Smith}
\affiliation{Steward Observatory, University of Arizona, 933 North Cherry Avenue, Tucson, AZ 85721, USA}

\begin{abstract}
The optical and ultra-violet broadband photometric and spectroscopic observations of the Type II supernova (SN) 2016gfy are presented. The $V$-band light curve (LC) shows a distinct plateau phase with a slope, $s_2$\,$\sim$\,0.12 \maghundred\ and a duration of 90\,$\pm$\,5 d. Detailed analysis of \sniip\ provided a mean $\rm^{56}Ni$ mass of 0.033\,$\pm$\,0.003 $\rm M_{\odot}$, a progenitor radius of $\sim$\,350--700 $\rm R_{\odot}$, a progenitor mass of $\sim$\,12--15 $\rm M_{\odot}$ and an explosion energy of 0.9--1.4$\rm \times 10^{51}\ erg\ s^{-1}$. The P-Cygni profile of H\,$\rm \alpha$ in the early phase spectra ($\sim$\,11--21 d) shows a boxy emission. Assuming that this profile arises from the interaction of the SN ejecta with the pre-existing circumstellar material (CSM), it is inferred that the progenitor underwent a recent episode (30--80 years prior to the explosion) of enhanced mass loss. Numerical modeling suggests that the early LC peak is reproduced better with an existing CSM of 0.15 $\rm M_{\odot}$ spread out to $\sim$\,70 AU. A late-plateau bump is seen in the $VRI$ LCs during $\sim$\,50--95 d. This bump is explained as a result of the CSM interaction and/or partial mixing of radioactive $\rm ^{56}Ni$ in the SN ejecta. Using strong-line diagnostics, a sub-solar oxygen abundance is estimated for the supernova \ion{H}{2} region (12 + log(O/H) = 8.50\,$\pm$\,0.11), indicating an average metallicity for the host of a Type II SN. A star formation rate of $\sim$\,8.5 $\rm M_{\odot}\ yr^{-1}$ is estimated for \hostglx\ using the archival \textit{GALEX} FUV data.

\end{abstract}

\keywords{supernovae: general $-$ supernovae: individual: \sniip\ $-$ galaxies: individual: \hostglx{}}

\section{Introduction}\label{sec:intro}

Core-Collapse Supernovae (CCSNe) are the result of gravitational core-collapse in massive stars with Zero Age Main Sequence (ZAMS) mass $\rm \gtrsim$\,8 $\rm M_{\odot}$ \citep{2003heger, 2009smartt}. Type II SNe (II-P and II-L) form the segment of CCSNe that display eminent P-Cygni profiles of hydrogen in their observed spectra \citep{1941minkowski, 1997filippenko} whereas the others belong to the class of stripped envelope SNe. Type II SNe have been a subject of extensive study due to their majority in the class of CCSNe and thus has resulted in unveiling various correlations between the physical parameters \citep{2003hamuy, 2014anderson, 2014spiro, 2015valenti}.

Type II SNe that retain a large hydrogen envelope at the epoch of explosion show a \lq \lq plateau\rq \rq\ in their light curve and form the most common sub-type, Type II-P SNe \citep{2011li}. On the other hand, the ones that show a \lq \lq linear\rq \rq\ decline past the maximum light belong to the sub-type, Type II-L SNe \citep{1979barbon, 1994patat, 2012arcavi}. The plateau is an optically-thick phase of almost constant luminosity characterized by the recombination of hydrogen, lasting an average of $\sim$\,84 d \citep[see optically-thick phase duration (OPTd) in][]{2014anderson}. \citet{1994patat} differentiated Type II-P and II-L SNe based on their decline rates in the $B$-band and classified Type II-P SNe as having $\beta_{100}^{B}$ $<$ 3.5 \maghundred. However, recent sample studies of \citet{2014anderson}, \citet{2015sanders} and \citet{2016valenti} have argued that the class of Type II-P and II-L SNe form a continuous distribution and do not belong to distinct classes. According to these authors, Type II-P and II-L SNe show a continual trend in decline rates and can be accredited to the differing hydrogen envelope mass \citep{2014bfaran, 2015valenti, 2018avinash}, which can be attributed to the higher mass-loss rate associated with the massive progenitors of Type II-L SNe in comparison with Type II-P SNe \citep[][and references therein]{2011elias}. 

Observational studies on metallicity of the host environment of CCSNe have helped in furnishing constraints on the progenitor properties \citep[][and references therein]{2008prieto,2013akuncarayakti,2013bkuncarayakti,2015taddia,2016anderson}. The modeling of Type II SN atmospheres have shown a palpable dependence of metal-line strengths on the metallicity of the progenitor \citep[][hereafter KW09 and D13, respectively]{2009kasen, 2013bdessart}. The temporal evolution of the photosphere during the plateau phase describes the composition of the progenitor and hence the metal lines can help constrain the metallicity of the progenitor \citep[][hereafter D14 and A16, respectively]{2014dessart, 2016anderson}. The increasing metallicity amidst the model progenitors (D13) of Type II SNe display stronger (large Equivalent-Width, EW) metal-line features at a given epoch.

An upper limit of 25 $\rm M_{\odot}$ has been predicted by hydrodynamical modeling of Red Supergiants (RSGs) to retain its hydrogen envelope and explode as Type II SNe \citep{2003heger,2011bersten,2015morozova}. The in-homogeneity in RSGs result from differences in initial masses, metallicity and mass-loss rates. Direct detection of progenitors in the nearby galaxies (distance $\leq$ 25 Mpc) have been possible in the recent past using the pre-explosion images obtained from the Hubble Space Telescope and other big telescopes \citep[][and references therein]{2019vandyk}. The inferred masses of progenitors from direct detection lie in the range of $\sim$\,9\,--\,17 $\rm M_{\odot}$ \citep{2009smartt}, which falls significantly short of the upper limit derived from modeling. This is termed as the RSG problem and has been explained as a result of \lq failed SNe\rq\ which occurs in the higher end of the RSG mass range \citep{2012woosley,2013lovegrove,2014horiuchi}. Alternatively, pre-SN mass loss can also affect the estimates of progenitor mass due to anomalous dust correction \citep{2012walmswell,2012kochanek}. \citet{2018davies} explains this as a result of uncertainties in the mass-luminosity relationship and small number statistics.

\begin{table}
\caption{Brief details of \sniip\ and its host \hostglx.}
\label{tab:host}
\setlength{\tabcolsep}{7pt}
\begin{tabular}{l l c}
\hline \noalign{\smallskip}
Parameters 			    & Value		 				                    & Ref.\\
\noalign{\smallskip} \hline \noalign{\smallskip}
\multicolumn{3}{l}{\textit{\underline \sniip}:}\\
RA (J2000) 			    & $\alpha=07^{\rm h} 26^{\rm m} 43\fs67$ 	    & 3 \\
DEC (J2000)             & $\delta=+85\degr 45\arcmin 51\farcs70$ 	    & 3 \\
Discovery date 		    & 2016 Sept 13.10 UT                             & 3 \\
Explosion date 		    & 2016 Sept 9.90 UT                              & 1 \\
Total reddening         & $E(B-V)$\,=\,0.21\,$\pm$0.05 mag                    & 1 \\

\noalign{\smallskip} \hline \noalign{\smallskip}
\multicolumn{3}{l}{\textit{\underline \hostglx}:}\\

Type                    & SAB(rs)c  				                    & 2 \\
RA (J2000) 			    & $\alpha=07^{\rm h} 27^{\rm m} 14\fs36$ 	    & 2 \\
DEC (J2000)             & $\delta=+85\degr 45\arcmin 16\farcs40$ 	    & 2 \\
Redshift                & z\,=\,0.008062\,$\pm$\,0.000013                       & 2 \\
Distance                & D\,=\,29.64$\pm$\,2.65 Mpc                          & 1 \\
Distance modulus        & $\mu$\,=\,32.36\,$\pm$\,0.18 mag 			            & 1 \\
\noalign{\smallskip} \hline
\end{tabular}
\newline 
(1) This paper;
(2) \citet{1991devac};
(3) \citet{2016discovery}
\end{table}

In the absence of direction detection, the progenitor properties of the SN can be inferred from the explosion properties such as explosion energy, $\rm ^{56}Ni$ mass etc. These estimates are dependent on the distance to the SN. Type II SNe have shown promise as a standard candle for estimating distances to extra-galactic sources. Due to increased star-formation rate with higher redshifts (up to $\sim$\,2, \citealp{2003dickinson}), the abundance of Type II SNe at higher redshifts than Type Ia SNe make them an important diagnostic for estimating distance and potentially determining cosmological parameters. However, Type II SNe being fainter than Type Ia SNe argues against their importance at higher redshifts although the different systematics of using them as distance indicators makes them important. The most commonly used techniques are the Expanding Photosphere Method \citep[EPM]{1974kirshner}, the Standard Candle Method \citep[SCM]{2002hamuy}, the Photospheric Magnitude Method \citep[PMM]{2014rodriguez} and Photometric Color Method \citep[PCM]{2015dejaeger}. The EPM is a geometrical technique used to derive distances using the angular and the photospheric radii of the SN. The SCM is built on the observed correlation of the expansion velocity and the luminosity at an epoch during the plateau phase of a Type II SN. The PMM employs the precise knowledge of the explosion epoch, expansion velocity and the extinction corrected magnitudes whereas the PCM utilizes the correlation between luminosity, color and the late-plateau decline rate, to compute the distance to a Type II SN.

The study of Type II SNe enables understanding the diversity among their progenitors and one such object is presented here. \sniip\ was discovered by Alessandro Dimai on 2016 September 13.10 UT in the galaxy \hostglx\ at an unfiltered apparent magnitude of $\sim$\,16.3 mag \citep{2016discovery}. It lies 18\arcsec E and 20\arcsec N from the nucleus of the host. A spectrum obtained by the NOT Unbiased Transient Survey (NUTS) on 2016 September 15.25 UT, displayed a blue continuum with broad Balmer emission lines classifying it as a young Type II SN \citep{2016classify}. Brief details on \sniip\ are given in Table~\ref{tab:host}.

We present here detailed photometric and spectroscopic analysis of the Type II-P \sniip. The temporal evolution of the SN is studied in detail and its explosion parameters are determined. The properties of the host galaxy \hostglx\ are also studied and the progenitor parameters estimated. The properties of \sniip\ are compared with Type II SNe from the literature whose details are presented in Table~\ref{tab:compsample}.

\section{Data Acquisition and Reduction}\label{sec:data}

\subsection{2\,m Himalayan Chandra Telescope}

The photometric and spectroscopic follow-up of \sniip\ with the Himalayan Faint Object Spectrograph Camera (HFOSC) mounted on the 2\,m Himalayan Chandra Telescope (HCT), Indian Astronomical Observatory (IAO), Hanle, India began on 2016 September 13.74 (JD 2457645.24), roughly $\sim$\,15 hrs from discovery. Broadband photometric monitoring was carried out in Bessell $UBVRI$ at 42 epochs and the spectroscopic observations\footnote{The slit orientation during the spectroscopic follow-up of \sniip\ was along the E-W direction.} were performed on 33 epochs using grisms Gr7 (3500--7800 \AA, R\,$\sim$\,500) and Gr8 (5200--9250 \AA, R\,$\sim$\,800). 

Landolt field PG0231+051 \citep{1992landolt} was observed on photometric nights of 2016 September 20, October 04 and December 05 for the photometric calibration of the SN field. Template subtraction was carried out due to significant contamination from the host galaxy, the details of which are given in Section~\ref{sec:tempsub}. The spectra from the two grisms were combined after scaling to a weighted mean using a common overlapping region in the vicinity of a flat continuum. A detailed description on the data reduction can be found in \citet{2018brajesh,2018sahu,2018avinash}.

\subsection{6.5\,m Multiple Mirror Telescope}

Medium-resolution spectra was obtained with the Bluechannel (BC) spectrograph mounted on the 6.5\,m Multiple Mirror Telescope (MMT, \citealp{1989schmidt}) using the 1200 line/mm grating centered at 6300 \AA. These spectra were reduced using standard techniques in PyRAF \citep{2012pyraf}, including bias subtraction, flat-fielding, wavelength calibration using arc lamps, and flux calibration using standard stars observed on the same nights at similar airmass. Observations were obtained with the slit aligned along the parallactic angle to minimize differential light losses \citep{1982filippenko}.

\subsection{SWIFT Ultraviolet/Optical Telescope}

\sniip\ was also observed with the Neil Gehrels Swift Observatory \citep{2004gehrels}. Observations with the Ultra-Violet Optical Telescope (UVOT; \citealp{2005roming}) began 2016 September 15 UT. Data reduction utilized the pipeline of the Swift Optical Ultraviolet Supernova Archive (SOUSA; \citealp{2014brown}) including the revised Vega-system zero-points of \citet{2011breeveld}. The underlying count rates from the host galaxy were measured from images obtained on 2018 March 19 and subtracted from the photometry.

\section{Host Galaxy - \hostglx}\label{sec:hostglx}

The host galaxy of \sniip, \hostglx\ is a face-on starburst spiral galaxy interacting with the elliptical galaxy NGC 2300 (d\,$\sim$\,30 Mpc, \citealp{2000mould}). Ram-pressure and viscous stripping form the basis for its distorted morphology and the increased star formation rate (SFR) in the galaxy \citep{1997davis, 2015wolter, 2018tomi}. Measurements of X-ray gas on the disc of \hostglx\ have yielded a low metallicity ($\sim$\,0.1\,$\rm Z_{\odot}$) with no appreciable differences between the edges of the galaxy \citep[towards or away from the interaction,][]{2006rasmussen}.

\begin{figure}
\centering
\resizebox{\hsize}{!}{\includegraphics{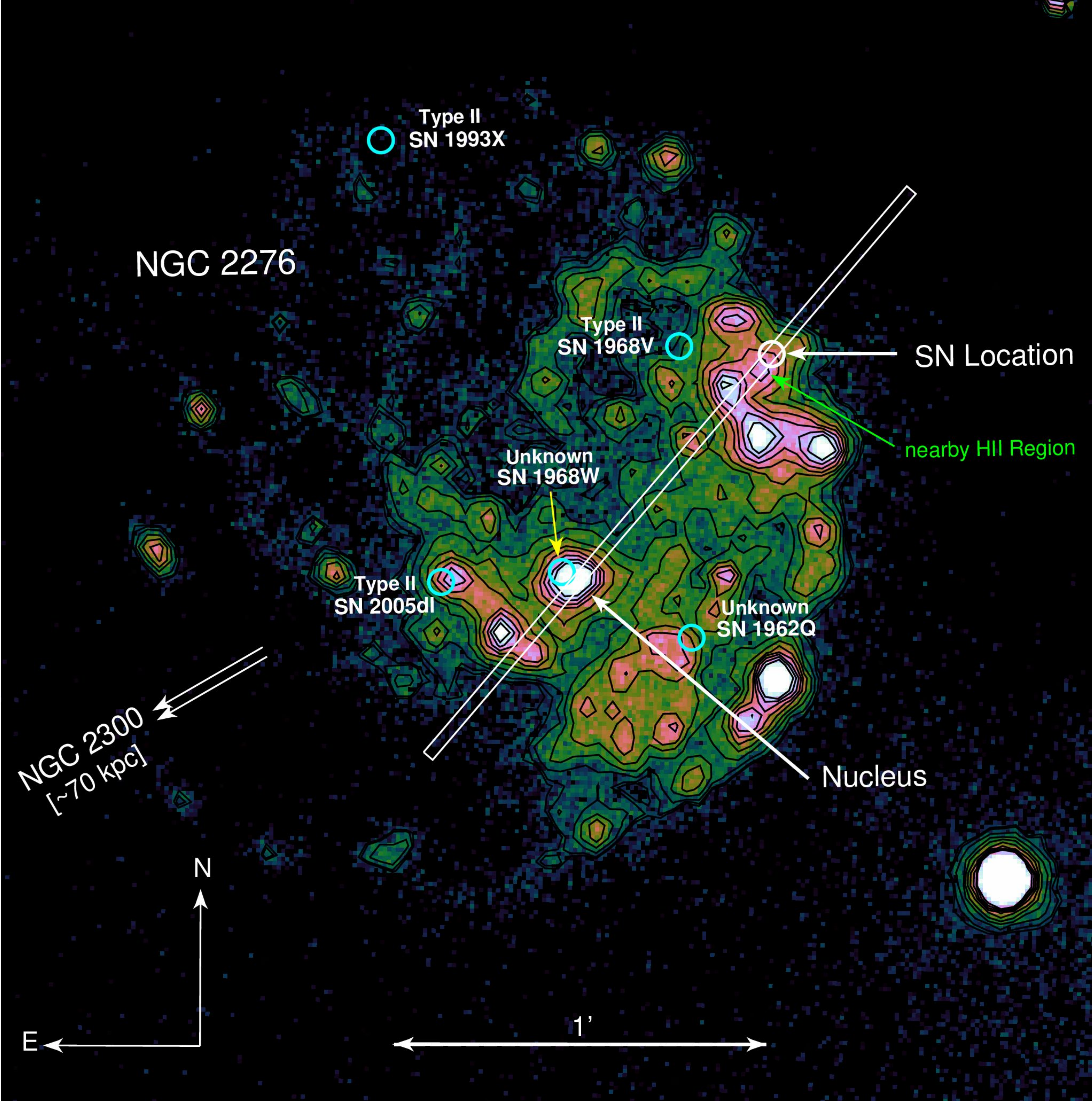}}
\caption{Narrow band H\,$\rm \alpha$ image of the host galaxy \hostglx\ obtained from \citet{2008epinat}. The interacting galaxy NGC 2300 \citep{1997davis} is located SE of \hostglx\ at a projected distance of $\sim$\,70 kpc. The nucleus and the location of \sniip\ is marked along with the five reported SNe in the galaxy and their sub-types (if known). Iso-intensity contours are shown in $black$ to reveal regions of enhanced H\,$\rm \alpha$ emission in the galaxy. The \ion{H}{2} region closest to the SN location is indicated. The size of the $circular$ markers depict the average seeing ($\sim$\,2$^{\prime\prime}$) at the site of HCT. The slit orientation for the host environment spectrum (Section~\ref{sec:hostspec}) is shown with a rectangular box. The image is shown in square-root intensity scale for clarity.}
\label{fig:hosthalpha}
\end{figure}

\begin{figure}
\centering
\resizebox{\hsize}{!}{\includegraphics{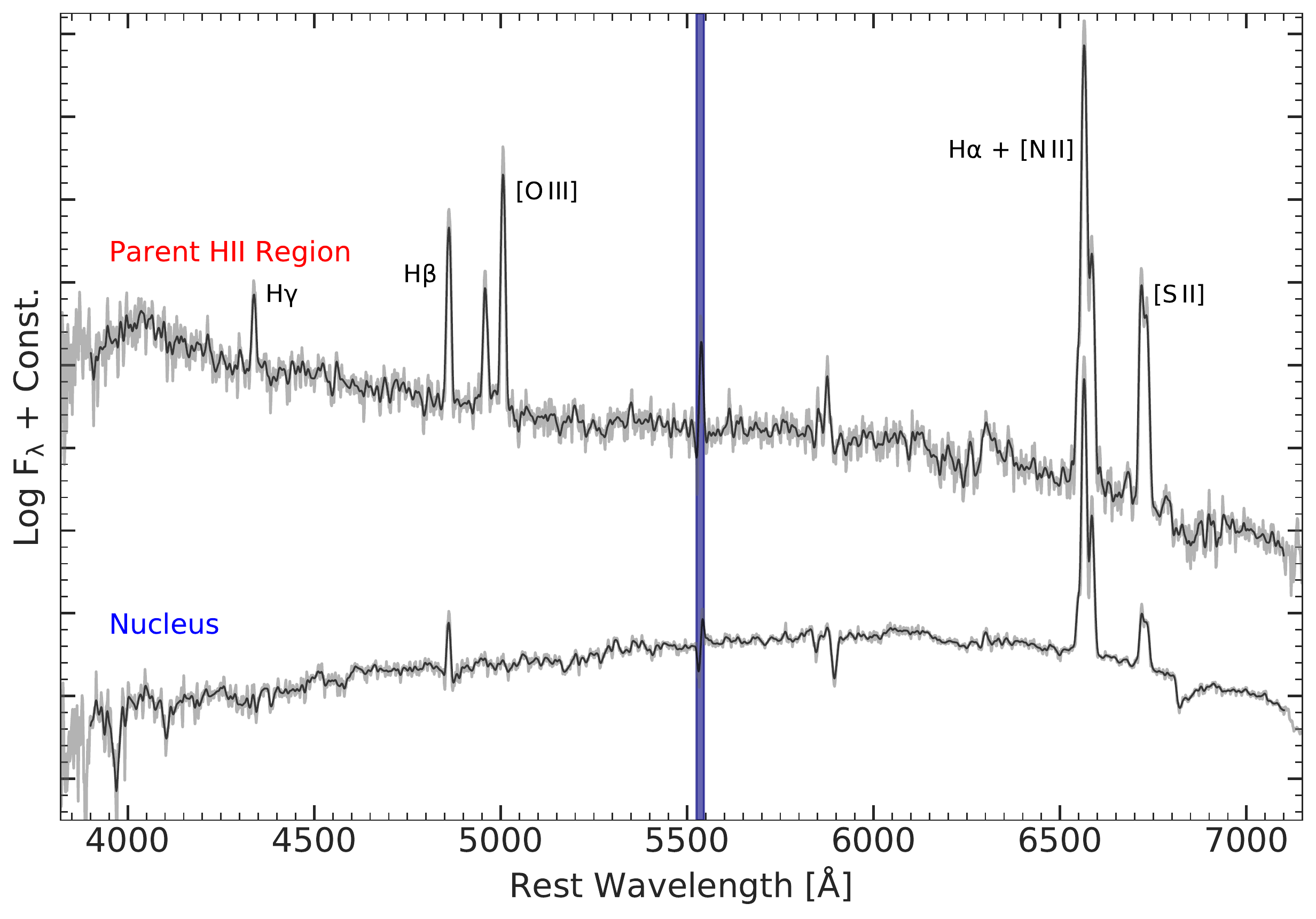}}
\resizebox{\hsize}{!}{\includegraphics{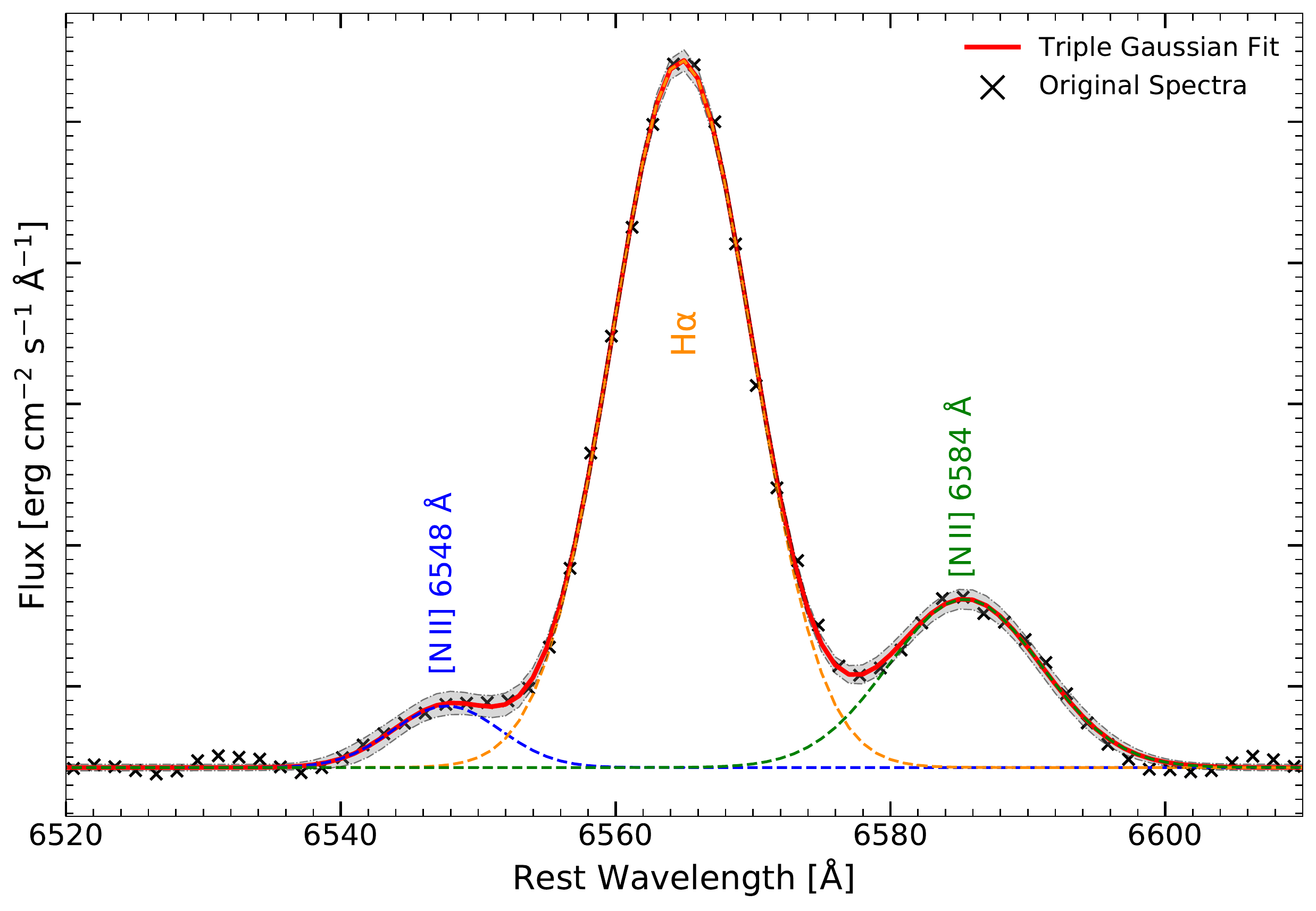}}
\caption{$Top\ panel$: Spectrum of the nucleus of the host galaxy \hostglx\ and the parent \ion{H}{2} region of the SN. Notable emission lines are labelled and the region shaded in $dark\ blue$ indicates the artifact in our spectra. $Bottom\ panel$: A triple Gaussian fit to the H\,$\rm \alpha$ (contaminated by the [\ion{N}{2}] doublet) profile to compute the individual line fluxes.}
\label{fig:hostspec}
\end{figure}

\subsection{Star formation rate from \textit{GALEX} archival image}\label{sec:hostsfr}

Flux-normalized and background-subtracted FUV intensity image of \hostglx\ was obtained from \textit{GALEX} Catalog Search\footnote{\url{http://galex.stsci.edu/GR6/?page=mastform}} and aperture photometry was performed with an elliptical aperture using the $photutils$ python package \citep{2017photutils}. The galaxy is surrounded by bright foreground stars and hence an aperture smaller than the isophotal diameter (at B\,=\,25 mag arcsec$^{-2}$) was used in computing the net flux from the galaxy. The flux obtained was converted into the AB magnitude system of \citet{1983oke} using the zero point in \citet{2007morrissey}. The correction for Galactic and internal extinction was applied assuming \citet{1999fitzpatrick} extinction law with the help of the York Extinction Solver \citep{2004mccall} to obtain the final FUV magnitude, $\rm m_{FUV}$\,$\sim$\,13.11 mag.

A star formation rate (SFR) of $\sim$\,8.5\,$\rm M_{\odot}\ yr^{-1}$ is estimated for \hostglx\ using its FUV magnitude \citep{2013karachentsev}. Using an H\,$\rm \alpha$ flux of 6.3\,$\rm \times$\,$\rm 10^{-12}\ erg\ cm^{-2}\ s^{-1}$ \citep{1997davis} for \hostglx\ and the relation by \citet{1998kennicutt}, an SFR of $\sim$\,5.2\,$\rm M_{\odot}\ yr^{-1}$ is determined. The SFR values obtained above are consistent with the values from the literature for \hostglx\ \citep{2018tomi}.

The galaxy has been a host to five reported SNe (prior to \sniip), namely SN~1962Q \citep{1967iskudarian}, SN~1968V\footnote{\label{typeII}Confirmed Type II SNe} \citep{1968shakhbazyan}, SN~1968W \citep{1968iskudarian}, SN~1993X\textsuperscript{\ref{typeII}} \citep{1993treffers} and SN~2005dl\textsuperscript{\ref{typeII}} \citep{2005dimai}. Of the six SNe, four are confirmed Type II (including \sniip), while the other two are unclassified. Hence, 4$^{+2}_{-0}$ CCSNe have occurred in the last 57 years leading up to 2019, giving us an observed supernova rate (SNR) of 0.070$^{+0.035}_{-0}$ CCSNe yr$^{-1}$.

The relation between SNR and SFR was estimated using the BPASS v2.2 catalogue \citep{2017eldridge,2018stanway} assuming the Chabrier initial mass function \citep{2003chabrier}. A mean SNR of $\sim$\,0.009 CCSNe yr$^{-1}$ is expected for an SFR of 1 $\rm M_{\odot}\ yr^{-1}$ for metallicities ranging from 0.1 (inferred from X-ray gas) to 0.8 (nuclear metallicity of \hostglx) $\rm Z_{\odot}$. This gives an SFR of $\sim$\,7.8 $\rm M_{\odot}\ yr^{-1}$ for \hostglx\ and is consistent with the photometric estimates of SFR.

\subsection{Parent H~II region}\label{sec:parenthii}

The observed H\,$\rm \alpha$ luminosity in spiral galaxies trace the ionised regions produced by the radiation from massive OB stars ($>$\,10 $\rm M_{\odot}$). Hence, the H\,$\rm \alpha$ line emission can help indicate the parent population of CCSNe \citep{1984kennicutt}. The H\,$\rm \alpha$ map of \hostglx\ obtained from \citet{2008epinat} is shown in Figure~\ref{fig:hosthalpha}. The nearest \ion{H}{2} region lies $\rm \sim$\,2$^{\prime\prime}$ away from \sniip\ signifying probable association and shares the property of the region.

\subsection{Host environment Spectroscopy}\label{sec:hostspec}

\begin{table*}
\centering
\caption{Distances derived from SCM analysis using $\rm H_0$ = 73.52\,$\pm$\,1.62 $\rm km\ s^{-1}\ Mpc^{-1}$.}
\label{tab:resultsscm}
\begin{tabular}{|c|c c c c c c c c |c|}
\toprule
Reference & Filter       &   $\alpha$       & $\beta$          & $\gamma$         &   Epoch         &  $V-I$          &   App. Mag.       &   $v_{Fe\,II}$    &  Distance         \\
&              &                  &                  &                  &   (d)           &   (mag)         &   (mag)           & ($\rm km\ s^{-1}$)&   (Mpc)           \\
\midrule
\multirow{2}[1]{*}{H04}
& V            &   6.25\,$\pm$\,1.35  & 1.46\,$\pm$\,0.15    & ---              & $t_0$+50        & ---             &  16.26\,$\pm$\,0.01   & 4272\,$\pm$\,53       &  29.14\,$\pm$\,3.49   \\
& I            &   5.45\,$\pm$\,0.91  & 1.92\,$\pm$\,0.11    & ---              & $t_0$+50        & ---             &  15.47\,$\pm$\,0.02   & 4272\,$\pm$\,53       &  29.19\,$\pm$\,2.35   \\
\midrule
\multirow{1}[0]{*}{N06}
& I            &   6.69\,$\pm$\,0.50  & -17.49\,$\pm$\,0.08  & 1.36             & $t_0$+50        & 0.68\,$\pm$\,0.02   &  15.47\,$\pm$\,0.02   & 4272\,$\pm$\,53       &  34.78\,$\pm$\,2.21$^*$  \\
\midrule
\multirow{1}[0]{*}{P09}
& I            &   4.4\,$\pm$\,0.6    & -1.76\,$\pm$\,0.05   & 0.8\,$\pm$\,0.3      & $t_0$+50        & 0.68\,$\pm$\,0.02   &  15.47\,$\pm$\,0.02   & 4272\,$\pm$\,53       &  31.29\,$\pm$\,3.40   \\
\midrule
\multirow{3}[1]{*}{O10}
& B            &   3.50\,$\pm$\,0.30  & -1.99\,$\pm$\,0.11   &  2.67\,$\pm$\,0.13   & $t_{PT}$--30     & 0.83\,$\pm$\,0.02   &  17.48\,$\pm$\,0.02   & 3022\,$\pm$\,42       &  27.00\,$\pm$\,2.68   \\
& V            &   3.08\,$\pm$\,0.25  & -2.38\,$\pm$\,0.09   &  1.67\,$\pm$\,0.10   & $t_{PT}$--30     & 0.83\,$\pm$\,0.02   &  16.29\,$\pm$\,0.01   & 3022\,$\pm$\,42       &  28.55\,$\pm$\,2.13   \\
& I            &   2.62\,$\pm$\,0.21  & -2.23\,$\pm$\,0.07   &  0.60\,$\pm$\,0.09   & $t_{PT}$--30     & 0.83\,$\pm$\,0.02   &  15.37\,$\pm$\,0.02   & 3022\,$\pm$\,42       &  27.50\,$\pm$\,1.59   \\
\midrule
Mean &         &                  &                  &                  &                 &                 &                   &                   &  29.64\,$\pm$\,2.65   \\   
\bottomrule
\multicolumn{4}{l}{$^*$Exception: $\rm H_0$\,=\,70 $\rm km\ s^{-1}\ Mpc^{-1}$}
\end{tabular}
\newline
Note: H04 - \citet{2004hamuy}, N06 - \citet{2006nugent}, P09 - \citet{2009pastorello} and O10 - \citet{2010olivares}    
\end{table*}

\begin{figure}
\centering
\resizebox{\hsize}{!}{\includegraphics{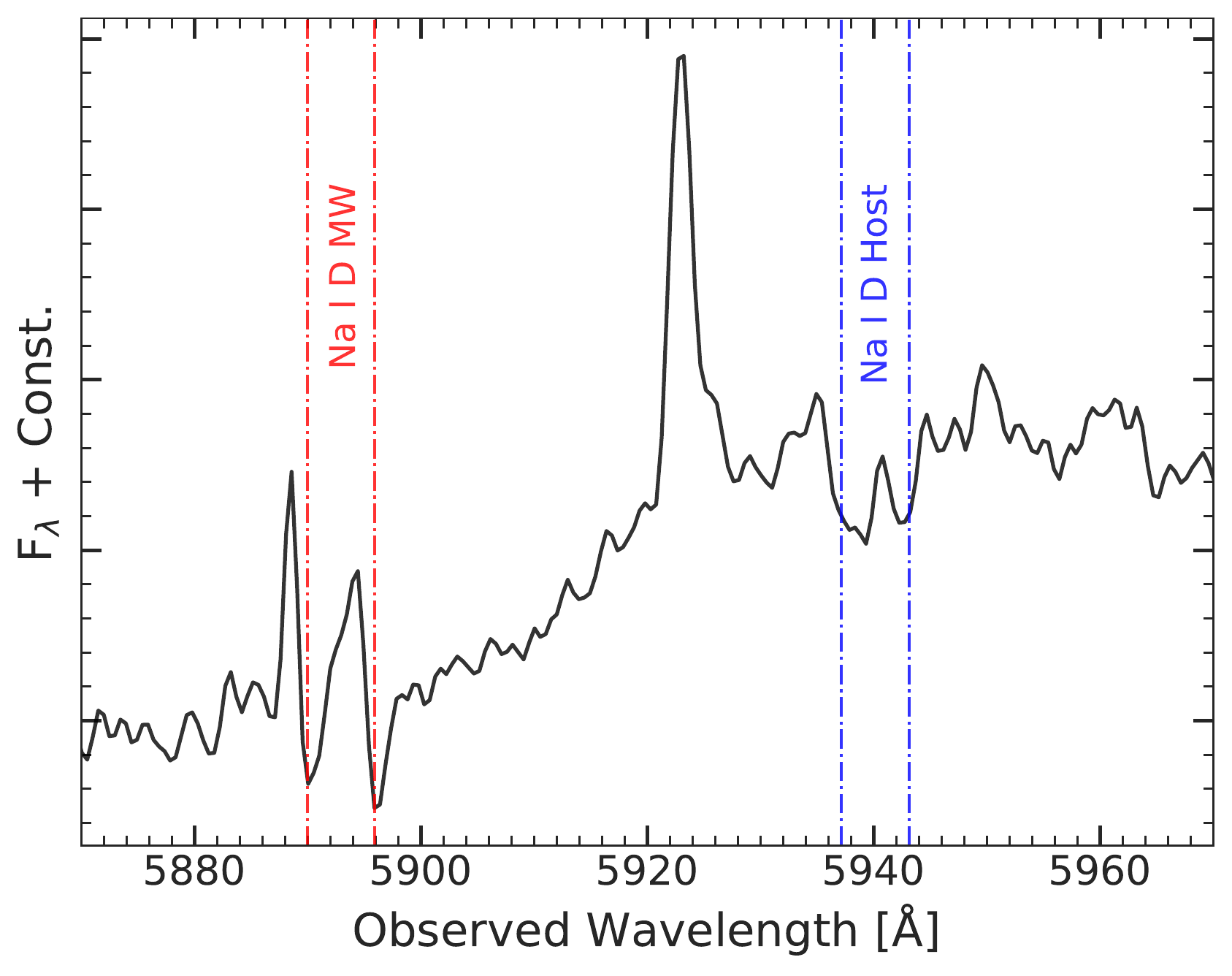}}
\caption{\ion{Na}{1}\,D from the MW and the host galaxy \hostglx\ in the spectrum of $\sim$\,175.5 d from MMT. The dash-dotted lines indicate the rest-wavelength of the features.}
\label{fig:naid}
\end{figure}

A Gr7 (3500--7800 \AA) spectrum of the parent \ion{H}{2} region of \sniip\ along with nucleus of the host galaxy \hostglx\ was obtained on 2018 Oct 31 by orienting the HFOSC slit across the two locations as shown in Figure~\ref{fig:hosthalpha}. Calibrated one-dimensional spectra corresponding to both the regions are shown in Figure~\ref{fig:hostspec}. The spectra of the two regions exhibited prominent emission lines of H\,$\rm \alpha$, H\,$\rm \beta$, [\ion{N}{2}] 6548, 6584 \AA, [\ion{S}{2}] 6717, 6731 \AA, whereas the [\ion{O}{3}] 4959, 5007 \AA\ lines were present only in the spectrum of the parent \ion{H}{2} region.

To be able to use emission line diagnostics for determining the metallicity of the nucleus, other ionizing sources such as AGN contamination and shock-excitation must be ruled out \citep{2015taddia}. The shorthand notation, N2\,$\equiv$\,log\,([\ion{N}{2}]\,$\rm \lambda$6584/H\,$\rm \alpha$), O3\,$\equiv$\,log([\ion{O}{3}]$\rm \lambda$5007/H\,$\rm \beta)$ and O3N2\,$\equiv$ log{([\ion{O}{3}]\,$\rm \lambda$5007/H\,$\rm \beta$)/[\ion{N}{2}]\,$\rm \lambda$6584/H\,$\rm \alpha$}) is used henceforth. The line ratios from the nucleus obey the relation, $\rm O3\,<\,0.61\,/\,((N2\,-\,0.05)\,+\,1.3)$ coined by \citet{2003kauffmann} based on the BPT diagram \citep{1981baldwin} and confirms the star-forming nature of the nucleus without any significant AGN contamination.

The gas-phase oxygen abundances of these regions were computed from the N2 and the O3N2 indices using the relations from \citet{2004pettini}. An oxygen abundance of 8.61\,$\pm$\,0.18 ($\sim$\,0.8\,$\rm Z_{\odot}$) was estimated for the nucleus of \hostglx\ using the N2 diagnostic and a mean oxygen abundance of 8.50\,$\pm$\,0.11 ($\sim$\,0.6\,$\rm Z_{\odot}$) was estimated for the parent \ion{H}{2} region using the N2 and O3N2 diagnostics. The lower metallicity of the parent \ion{H}{2} region in comparison to the nucleus is consistent with radially decreasing metallicity gradients seen in galaxies \citep{1999henry}. The abundance of the parent \ion{H}{2} region indicates a sub-solar oxygen abundance adopting a solar abundance of 8.69\,$\pm$\,0.05 \citep{2009asplund}. The use of emission line ratios in these diagnostics minimizes the need for precise extinction correction and flux calibration.

A mean oxygen abundance of $\sim$8.49 was estimated by \citet{2016anderson} for an unbiased sample of Type II SNe host \ion{H}{2} regions. This indicates that the parent \ion{H}{2} region of \sniip\ has an average oxygen abundance for the host of a Type II SN.

\begin{table*}
\centering
\renewcommand{\arraystretch}{1.1}
\caption{Parameters extracted from the fit (Equation~\ref{eqn:earlyrise}) to the early time LC of \sniip.}
\label{tab:fitrisetime}
\begin{tabular}{|c| c c c c c c|}
\toprule
Filter  &  $a1$           	    &   $a_2$           &  $a_3$         	    &  $t_{Max}$         &  $t_{Max}-t_{0}$     &  $t_{Max}-t_{0}^{Mean}$   \\
        & ($\times 10^{-15}$)   &                   & ($\times 10^{-18}$)   &     (JD)           &       (d)            &       (d)                 \\
\midrule
$U$     &  2.75\,$\pm$\,0.34  	    &  1.224\,$\pm$\,0.042  & -0.60  		        &   2457648.2        &   6.8\,$\pm$\,0.5      &  6.8\,$\pm$\,0.5                      \\
$B$     &  1.28\,$\pm$\,0.07  	    &  1.057\,$\pm$\,0.016  &  0.50  		        &   2457649.2        &   9.4\,$\pm$\,0.3      &  7.8\,$\pm$\,0.3                      \\
$V$     &  0.90\,$\pm$\,0.11  	    &  1.155\,$\pm$\,0.042  &  1.50  		        &   2457650.5        &   8.5\,$\pm$\,0.4      &  9.1\,$\pm$\,0.4                      \\
$R$     &  0.42\,$\pm$\,0.11  	    &  1.028\,$\pm$\,0.080  &  1.00  		        &   2457653.2        &  12.1\,$\pm$\,0.8      &  12.0\,$\pm$\,0.8                     \\
$I$     &  0.35\,$\pm$\,0.06  	    &  1.148\,$\pm$\,0.059  &  0.80  		        &   2457651.6        &   9.1\,$\pm$\,0.4      &  10.3\,$\pm$\,0.4                     \\
\bottomrule
\end{tabular}
\end{table*}


\section{Estimate of Total Extinction and Distance}\label{sec:reddening}

Extinction along the line-of-sight (LOS) of \sniip\ is composed of reddening from the dust in the Milky Way (MW) and the host galaxy \hostglx. A Galactic reddening of \ebv\,=\,0.0865\,$\pm$\,0.0018 mag is obtained from the dust-extinction map of \citet{2011schlafly}, which assumes the \citet{1999fitzpatrick} extinction law. To determine the strength of the \ion{Na}{1}\,D feature, four early phase spectra (4--18 days from the date of explosion, c.f. Section~\ref{sec:risetime}) of \sniip\ were co-added. The equivalent width (EW) of \ion{Na}{1}\,D as measured from the combined spectrum is 0.44\,$\pm$\,0.08 \AA\ and gives an \ebv\,=\,0.06\,$\pm$\,0.01 mag \citep{2003turatto} and 0.05\,$\pm$\,0.01 mag \citep{2012poznanski}. Hence, a mean Galactic reddening of \ebv\,=\,0.07\,$\pm$\,0.01 mag is adopted in the direction of \sniip.

A weak \ion{Na}{1}\,D is also identified at the redshift of the host galaxy and is seen superimposed over the P-Cygni profile from the SN (\ion{He}{1} in the early phase and \ion{Na}{1}\,D in the late phase). The composite spectra yields an \ion{Na}{1}\,D EW of 0.89\,$\pm$\,0.13 \AA\ which corresponds to an \ebv\,=\,0.13\,$\pm$\,0.02\,mag \citep{2003turatto} and 0.16\,$\pm$\,0.07 mag \citep{2012poznanski}. Host galaxy reddening was further confirmed using the \lq \lq colour method\rq \rq\ proposed by \citet{2010olivares} which postulates that the intrinsic $(V-I)$ colour is constant for Type II-P SNe (i.e. $(V-I)_0$\,=\,0.656 mag) at the end of the plateau phase. Using the Galactic reddening corrected $(V-I)$ colour prior to the end of the plateau phase ($\sim$\,80.8 d), an $E(B-V)_{host}$\,=\,0.14\,$\pm$\,0.11 mag was obtained assuming a total-to-selective extinction ratio, $R_V$\,=\,3.1. A mean reddening of \ebv\,=\,0.14\,$\pm$\,0.05 mag is estimated for the host galaxy \hostglx. These measurements were verified with the resolved \ion{Na}{1}\,D in the medium-resolution spectrum obtained from MMT (see Figure~\ref{fig:naid}).

\begin{figure}
\centering
\resizebox{\hsize}{!}{\includegraphics{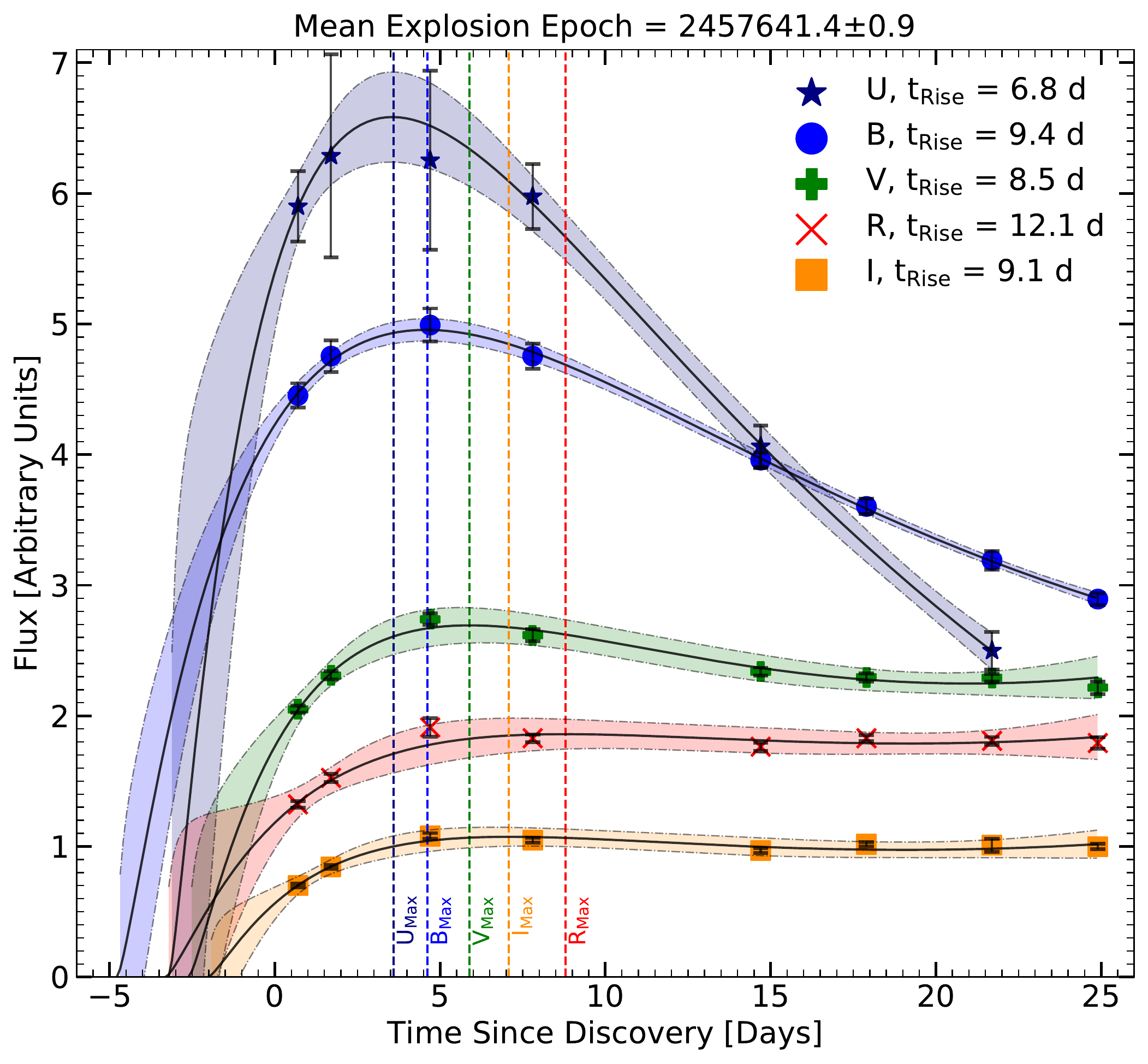}}
\caption{Fit to the early time LC ($<$\,25 d) of \sniip\ in Bessell $UBVRI$ bands. The fit was performed using the relation in \citet{2010cowen} and is shown with a solid line. 3$\sigma$ confidence interval of the fits in different bands are shown in shaded colours.}
\label{fig:fitrisetime}
\end{figure}

The host extinction estimate was also verified using Balmer decrement \citep[H\,$\rm \alpha$/H\,$\rm \beta$ ratio,][]{1989osterbrock}. Using Equation 4 from \citet{2013dominguez} which assumes Case B recombination (T $\sim$\,10$^4$ K and a large $\rm \tau$), the emission line flux ratios from the spectrum of the parent \ion{H}{2} region gives an $E(B-V)_{host}$\,$\sim$\,0.13 mag, confirming the estimate for the host reddening by other methods. A total reddening of \ebv\,=\,0.21\,$\pm$\,0.05 mag is adopted for \sniip.

The distance to \sniip\ is estimated using various SCM techniques and is mentioned in Table~\ref{tab:resultsscm}. A mean SCM distance of 29.64\,$\pm$\,2.65 Mpc ($\rm \mu$\,=\,32.36\,$\pm$\,0.18) is inferred for the host galaxy \hostglx. The redshift (z\,=\,0.008062) of \hostglx\ obtained from \citet{2008epinat} corresponds to a luminosity distance estimate of 33.1 Mpc, with H\,$\rm _0$\,=\,73.52 \kms\ $\rm Mpc^{-1}$, $\omega_M$\,=\,0.286 and $\omega_{\Lambda}$\,=\,0.714 (see Table~\ref{tab:dist_host}) and is slightly higher in comparison with the SCM distance. The uncertainty inferred in measuring SCM distances is 6\%-9\%\ \citep{2010olivares}. It is to be noted that the SCM technique is sensitive to the progenitor mass and metallicity which directly influence the mass of the hydrogen envelope (KW09).

\section{Photometric Evolution}\label{sec:lightcurve}

\subsection{Epoch of explosion and Rise time}\label{sec:risetime}

\begin{figure*}
\centering
\resizebox{\hsize}{!}{\includegraphics{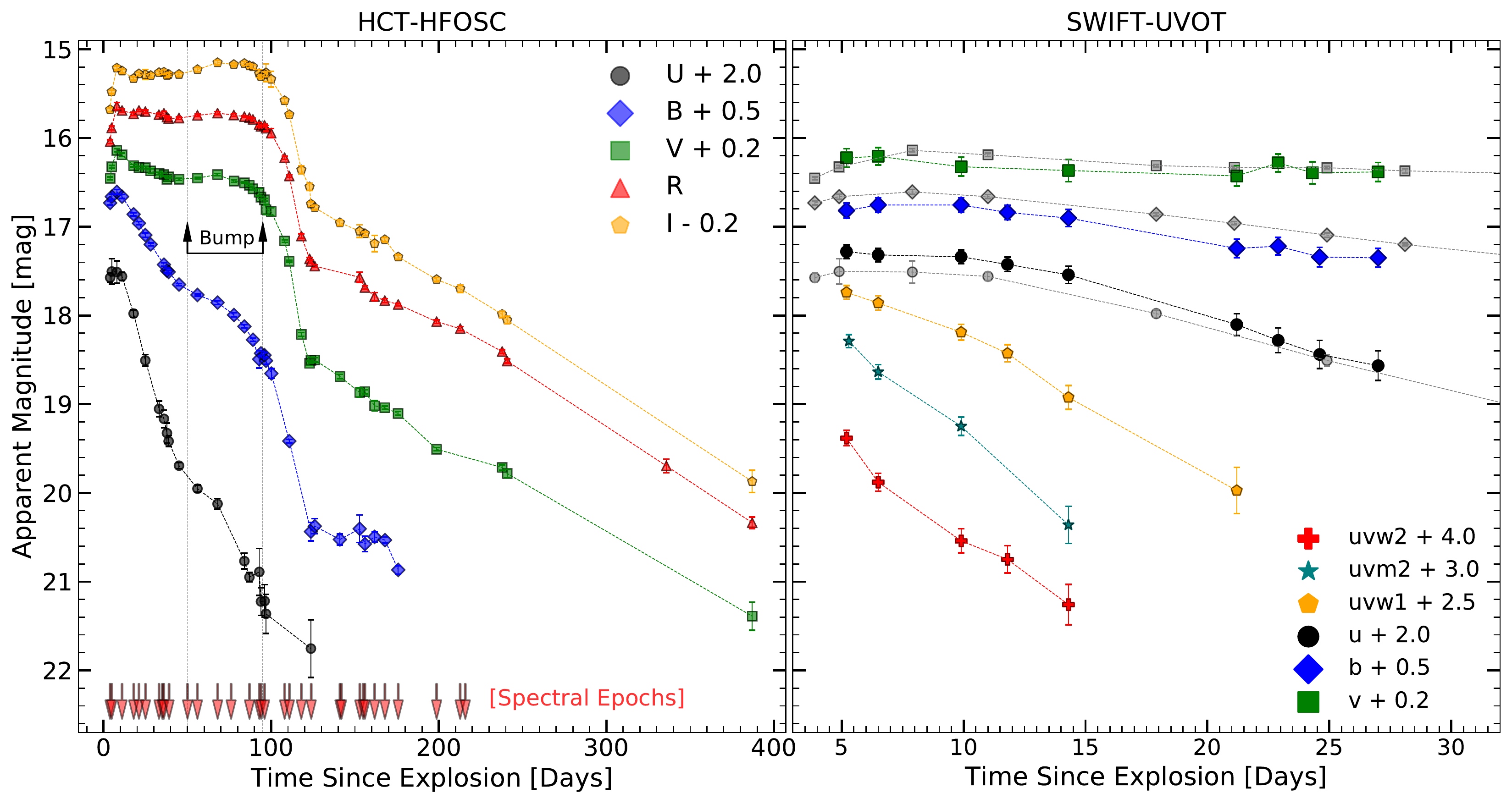}}
\caption{Apparent magnitude light curves of \sniip\ obtained from HCT-HFOSC and SWIFT-UVOT in the $left$ and $right$ panels respectively. Additionally, data from $UBV$ filters (in $grey$) are also shown in the $right\ panel$. The \lq \lq bump\rq \rq\ in the $VRI$ LC has been indicated. Offsets have been applied for clarity.}
\label{fig:apparentlc}
\end{figure*}

\begin{figure}
\centering
\resizebox{0.95\hsize}{!}{\includegraphics{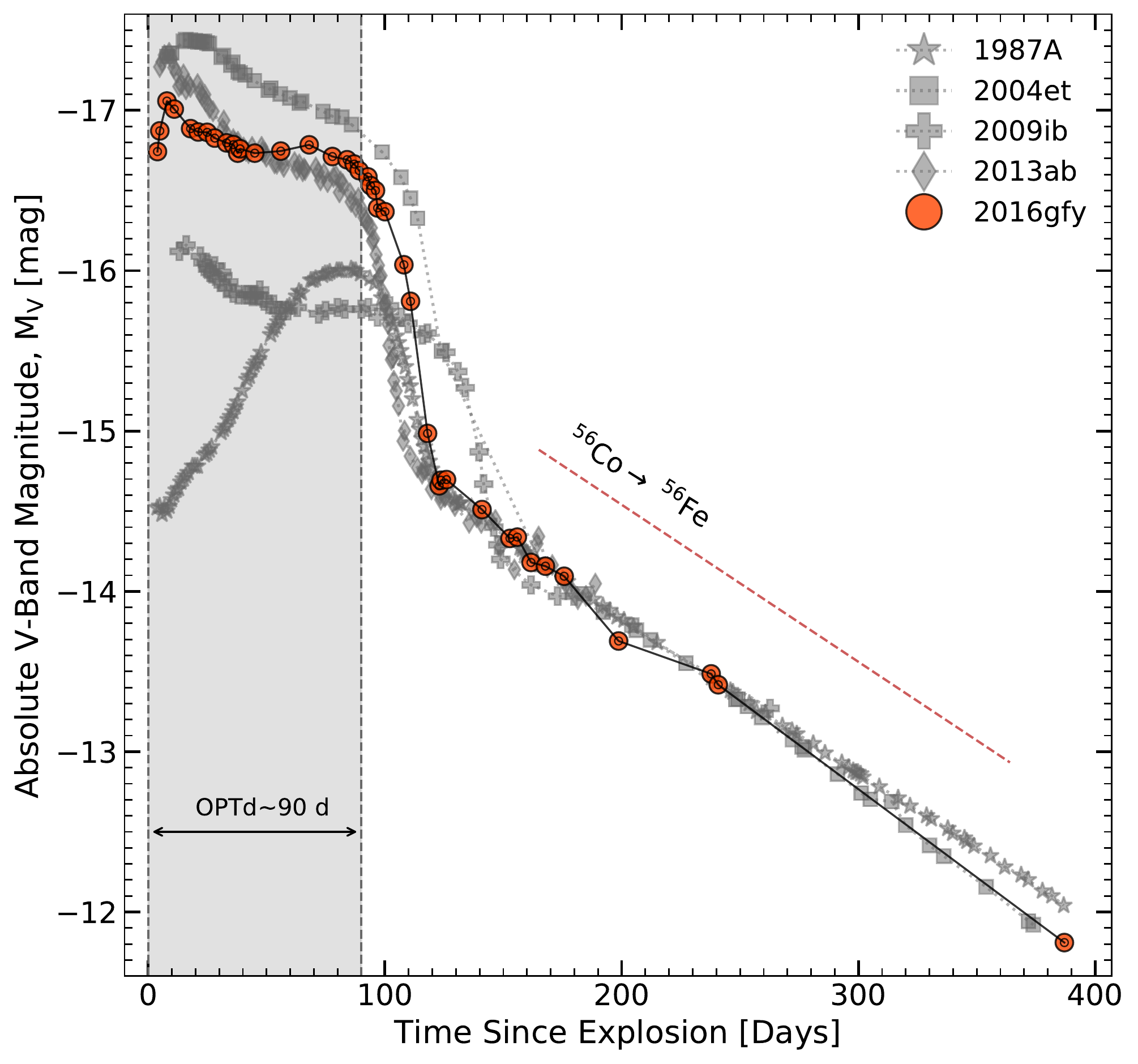}}
\caption{$V$-band absolute magnitude LC of \sniip\ in comparison with other Type II-P SNe. References: 1987A \citep{1990hamuy}, 2004et \citep{2006sahu}; 2009ib \citep{2015takats}; 2013ab \citep{2015boseab}.}
\label{fig:vabslc}
\end{figure}

\begin{figure}
\centering
\resizebox{\hsize}{!}{\includegraphics{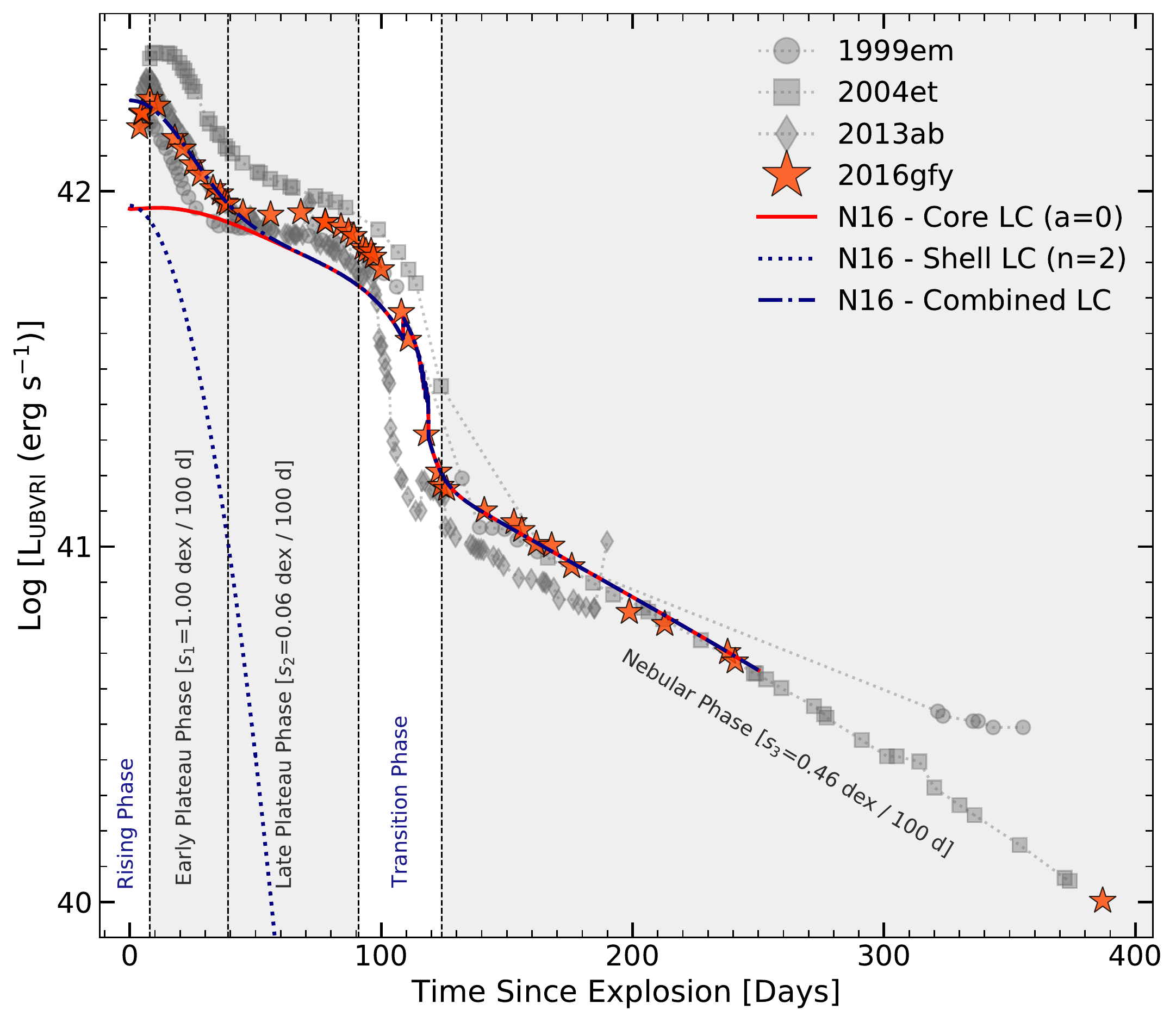}}
\caption{Pseudo-bolometric light curve of \sniip\ along with other Type II SNe. $Dashed$ vertical lines show the epochs of transition between the different phases. The two-component fit from the analytic model of \citet{2016nagy} is also shown. References: 1999em \citep{2002bleonard}, 2004et \citep{2006sahu}; 2013ab \citep{2015boseab}.}
\label{fig:bolometriclc}
\end{figure}

\begin{figure*}
\centering
\resizebox{0.85\hsize}{!}{\includegraphics{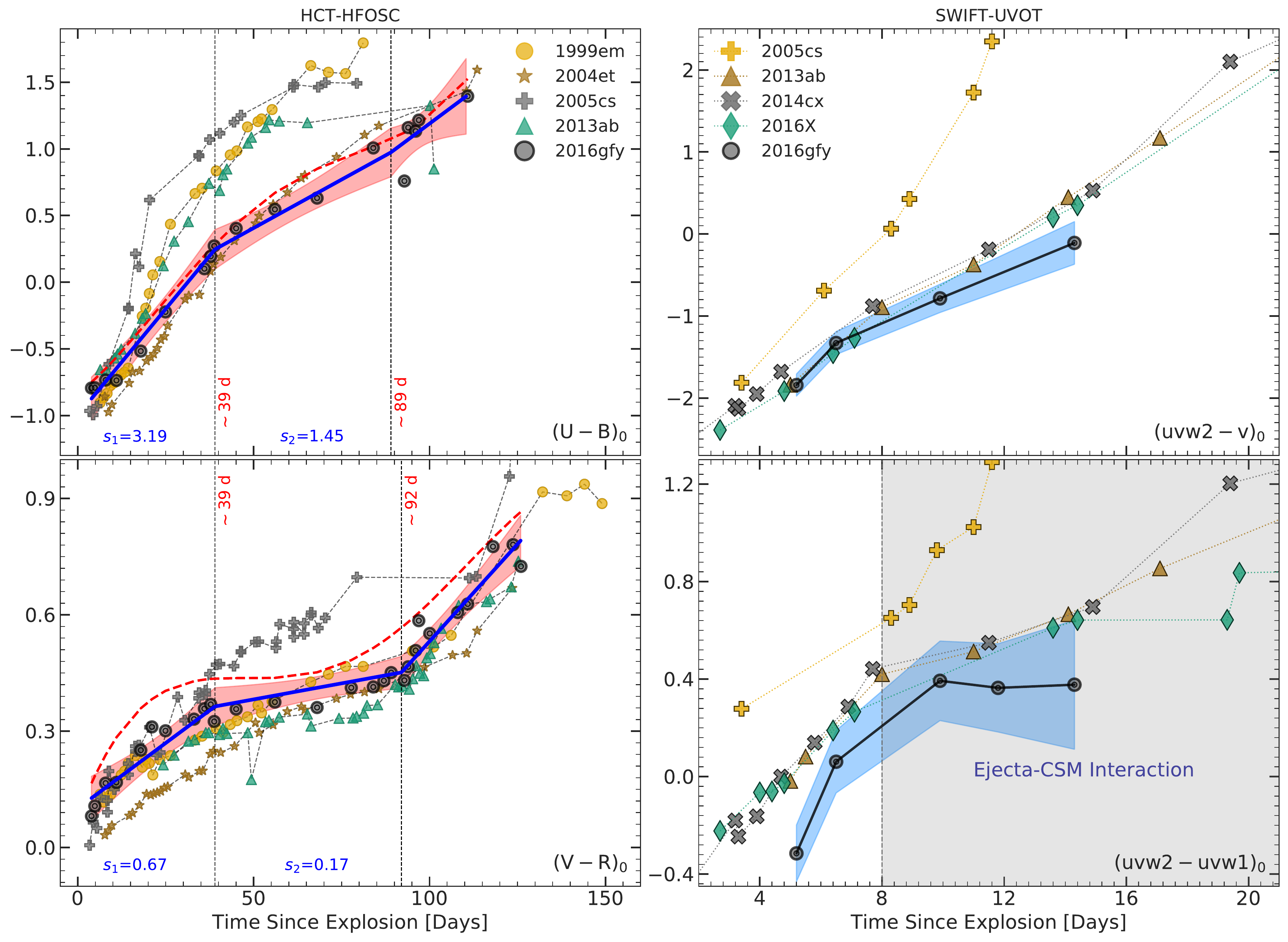}}
\caption{Intrinsic colour evolution of \sniip\ in comparison with other Type II SNe. The fit to the $U-B$ and $V-R$ colours with a linear piece-wise polynomial is shown with a $solid\ blue$ line and the slopes, $s_1$ and $s_2$ (see \citealp{2018adejaeger}) are inferred from the fit indicated in $blue$. $Dashed$ vertical lines mark the epoch of transition in decline rates and 3$\rm \sigma$ confidence interval of the fits are shaded in $pink$. The fits to the Galactic extinction corrected colour with the Legendre polynomial are shown in $red$. The 1$\rm \sigma$ uncertainty for the SWIFT colours are shaded in $light\ blue$.}
\label{fig:colour}
\end{figure*}

\begin{figure}
\centering
\resizebox{\hsize}{!}{\includegraphics{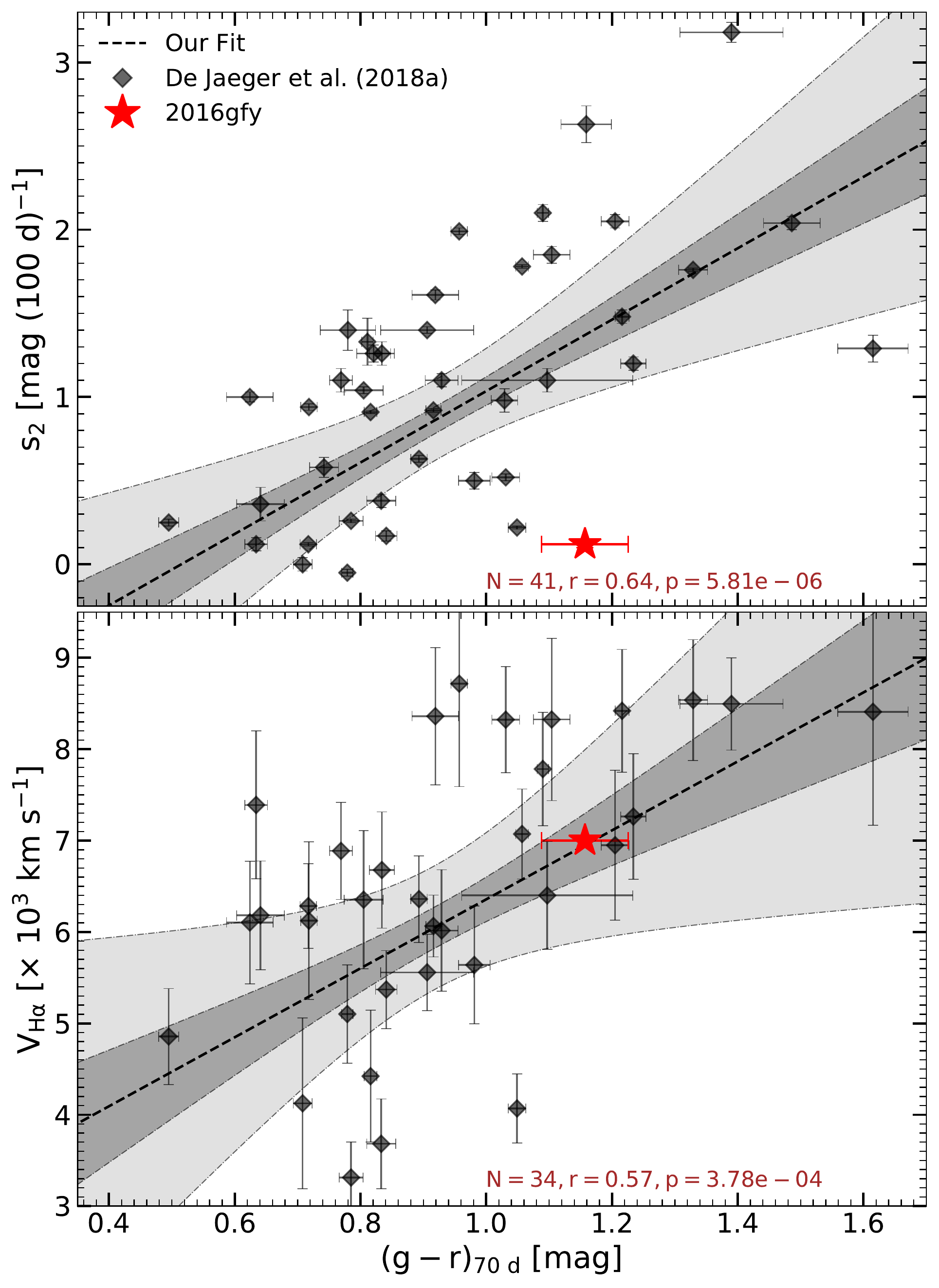}}
\caption{Correlation of $s_{2,(B-V)}$ and $\rm H\alpha$ velocity at 70 d with $\rm(g-r)_{70\ d}$ indicated by \citet{2018adejaeger}. 1$\rm \sigma$ and 3$\rm \sigma$ confidence interval of the fits are shaded in $dark-grey$ and $light-grey$, respectively.}
\label{fig:colourcomp}
\end{figure}

The first glimpse of light in Type II SNe is seen shortly after the shock breakout from the stellar surface \citep{1974colgate, 1977falk}. The flux in the early phase is governed by the rapid cooling of the SN ejecta and its expansion. To investigate the rise time, the shock breakout formulation from \citet{2007waxman} was used as it approximates the ejecta as a blackbody emitting at a fixed wavelength with a dependence on the SN radius, $r$\,$\propto$\,$(t-t_0)^{0.8}$ and shock breakout temperature, $T$\,$\propto$\,$(t-t_{0})^{0.5}$, where $t_0$ is the explosion epoch and $(t-t_0)$ denotes the time since the explosion epoch. The time-dependent diffusion relation from \citet{1982arnett} was used to account for the expansion phase, which represents the SN photosphere as a constant temperature blackbody expanding with a constant velocity and has a $(t-t_0)^2$ dependence. Together,

\begin{equation}
\label{eqn:earlyrise}
    f(t) = \underbrace{\frac{a_1}{e^{a_2(t - t_{0})^{0.5}} - 1} (t - t_0)^{1.6}}_\text{\citet{2007waxman}} + \underbrace{a_3(t - t_0)^2,}_\text{\citet{1982arnett}}
\end{equation}

where $a_1$, $a_2$ and $a_3$ are free parameters. Trends in the past observational studies have obtained longer rise time for Type II-L SNe (possibly larger radii) than Type II-P SNe \citep{1993blinnikov} as the photons take longer time to diffuse through the ejecta and is in coherence with the hydrodynamical simulations of \citet{1991swartz}. However, the longer rise time seen in Type II-L SNe can also be due to a higher E/M ratio and not necessarily larger radii \citep{2011rabinak}.

The fits to the early time LC in $UBVRI$ bands are shown in Figure~\ref{fig:fitrisetime}. A mean explosion epoch ($t_0^{Mean}$) of JD 2457641.4\,$\pm$\,0.9 is estimated for \sniip\ from the functional fit in $UBVRI$ bands. The rise times inferred are mentioned in Table~\ref{tab:fitrisetime} and are intermediate to those of Type II-P (7.0\,$\pm$\,0.3 d) and Type II-L (13.3\,$\pm$\,0.6 d) SNe \citep{2015gall}. The increase in rise time with wavelength seen for \sniip\ is consistent with the inference of \citet{2015gonzalez}. For comparison, the rise times in $UBVRI$ match the quintessential Type II-P SN~1999em in $UBV$ \citep[6, 8 and 10 d,][]{2002aleonard}, but are significantly faster when compared with the bright Type II-P SN~2004et in $UBVRI$ \citep[9, 10, 16, 21 and 25 d,][]{2006sahu}. Faster rise times in Type II SNe can be attributed to the presence of an immediate CSM \citep[][see Figure~\ref{fig:modelearlylc}]{2017moriya,2018moriya,2017morozova,2018forster}.

\subsection{Optical light curves}

The $UBVRI$ light curves of \sniip\ span $\sim$\,4--387 d from the date of explosion and are shown in the left\ panel of Figure~\ref{fig:apparentlc}. The $VRI$ LCs of \sniip\ show four visually distinguishable phases: the rising phase ($\sim$\,10 d), the plateau phase ($\sim$\,10--90 d), the transition phase ($\sim$\,90--115 d) and the nebular phase ($>$115 d). The rise to the maximum seen in $UBVRI$ bands has been used in estimating the date of explosion in Section~\ref{sec:risetime}. The $B$-band magnitude declines by $\sim$\,1.8 mag in 100 d which is well within the value quoted (i.e. $\beta^{B}_{100}$\,$<$\,3.5 mag) for Type II-P SNe by \citet{1994patat}. The late-plateau decline rates in $UBVRI$ for \sniip\ are 2.94, 1.15, 0.12, -0.01 and -0.27 \maghundred. The $V$-band decline rate for \sniip\ is much lower in comparison to the luminous Type II-P SNe like SN~2013ab (0.92, \citealp{2015boseab}), SN~2013ej (1.53, \citealp{2015boseej}) and ASASSN-14dq (0.96, \citealp{2018avinash}), and is shown in Figure~\ref{fig:vabslc}.

\subsection{\textit{Swift} UVOT light curves}\label{sec:swiftuvlc}

$Swift$ UVOT LCs shown in Figure~\ref{fig:apparentlc} span the epochs $\sim$\,5--27 d from the date of explosion and show faster decline in bluer bands as is expected for a Type II SN \citep{2007brown,2014pritchard}. The UV bands $uvw2$, $uvm2$ and $uvw1$ do not cover the peak as the observations were triggered more than 2 days from discovery ($>$5 days from explosion). The decline rates in $uvw2$, $uvm2$ and $uvw1$ are 0.21, 0.23 and 0.14 \magday\ and are similar to the decline rates observed in other Type II SNe. The decline rate in $uvm2$ is higher than in $uvw2$, contrary to the general anti-correlation of decline rates with wavelength. The faster $uvm2$ decay results from the higher density of \ion{Fe}{2} lines in the $uvm2$ band-pass which absorbs more effectively as the SN cools (see Figure 5 in \citealp{2007brown}).

\subsection{Bolometric light curve}

The pseudo-bolometric light curve of \sniip\ was generated following the prescription in \citet{2018avinash} and integrating over the wavelength range 3100\,--\,9200 \AA. A comparison of \sniip\ with other Type II SNe is shown in Figure~\ref{fig:bolometriclc}. The peak bolometric luminosity of \sniip\ is $\rm \sim1.8\times10^{42}\ erg\ s^{-1}$. The bolometric luminosity declines at the rate of 1.00 and 0.06 dex $\rm (100\ d)^{-1}$, respectively during the early and late plateau phases, and 0.46 dex $\rm (100\ d)^{-1}$ during the nebular phase. The late-plateau phase shows a moderate bump (increase in flux) which is seen mostly in low-luminosity Type II SNe \citep[e.g. SN~2005cs,][]{2009pastorello}. The behaviour of \sniip\ during the late-plateau phase is discussed in Section~\ref{sec:nimix}. The slow decline rate during the late-plateau phase is also a signature of low-mass progenitors in the modelled explosions of \citet[][hereafter S16]{2016sukhbold}, indicating a low-mass progenitor of \sniip.

\subsection{Colour Evolution}\label{sec:colourevolution}

Evolution of intrinsic colour terms $(U-B)_0$, $(V-R)_0$, $(uvw2-uvw1)_0$ and $(uvw2-v)_0$ of \sniip\ in comparison with other Type II SNe is shown in Figure~\ref{fig:colour}. The $(U-B)_0$ evolution of \sniip\ doesn't follow other Type II SNe (except SN~2004et), however, $(V-R)_0$ evolution shows no such differences. The significantly bluer colour evolution in $U-B$ during the plateau is indicative of lower line blanketing in the blue region, consequently implying lower metallicity of the progenitor (D14) similar to SN~2004et \citep{2012jerkstrand}. Even with a clear observable difference in the $V$-band LC of SN~2013ab and \sniip\ (due to the presence of a late-plateau bump in \sniip), we see insignificant differences in their $V-R$ colour evolution. The $(uvw2-uvw1)_0$ and $(uvw2-v)_0$ colours of \sniip\ also show a bluer evolution in comparison to other Type II SNe. Also, the former colour becomes flatter during the epoch in which signatures of ejecta-CSM interaction are seen in the spectra (see Section~\ref{sec:earlycsm}).

The Bessell colours of \sniip\ were converted to SDSS colours using the transformation equations from \citet{2006jordi} for comparison with the sample of \citet[][hereafter DJ18a]{2018adejaeger}. During the late-plateau phase, DJ18a inferred that the redder Type II SNe display higher $\rm H\alpha$ velocities ($\sim$\,70 d) and, steeper decline rates ($s_{2,V}$). \sniip\ lies outside the 3$\sigma$ dispersion of the latter correlation as seen in Figure~\ref{fig:colourcomp}.


\section{Spectroscopic Evolution}\label{sec:specevol}

\subsection{Early phase (\texorpdfstring{$\rm <$~30 d)}{Lg}}\label{sec:specearlyplat}

\begin{figure*}
\centering
\resizebox{\hsize}{!}{\includegraphics{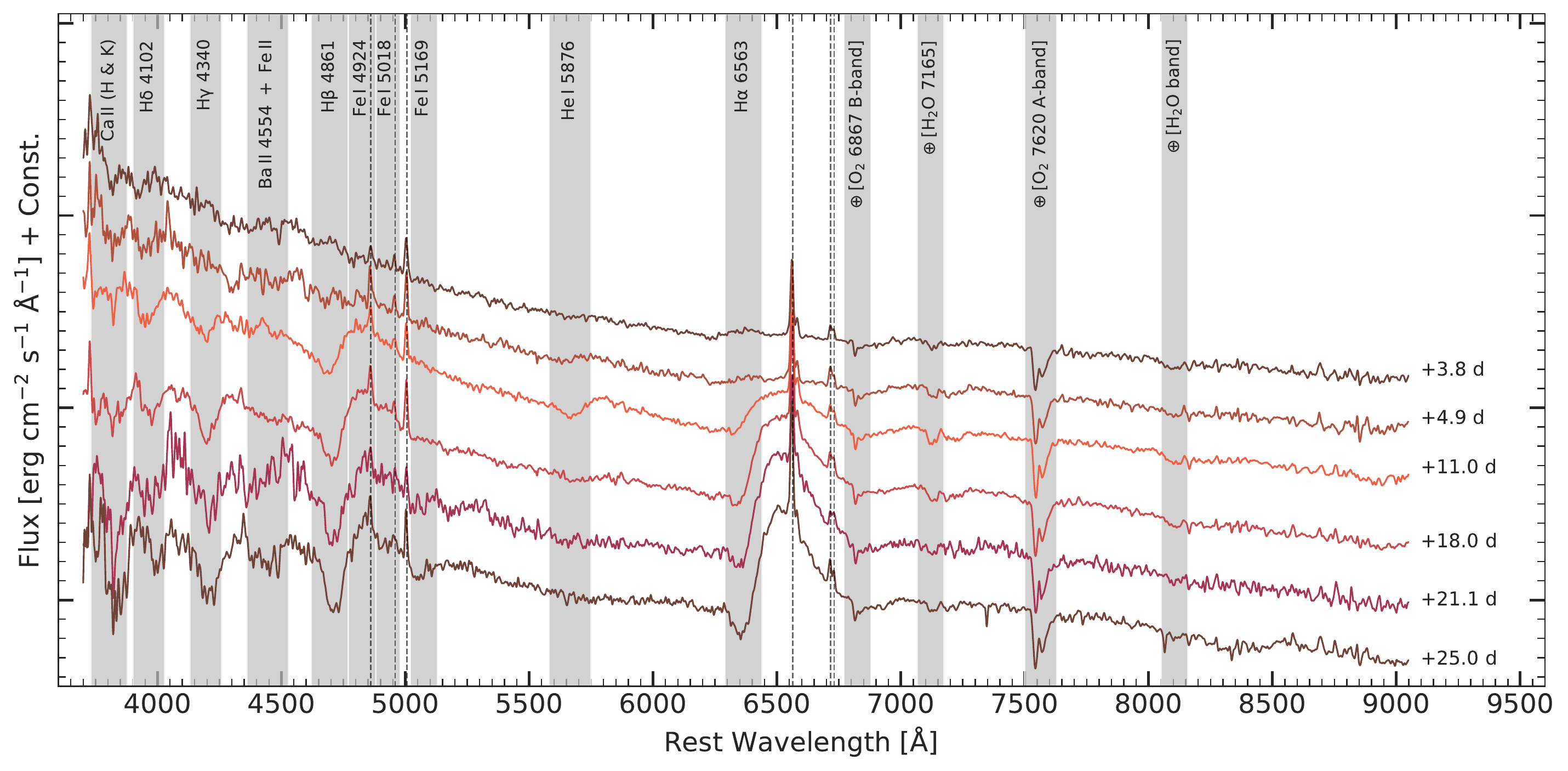}}
\caption{Early phase ($<$\,30 d) of \sniip\ from HCT-HFOSC. The spectra are corrected for the redshift of the host galaxy. Prominent emission lines of H\,$\rm \beta$ 4861 \AA, [\ion{O}{3}] 4959 \AA, [\ion{O}{3}] 5007 \AA, H\,$\rm \alpha$ 6563 \AA, [\ion{N}{2}] 6584 \AA\ and [\ion{S}{2}] 6717, 6731 \AA\ from the parent \ion{H}{2} region are marked with $dashed$ vertical lines.}
\label{fig:specearlyplat}
\end{figure*}

\begin{figure}
\centering
\resizebox{\hsize}{!}{\includegraphics{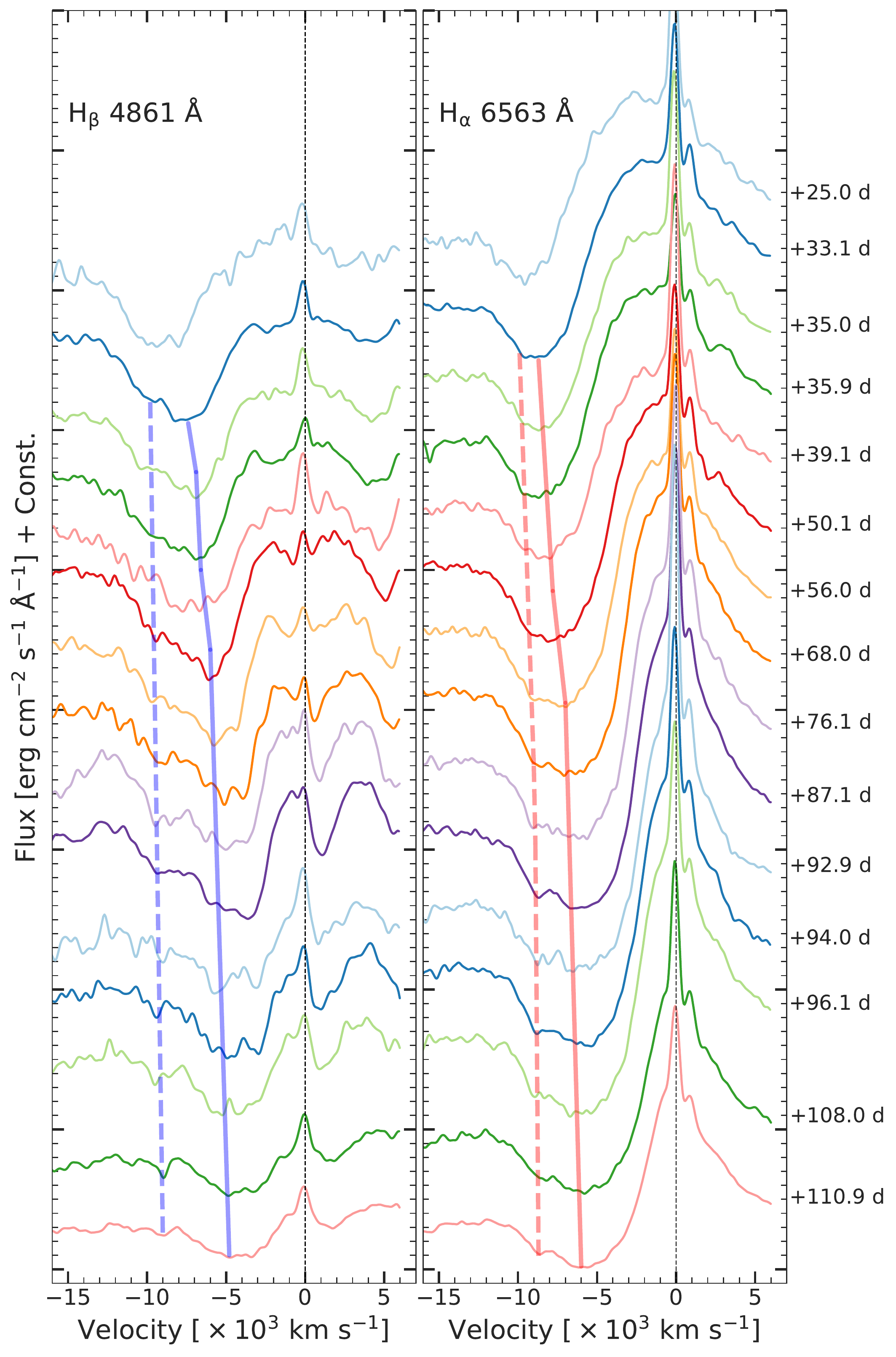}}
\caption{Temporal evolution of the Balmer features, H\,$\rm \alpha$ and H\,$\rm \beta$ in \sniip. The $solid$ lines depict the evolution of the normal velocity component whereas the dashed lines show the evolution of the high-velocity features.}
\label{fig:balmerevol}
\end{figure}

\begin{figure*}
\centering
\resizebox{0.9\hsize}{!}{\includegraphics{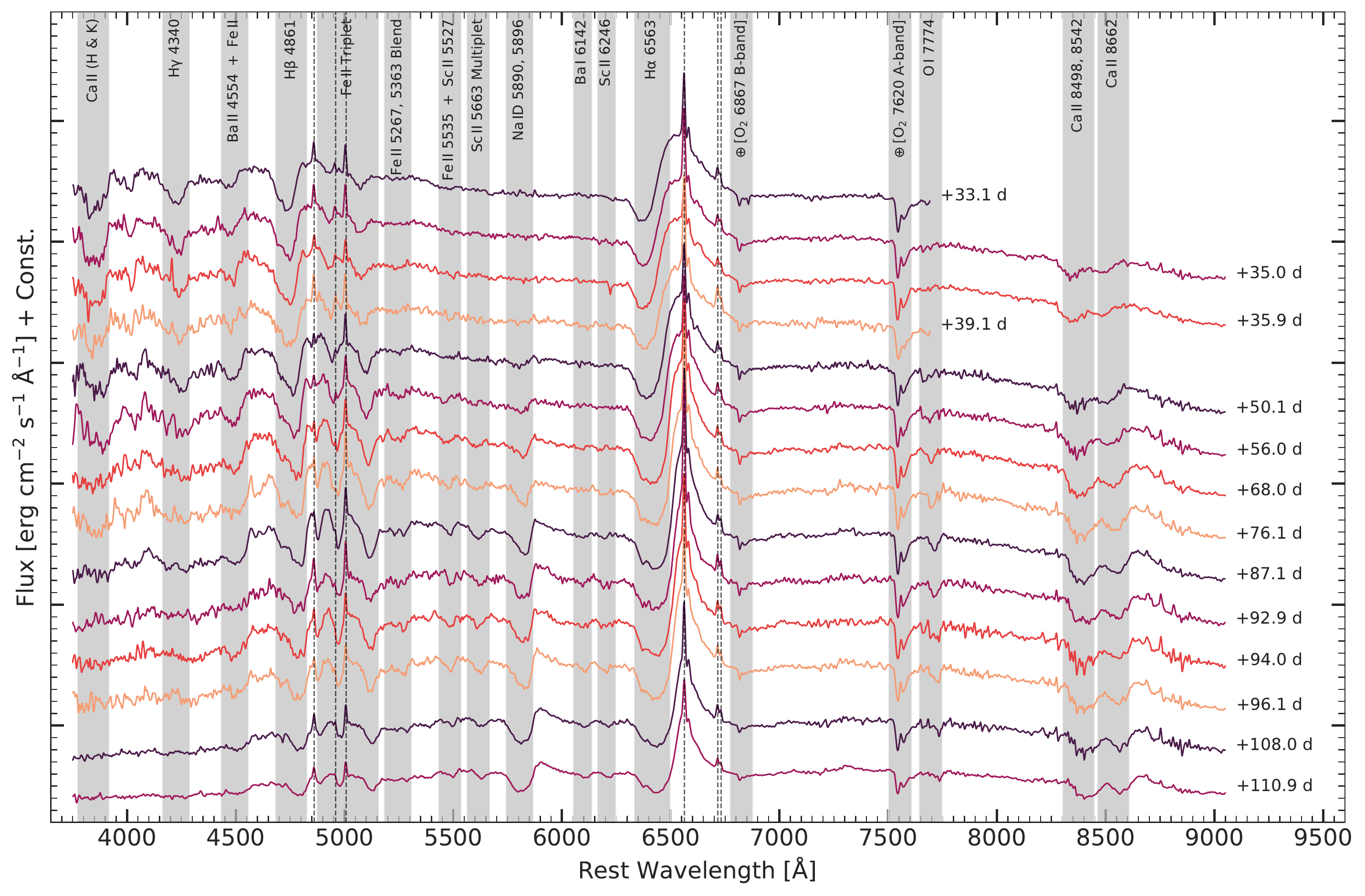}}
\caption{Plateau and transition phase ($<$\,115 d) spectra of \sniip\ from HCT-HFOSC. The plot description is same as Figure~\ref{fig:specearlyplat}.}
\label{fig:speclateplat}
\end{figure*}

\begin{figure*}
\centering
\resizebox{0.9\hsize}{!}{\includegraphics{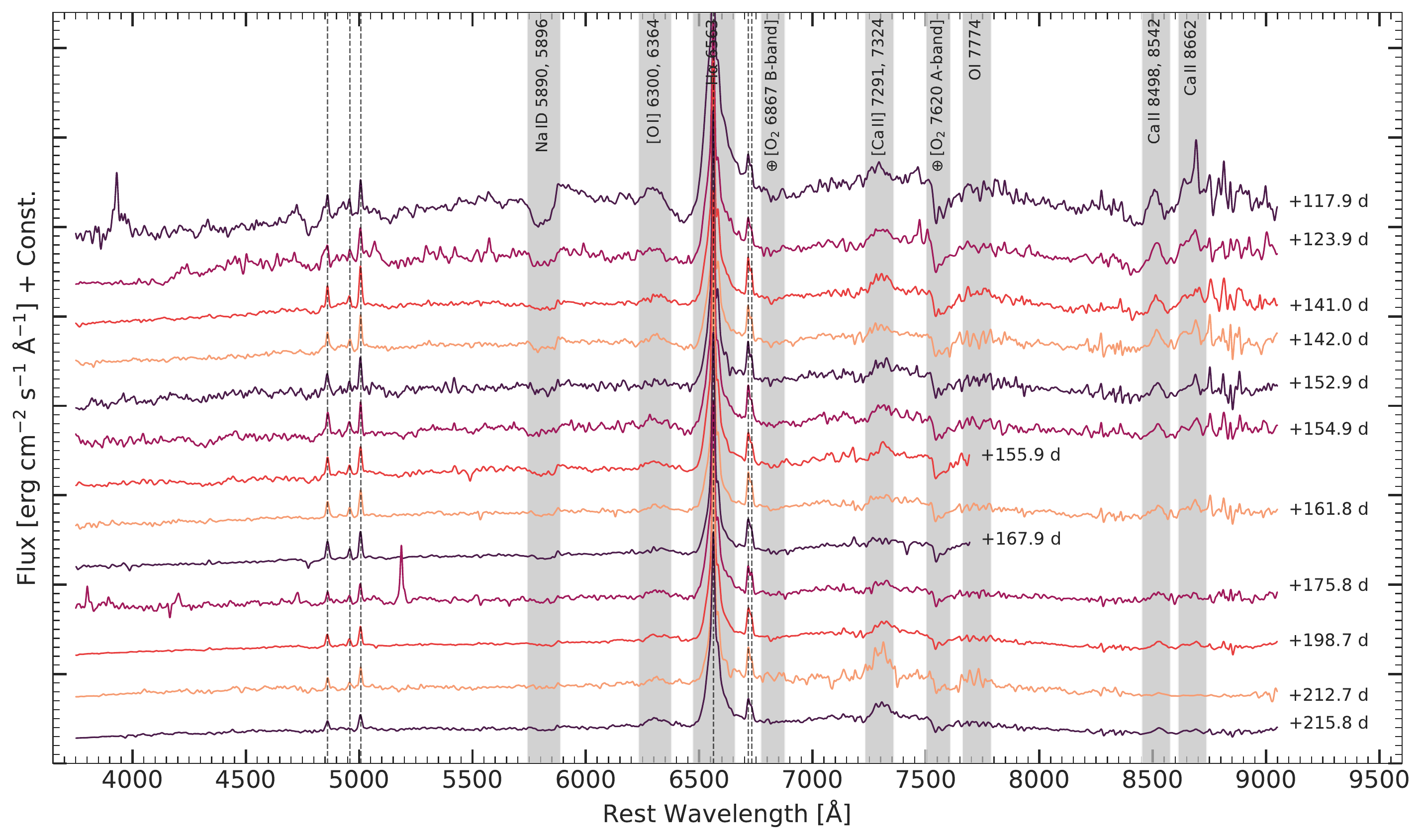}}
\caption{Nebular phase ($>$\,115 d) spectra of \sniip\ from HCT-HFOSC. The spectra are emission dominated with a flat continuum. The plot description is same as Figure~\ref{fig:specearlyplat}.}
\label{fig:specnebular}
\end{figure*}

\begin{figure}
\centering
\resizebox{\hsize}{!}{\includegraphics{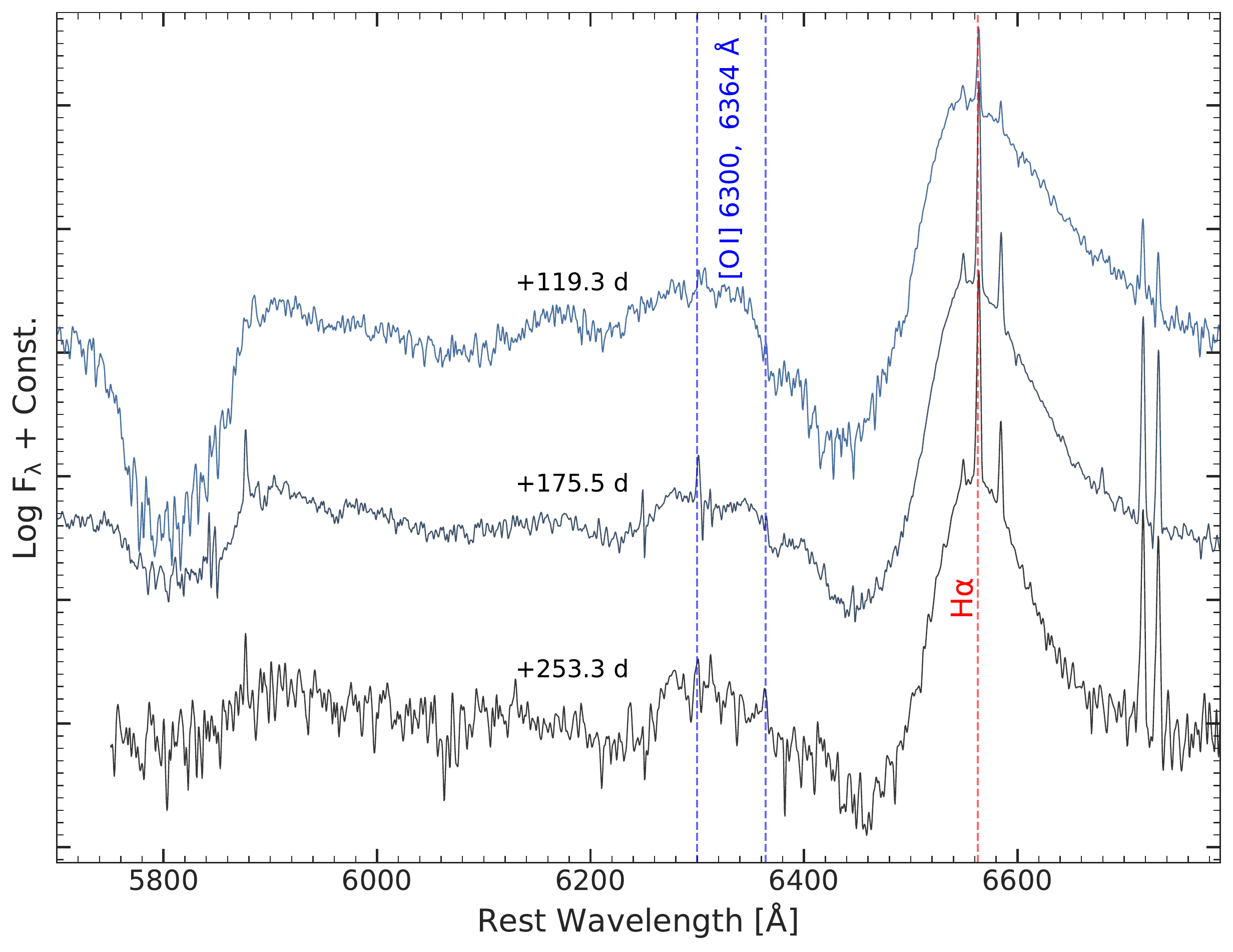}}
\caption{Medium-resolution nebular spectra of \sniip\ from the BC spectrograph mounted on the 6.5\,m MMT. The rest-wavelengths of H\,$\rm \alpha$ and [\ion{O}{1}] doublet are indicated with $dashed$ lines.}
\label{fig:specnebmmt}
\end{figure}

Figure~\ref{fig:specearlyplat} shows the early phase spectral evolution of \sniip. The first spectrum ($\sim$\,4 d) shows narrow emission lines superposed over a blue continuum. However, the narrow features possibly owe their origin to the parent \ion{H}{2} region as they do not cease to exist in the later epochs. No signatures of \ion{He}{2} emission lines are seen in the very early spectra ($<$\,5 d) of \sniip\ which results from a strong progenitor wind \citep{2016khazov}. A boxy shape of H\,$\rm \alpha$ is seen in the spectrum between $\sim$\,11--21 d (see discussion in Section~\ref{sec:earlycsm}) implying an interaction of the ejecta with the CSM \citep{2010andrews,2011inserra,2018bdejaeger}.

The spectrum of $\sim$\,11 d shows the emergence of P-Cygni Balmer features (H\,$\rm \alpha$, H\,$\rm \beta$, H\,$\rm \gamma$ and H\,$\rm \delta$) along with \ion{He}{1} $\rm \lambda$5876 (which is a result of the high ejecta temperature) and the \ion{Ca}{2} H\&K doublet. The \ion{He}{1} $\rm \lambda$5876 feature fades away in the spectra beyond $\sim$\,18 d as the ejecta temperature drops with time. The signature of \ion{Fe}{2} $\rm \lambda$5169 is seen in the spectrum of $\sim$\,21 d and becomes prominent around $\sim$\,25 d, which also marks the emergence of \ion{Fe}{2} $\rm \lambda \lambda$4924, 5018. The \ion{Ba}{2} $\rm \lambda$4554 blend is evident in the spectrum of $\sim$\,21 d. It is important to note here that the emission features of H\,$\rm \alpha$ and H\,$\rm \beta$ are contaminated by narrow emission lines from the parent \ion{H}{2} region throughout the evolution of the SN.

\subsection{Plateau phase}\label{sec:speclateplat}

The temporal evolution of Balmer features H\,$\rm \alpha$ and H\,$\rm \beta$ is shown in Figure~\ref{fig:balmerevol}. The P-Cygni absorption troughs of both the features show a distinct peculiar notch \citep[referred as \lq \lq Cachito\rq \rq\ in][]{2017gutierrez} starting $\sim$\,33 d. The solid line in the figure depicts the evolution of the normal velocity component (NC), which follows a power-law decline trend in velocities. However, the dashed line depicts a slow temporal evolution (1000 \kms decline in $\sim$\,80 d) of the \lq \lq Cachito\rq \rq, which could either be \ion{Si}{2} 6355 \AA\ or a high-velocity (HV) feature of hydrogen \citep{2017gutierrez}. The presence of a H\,$\rm \beta$ counterpart to the H\,$\rm \alpha$ Cachito at a similar velocity ($\sim$\,9500 \kms) strengthens its presence as a HV feature of hydrogen.

Theoretical investigation in \citet{2007chugai} argued that the enhanced excitation of the outer ejecta can result in a \lq \lq Cachito\rq \rq\ near H\,$\rm \alpha$, but they denied the presence of a H\,$\rm \beta$ \lq \lq Cachito\rq \rq\ due to its low optical depth. This is in contrast with 63\% of the Type II SNe in the sample study of \citet{2017gutierrez} that displayed \lq \lq Cachito\rq \rq\ near both the balmer features. However, \citet{2007chugai} also suggested that the \lq \lq Cachito\rq \rq\ can form behind the reverse shock, in the cold dense shell (CDS). This advocates that the HV features were produced from the interaction of the ejecta with the Red Supergiant (RSG) wind (see Section~\ref{sec:earlycsm}). 

The \ion{Fe}{2} $\rm \lambda \lambda \lambda$4924,\,5018,\,5169 triplet strengthens as the photosphere of the SN traverses deep inside the ejecta. A weak imprint of \ion{Na}{1}\,D from the SN appears in the spectrum of $\sim$\,39 d and becomes clearly discernible in the spectrum of $\sim$\,50 d as seen in Figure~\ref{fig:speclateplat}. The metal features of \ion{Fe}{2} $\rm \lambda \lambda$5267, 5363, \ion{Sc}{2} $\rm \lambda$5663 multiplet, \ion{Ba}{1} $\rm \lambda$6142 and \ion{Sc}{2} $\rm \lambda$6246 can be clearly sighted in the spectra past $\sim$\,50 d. \ion{O}{1} $\rm \lambda$7774 is seen during the plateau phase but becomes increasingly fainter as the SN enters the transition phase. A hint of [\ion{O}{1}] $\rm \lambda \lambda$6300, 6364 and [\ion{Ca}{2}] $\rm \lambda \lambda$7291, 7324 can be spotted at in the transition phase ($\sim$\,110 d).

\subsection{Nebular phase (\texorpdfstring{$\rm >$~115 d)}{Lg}}\label{sec:specnebular}

The nebular phase of a SN unmasks the progenitor structure as the outer ejecta becomes optically thin. The low-resolution spectra of \sniip\ from HCT during this phase are shown in Figure~\ref{fig:specnebular} and the medium-resolution spectra from MMT in Figure~\ref{fig:specnebmmt}. The spectrum of $\sim$\,118 d depicts a flat continuum and emission-dominated spectral features of \ion{Na}{1}\,D, [\ion{O}{1}] $\rm \lambda \lambda$6300,\,6364, H\,$\rm \alpha$, [\ion{Ca}{2}] $\rm \lambda \lambda$7291,\,7324 and the \ion{Ca}{2} $\rm \lambda \lambda \lambda$8498,\,8542,\,8662 NIR triplet.\\

The width of narrow H\,$\rm \alpha$ from the parent \ion{H}{2} region present in our medium-resolution spectra taken with MMT at $\sim$\,119, 175 and 253 d shows no indication of broadening and stays at the resolution of the instrument $\sim$\,2 \AA\ (see Figure~\ref{fig:specnebmmt}). This strengthens the proposition that the emission is purely from the host \ion{H}{2} region and shows no definitive signature of CSM interaction during this phase in \sniip. Further, the broad H\,$\rm \alpha$ feature in the spectrum of $\sim$\,253 d is symmetric and does not show any peculiarity. However, this does not indicate an absence of CSM (see Section\ref{sec:earlycsm}).


\section{Physical Parameters of \sniip}\label{sec:physparam}

\subsection{Ejected \texorpdfstring{$\rm ^{56}Ni$ Mass}{Lg}}\label{sec:calcnickel}

Radioactive $\rm ^{56}Ni$ is produced in the explosive nucleo-synthesis of Si and O in CCSNe \citep{1980arnett}. The radioactive decay of $\rm ^{56}Ni \rightarrow\ ^{56}Co \rightarrow\ ^{56}Fe$ thermalizes the SN ejecta and powers the late-phase (nebular) light curve in Type II SNe through the emission of $\gamma$-rays and positrons. The ejected $\rm ^{56}Ni$ mass for \sniip\ is estimated in the ensuing subsections.

\subsubsection{Estimate from the tail bolometric luminosity}

\citet{2003hamuy}, in his study of 24 Type II-P SNe postulated a relation between the nebular-phase bolometric luminosity and the nickel mass synthesized in the explosion assuming that the $\gamma$-rays released in the radioactive decay completely thermalizes the SN ejecta. The mean tail luminosity, $L_t$ of \sniip\ computed over 4 epochs ($\sim$\,199--241 d) with a mean phase of $\sim\,$223 d is $\rm 5.6\pm1.0\times 10^{40} erg\ s^{-1}$ and yields a $\rm ^{56}Ni$ mass of $\rm 0.031\pm0.006\ M_{\odot}$.

\subsubsection{Comparison with SN~1987A light curve}

$\rm ^{56}Ni$ mass can also be procured from the fact that highly energetic explosions yield more $\rm ^{56}Ni$ \citep{2003hamuy} under the assumption that the $\rm \gamma$-ray deposition is similar for the SNe in comparison. \citet{1998turatto} obtained a $\rm ^{56}Ni$ mass estimate of 0.075\,$\pm$\,0.005 $\rm M_{\odot}$ for SN~1987A. A $\rm ^{56}Ni$ mass of $\rm 0.033\pm 0.008\ M_{\odot}$ is estimated after comparing its bolometric luminosity with SN~1987A at $\sim$\,241 d.

\subsubsection{Fitting late-phase light curve}\label{sec:fitnilc}

In Section~\ref{sec:lightcurve}, a decline rate marginally higher ($\sim$\,1.00 \maghundred) than the radioactive decay rate of $\rm ^{56}Co$ (i.e. 0.98 \maghundred) with complete $\gamma$-ray trapping, was determined from the $V$-band light curve of \sniip. To account for the $\gamma$-ray leakage, Equation 3 from \citep{2016yuan} was fit to the late phase light curve beyond 140 d, where $t_c$ is the characteristic time-scale for the optical depth of $\gamma$-rays to become one. A $\rm ^{56}Ni$ mass of 0.031\,$\pm$\,0.006\ $\rm M_{\odot}$ and a $t_c$ of $\sim$\,486 d was obtained for \sniip. The high value of $t_c$ here is similar to that of SN~1987A ($\sim$\,530 d) and signifies insignificant $\gamma$-ray leakage in \sniip. 

\subsubsection{Correlation with \lq \lq Steepness parameter\rq \rq}

An empirical relation between the $V$-band decline rate during the transitional phase (\lq \lq steepness parameter\rq \rq, $S$\,=\,--$dM_V/dt$) and the ejected $\rm ^{56}Ni$ mass was reported by \citet{2003elmhamdi} using 10 Type II SNe, which was later improved upon in \citet{2018avinash} using a sample of 39 Type II SNe. A steepness parameter of 0.121 was determined for \sniip\ using its $V$-band light curve, which yields a $\rm ^{56}Ni$ mass of 0.036\,$\pm$\,0.004\ $\rm M_{\odot}$ using the relation from \citet{2018avinash}.

The slightly higher $\rm ^{56}Ni$ mass estimate obtained using the steepness parameter in comparison with other techniques is an indication that the LC of \sniip\ exhibits a non-negligible degree of $\rm ^{56}Ni$-mixing, which can decrease the steepness during the transition phase \citep{2019kozyreva}.

\subsubsection{Mean estimate of \texorpdfstring{$^{56}Ni$}{Lg} mass}

All the above methods return a mean ejected $\rm ^{56}Ni$ mass of $\rm 0.033\pm0.003\ M_{\odot}$ for \sniip. The use of the pseudo-bolometric ($UBVRI$) light curve in determining the $\rm ^{56}Ni$ mass, makes the inferred estimate a lower limit for the $\rm ^{56}Ni$ synthesized in \sniip.


\subsection{Ejecta Velocity}

\begin{figure}
\centering
\resizebox{\hsize}{!}{\includegraphics{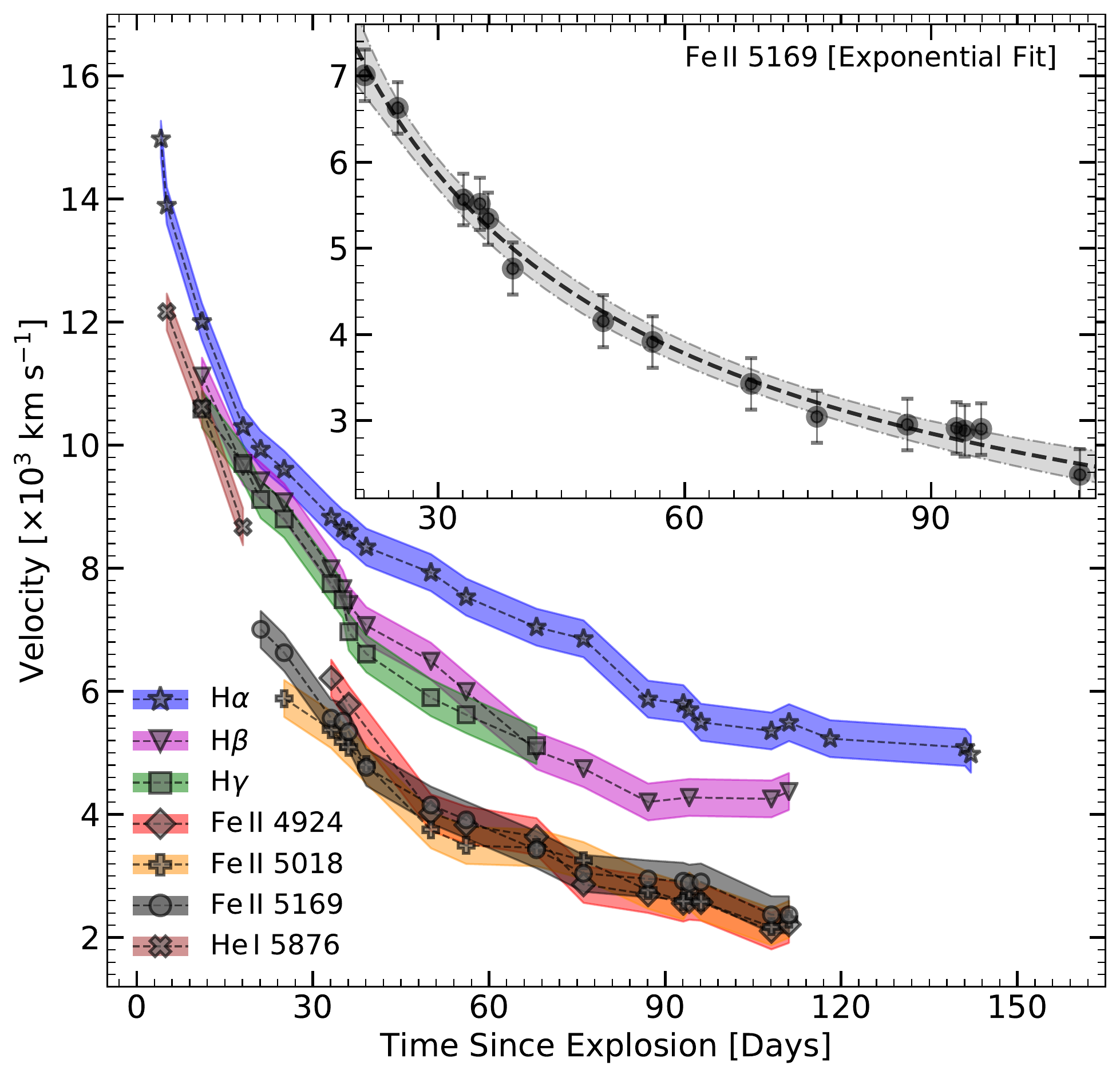}}
\caption{Line velocity evolution of spectral features in \sniip: H\,$\rm \alpha$, H\,$\rm \beta$, H\,$\rm \gamma$, \ion{He}{1} 5876 \AA\ and the \ion{Fe}{2} triplet. The velocities were determined from the blue-shifted absorption minima of the P-Cygni profile.}
\label{fig:photvel}
\end{figure}

\begin{figure}
\centering
\resizebox{\hsize}{!}{\includegraphics{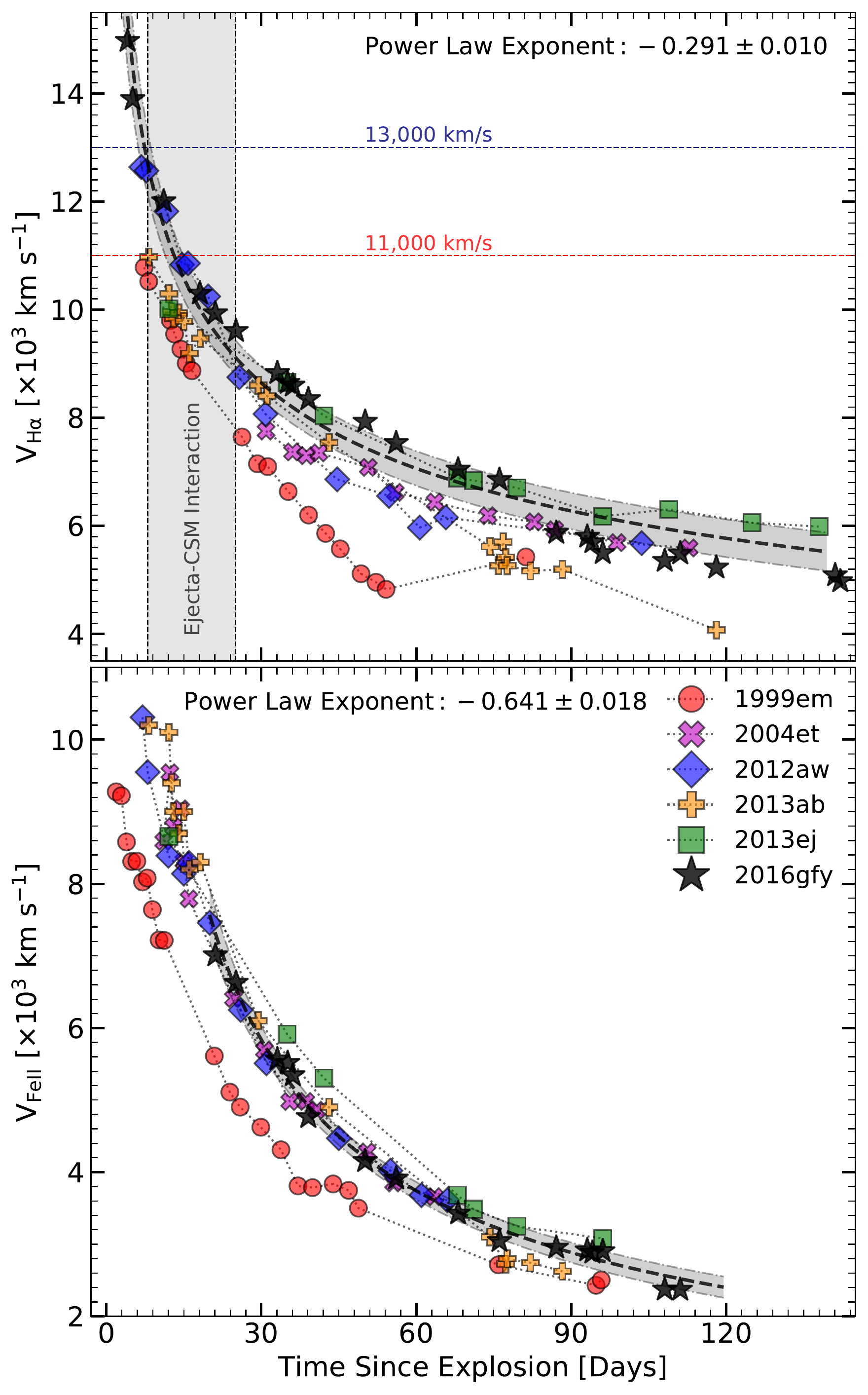}}
\caption{Comparison of line velocity evolution of \sniip\ with Type II SNe from the literature. $Top$ and $bottom$ panel displays the velocity evolution of H\,$\rm \alpha$ and \ion{Fe}{2} $\lambda$5169, respectively. The data adopted for comparison is from \citet{2014bose} and references therein.}
\label{fig:compphotvel}
\end{figure}

Progenitors of CCSNe have an \lq \lq onion-ring\rq \rq\ structure of elements, whose spectral features show up at varied velocities in a SN as they originate at different heights (and time). The line velocities were inferred from the blue-shifted minima of the P-Cygni absorption profiles in the redshift corrected spectra. The velocity evolution of H\,$\rm \alpha$, H\,$\rm \beta$, H\,$\rm \gamma$, \ion{He}{1} $\rm \lambda$5876, \ion{Fe}{2} $\rm \lambda \lambda \lambda$ 4924,\,5018,\,5169 is shown in Figure~\ref{fig:photvel}. The Balmer lines show faster velocities as their integrated extent of line formation has a higher radii than the radius of the photosphere \citep[optical depth $\sim2/3$,][]{2002aleonard}. During the plateau phase, the velocities computed from the \ion{Fe}{2} acts as a good proxy for the photospheric velocity \citep{2005adessart}.

The line velocity in Type II SNe is known to decrease as a power law \citep{2001hamuy}. A power law fit returned exponents of --0.291\,$\pm$\,0.010 and --0.641\,$\pm$\,0.018 for the H\,$\rm \alpha$ and \ion{Fe}{2} $\lambda$5169 features in \sniip, respectively. The comparison of line velocity evolution of \sniip\ with other Type II SNe along with the power law fits is shown in Figure~\ref{fig:compphotvel}. The H\,$\rm \alpha$ velocity evolution matches the bright Type II SN~2013ej \citep{2015boseej} and stays faster compared to other Type II SNe. The \ion{Fe}{2} $\lambda$5169 evolution is similar to the average value derived from samples of Type II-P SNe (--0.581\,$\pm$\,0.034) in \citet{2014afaran} and Type II SNe (--0.55\,$\pm$\,0.20) in \citet{2015dejaeger}. In the case of \sniip, the velocities measured during the mid-plateau phase ($\sim$\,50 d) are $\sim$\,7900 \kms\ and $\sim$\,4150 \kms\ for H\,$\rm \alpha$ and \ion{Fe}{2} $\lambda$5169, respectively. This is faster than the mean values of $\sim$\,6500 \kms\ and $\sim$\,3500 \kms\ inferred for Type II SNe from the sample of \citet{2017gutierrez}.

\subsection{Temperature and Radius Evolution}

\begin{figure}
\centering
\resizebox{\hsize}{!}{\includegraphics{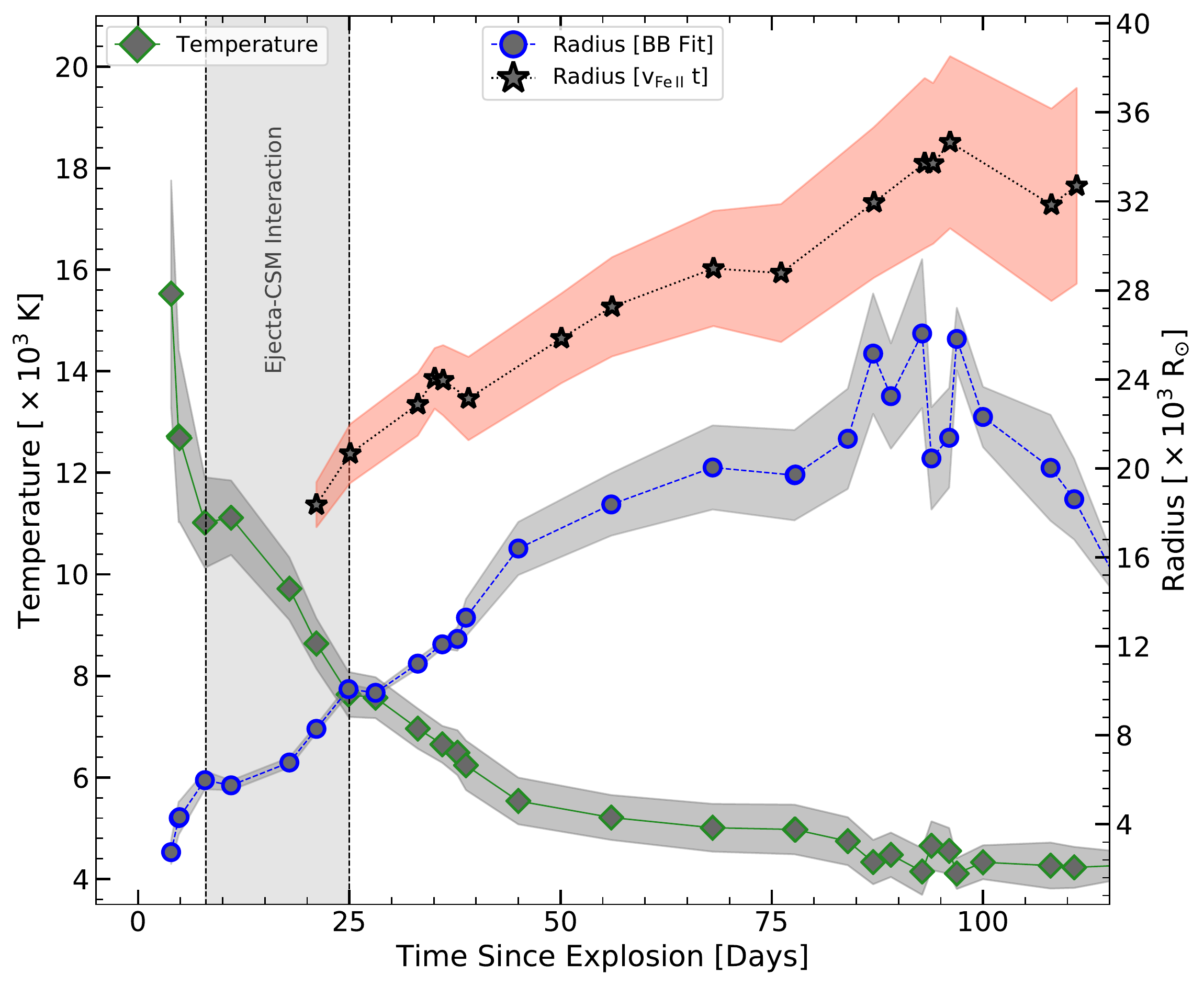}}
\caption{Temperature evolution of \sniip\ estimated from blackbody fits to the $UBVRI$ fluxes. Radius is calculated using Stefan-Boltzmann law (R\,=\,$\rm \sqrt{L / 4\pi \sigma T^{4}}$).}
\label{fig:tempradevol}
\end{figure}

The evolution of the observed color temperature ($\rm T_c$) of \sniip\ is estimated with a blackbody fit to Spectral Energy Distribution (SED) constructed using the extinction corrected $UBVRI$ fluxes and is shown in Figure~\ref{fig:tempradevol}. The radius is estimated from the Stefan-Boltzmann law for a spherical blackbody, i.e. R\,=\,($\rm L / 4\pi \sigma T^{4})^{0.5}$. The temperature drops swiftly in the first 40 days and is ascribable to the rapid cooling of the expanding envelope. Also plotted is the radius of the line-forming region estimated from the velocity evolution of the \ion{Fe}{2} $\rm \lambda$5169 feature. This radius is similar to the photospheric radii in Type II SNe inferred from the blackbody fits to the SED within an order-of-magnitude \citep{2005adessart,2017arcavi}, as is seen in the case of \sniip.


\subsection{Progenitor properties}\label{sec:progenitorprop}

To understand the relation of the observable parameters to the progenitor properties, \citet[][hereafter LN85]{1985litvinova} performed hydrodynamical modeling on a grid of Type II SNe. Using their empirical relations with a plateau length, $t_p$\,=\,90\,$\pm$\,5 d, mid-plateau photospheric velocity, $v_{ph}$\,=\,4272\,$\pm$\,53 \kms\ and a mid-plateau $V$-band absolute magnitude, $M_V^{50}$\,=\,--16.74\,$\pm$\,0.22\,mag, a progenitor radius of 310\,$\pm$\,70 $\rm R_{\odot}$, an explosion energy of 0.90\,$\pm$\,0.15 foe (1 foe = $\rm 10^{51}$ erg) and an ejecta mass of 13.2\,$\pm$\,1.2 $\rm M_{\odot}$ is inferred for \sniip.

Approximate physical properties of Type II SNe and the progenitor parameters can also be obtained from the semi-analytical formulation of \citet{2014nagy}. Their revised two-component framework \citep[][hereafter N16]{2016nagy} comprising of a dense inner core with an extended massive outer envelope was used to model the light curve of \sniip. The late-plateau bump in \sniip\ was not reproduced by the two-component fit (see Figure~\ref{fig:bolometriclc}). However, the early-plateau phase, transition phase and the nebular phase were reproduced well. A radius of $\rm \sim\,350\ R_{\odot}$, an ejecta mass of 11.5 $\rm M_{\odot}$ and an explosion energy of 1.4 foe were estimated for \sniip\ from the best fit.

Using the characteristic time-scale $t_c$, estimated in Section~\ref{sec:fitnilc}, a uniform density profile, $\gamma$-ray opacity of 0.033 $\rm cm^2 g^{-1}$ and a kinetic energy of 0.9 foe (from LN85) for the ejecta, an ejecta mass of $\sim$\,13.0 $\rm M_{\odot}$ is inferred for \sniip\ utilizing the diffusion equation from \citet[][Eqn. 3]{2016terreran}.

\begin{figure}
\centering
\resizebox{\hsize}{!}{\includegraphics{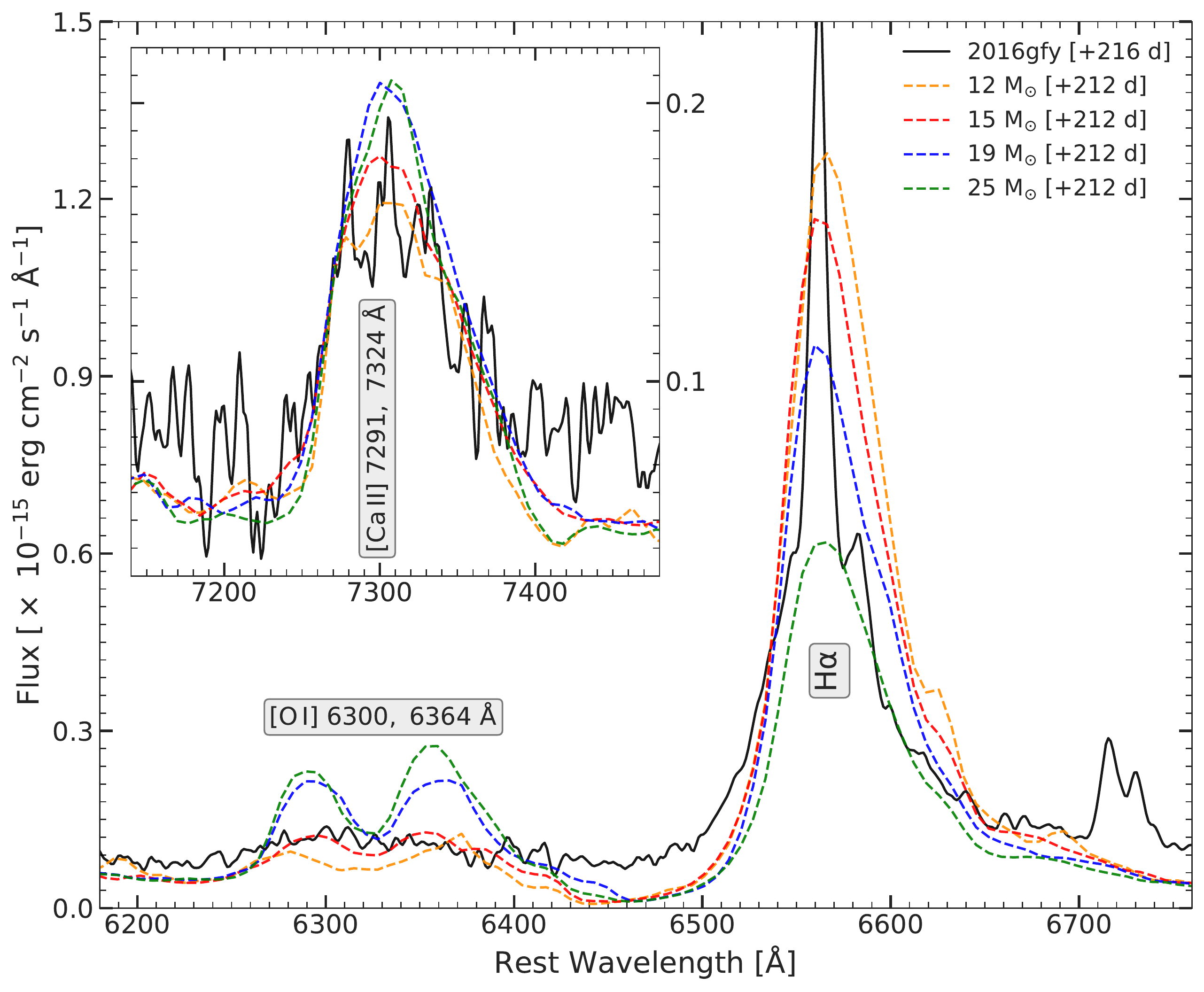}}
\caption{Comparison of the nebular spectra ($\sim$\,216 d) of \sniip\ with synthetic spectra from \citet{2014jerkstrand} at $\sim$\,212 d past the explosion.}
\label{fig:compjerkspecha}
\end{figure}

The mass of the progenitor was also constrained using the nebular phase spectra ($>$\,150 d) as the SN ejecta becomes transparent, revealing the dense inner core. The intensities of prominent emission lines during this phase help constrain the elemental abundances and hence indicate the ZAMS mass of the progenitor \citep[][hereafter J14]{2014jerkstrand}. The [\ion{Ca}{2}]/[\ion{O}{1}] flux ratio in the late nebular phase ($>$\,200 d) remains constant because the mass of calcium produced in the explosion is insensitive to the progenitor mass whereas the oxygen mass depends on it \citep{1989fransson}. A higher-mass progenitor has a stronger [\ion{O}{1}] $\rm \lambda \lambda$6300, 6364 feature in comparison with H\,$\rm \alpha$ and [\ion{Ca}{2}] $\rm \lambda \lambda$7291, 7324. Hence, the [\ion{Ca}{2}]/[\ion{O}{1}] flux ratio of $\sim$\,1.2 in the spectrum of $\sim$\,216 d indicates a low-mass progenitor of \sniip\ \citep{2010maguire,2013sahu}.

Also, the strength of the [\ion{O}{1}] $\rm \lambda \lambda$6300, 6364 emission feature in the nebular phase is relatively insensitive to the explosive nucleosynthesis in a SN and exhibits the progenitor's oxygen abundance, which tightly correlates with the $\rm M_{ZAMS}$ of the progenitor \citep{1995woosley}. In order to perform an accurate comparison, the modelled spectra from J14 were scaled to the estimated distance (see Section~\ref{sec:scm}) and the amount of $\rm ^{56}Ni$ synthesized (see Section~\ref{sec:calcnickel}) for \sniip. The spectra were also corrected for differences in $\gamma$-ray leakage across the models and our spectrum using the following equation:

\begin{equation}
    F_{obs} = F_{decay} * (1 - e^{-t_c^2/t^2})
\label{eqn:fitnilc}
\end{equation}

where, $t_c$ is the characteristic time-scale (see Section~\ref{sec:fitnilc}), $\rm F_{decay}$ is the flux from the radioactive decay and $\rm F_{obs}$ is the observed flux. However, phase correction was not applied as the observed and the synthetic spectrum were just separated by $\sim$\,4 d. The modelled nebular spectra from J14 for progenitors of masses 12, 15, 19 and 25 $\rm M_{\odot}$ are compared with the $\sim$\,216 d spectrum of \sniip\ in Figure~\ref{fig:compjerkspecha}. The [\ion{O}{1}] doublet of \sniip\ matches closely to the J14 model of 15 $\rm M_{\odot}$ and is backed by the findings of S16, who showed that models with a $\rm M_{ZAMS}$ below 12.5 $\rm M_{\odot}$ are inefficient at producing oxygen.

The net amount of oxygen varies from 0.2--5 $\rm M_{\odot}$ for CCSNe progenitors in the mass range of 10--30 $\rm M_{\odot}$ \citep{2007woosley}. During the late nebular phase ($>$200 d), the luminosity of the [\ion{O}{1}] doublet is powered by $\gamma$-ray deposition in the oxygen content of the SN, and hence correlates with the oxygen mass \citep{2003aelmhamdi}. Using an [\ion{O}{1}] flux of $\rm 7.81\times10^{-15}\ erg\ s^{-1}$ in the spectrum of $\sim$\,216 d for \sniip\ and an oxygen mass of 1.2--1.5 $\rm M_{\odot}$ for SN~1987A \citep{1994chugai}, an oxygen mass of 0.8--1.0 $\rm M_{\odot}$ is inferred for \sniip, assuming similarity with SN~1987A in the efficiency of energy deposition and the excited mass. 

\citet{2016morozova} modelled the early phase light curves of Type II SNe using the SuperNova Explosion Code \citep[\textsc{SNEC},][]{2015morozova} and showed that the rise-time depends on the progenitor radii. Using their relation between the progenitor radius at the time of explosion and the $V$-band rise time (instead of $g$-band in their work) of 9.07\,$\pm$\,0.36 d, a progenitor radius of 733\,$\pm$\,36 $\rm R_{\odot}$ is estimated for \sniip. However, due to the effect of CSM on the early LC (and the rise time) of \sniip, the above technique may not truly reflect the progenitor radius. The progenitor parameters estimated for \sniip\ using various techniques are summarized in Table~\ref{tab:modelprogprop}.

\begin{figure}
\centering
\resizebox{\hsize}{!}{\includegraphics{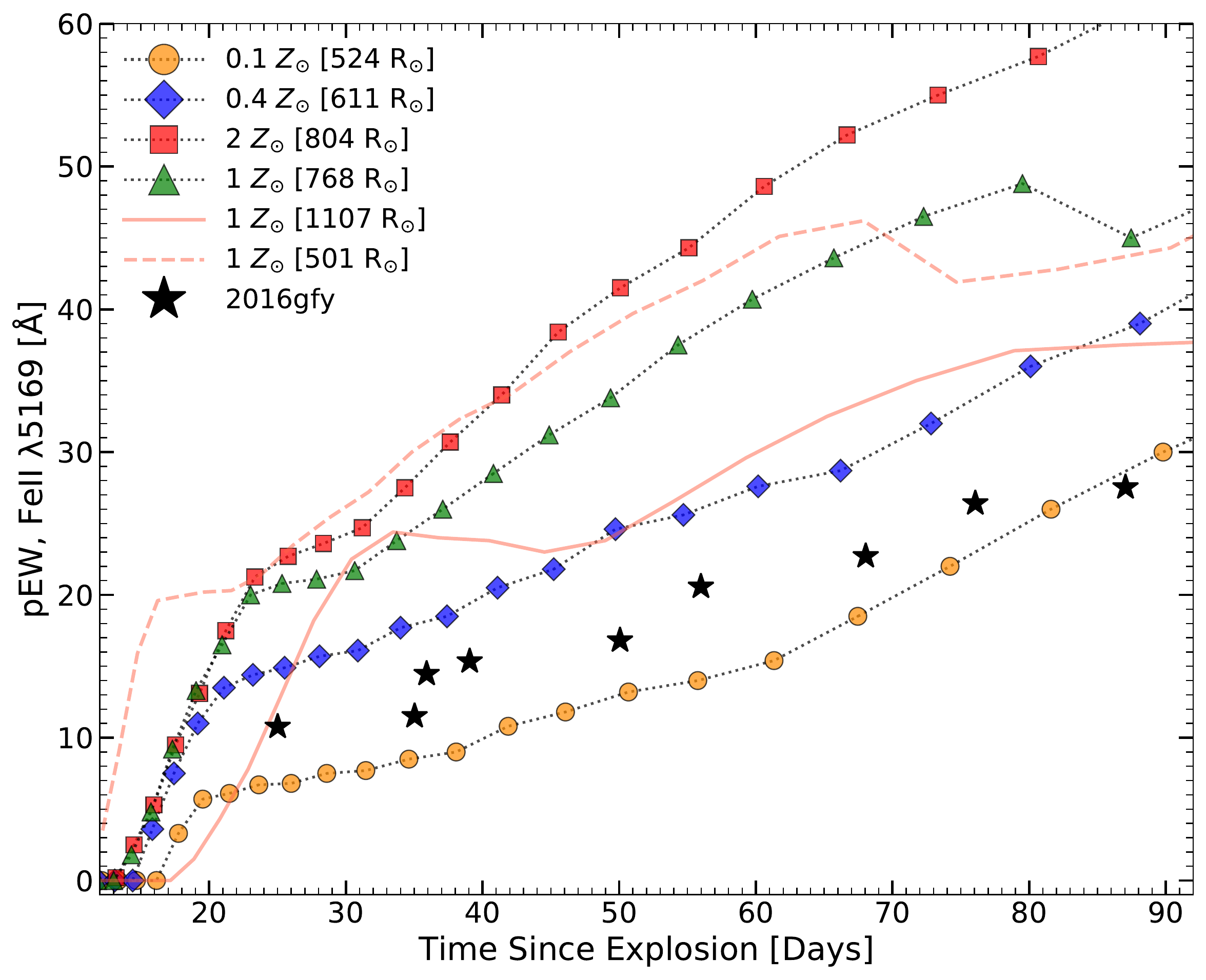}}
\caption{Temporal evolution of pEW of \ion{Fe}{2} $\rm \lambda$5169 feature estimated from the observed spectra of \sniip\ during the plateau phase. The comparison with the theoretical models (shown with a $dotted$ line) of a 15 $\rm M_{\odot}$ progenitor with different metallicities from D13 is also shown. The $dashed$ and $solid$ lines differ from the other models in the mixing length parameter.}
\label{fig:compeqwdes}
\end{figure}

The effect of progenitor metallicity on the spectra of Type II SNe was first indicated in the theoretical modeling of SN atmospheres (D14). This conjecture was further strengthened in the study of A16, who provided observational evidence for the correlation between the metallicity of the host \ion{H}{2} region and the pseudo-Equivalent Widths (pEW) of metal lines during the photospheric phase (plateau) of Type II SNe. The pEW was measured from a Gaussian fit after defining a pseudo-continuum on the either side of the absorption feature. In order to determine the progenitor metallicity of \sniip, the pEW of \ion{Fe}{2} $\rm \lambda$5169 feature was measured in the plateau phase and was compared with the 15 $\rm M_{\odot}$ progenitor models (D13) of different metallicities (0.1, 0.4, 1.0, 2.0 $\rm Z_{\odot}$) in Figure~\ref{fig:compeqwdes}.

The estimated pEW for \sniip\ lies between the 0.1 and 0.4 $\rm Z_{\odot}$ models of D13 and is consistent with the weak presence of [\ion{Ca}{2}] $\rm \lambda \lambda$7291, 7324 during the plateau phase, indicating a low progenitor metallicity. This can also possibly explain the disappearance of \ion{Ca}{2} NIR triplet in the nebular spectra due to its low abundance in the progenitor. This result is in coherence with the sub-solar oxygen abundance estimated for the parent \ion{H}{2} region in Section~\ref{sec:hostspec}. It should be noted here that the mixing-length (mlt) parameter in the theoretical models could significantly alter the pEWs of metal lines as a result of differences in the progenitor radii along with the fact that D13 models are not tailored for the progenitor of \sniip.

\begin{table*}
\centering
\caption{Progenitor parameters estimated for \sniip\ using various techniques.}
\label{tab:modelprogprop}
\begin{tabular}{|c|cccc|}
\toprule
Technique       & $\rm M_{Ni}$      	&   $\rm M_{ej}$		& $\rm E_k$     		& $\rm Radius$		\\
                & ($\rm M_{\odot}$) 	& ($\rm M_{\odot}$)	    & ($\rm 10^{51} erg$)	& ($\rm R_{\odot}$) \\
\midrule
Empirical relation \citep{1985litvinova}    & --- 			& 13.2\,$\pm$\,1.2      & 0.90\,$\pm$\,0.15     & 310\,$\pm$\,70    \\
Two-component model \citep{2016nagy}        & 0.029 		& $\sim$\,11.5          & $\sim$\,1.4           & $\sim$350         \\
Diffusion relation \citep{2016terreran}     & --- 			& $\sim$\,13.0          & ---                   & ---               \\
Comparison of nebular spectra \citep{2014jerkstrand}    & --- 			& $\sim$\,15 ($\rm M_{ZAMS}$)        & ---                   & ---    \\
Correlation of rise-time and progenitor radii \citep{2016morozova} & --- 			& ---        & ---                   & 733\,$\pm$\,36    \\
\bottomrule
\end{tabular}
\end{table*}


\begin{figure}
\centering
\resizebox{\hsize}{!}{\includegraphics{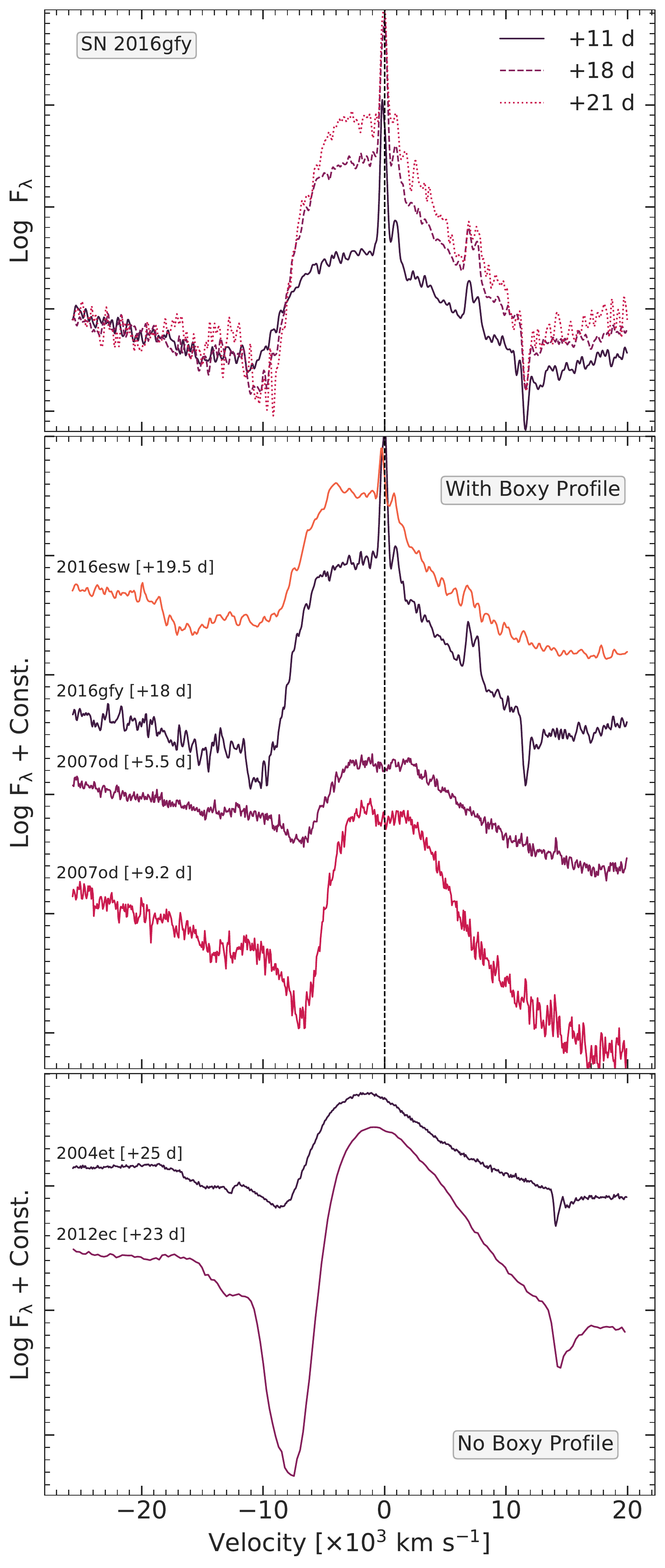}}
\caption{$Top\ panel$: Boxy profile of H\,$\rm \alpha$ seen in the spectra of \sniip\ from $\sim$\,11--21 d. $Middle\ panel$: Comparison with Type II SNe that show boxy emission profile: SN~2007od \citep{2011inserra} and SN~2016esw \citep{2018bdejaeger}. $Bottom\ panel$: Type II SNe that show no boxy profile have been shown for reference: 2004et \citep{2006sahu} and 2012ec \citep{2015barbarino}.}
\label{fig:specboxyha}
\end{figure}

\begin{figure}
\centering
\resizebox{\hsize}{!}{\includegraphics{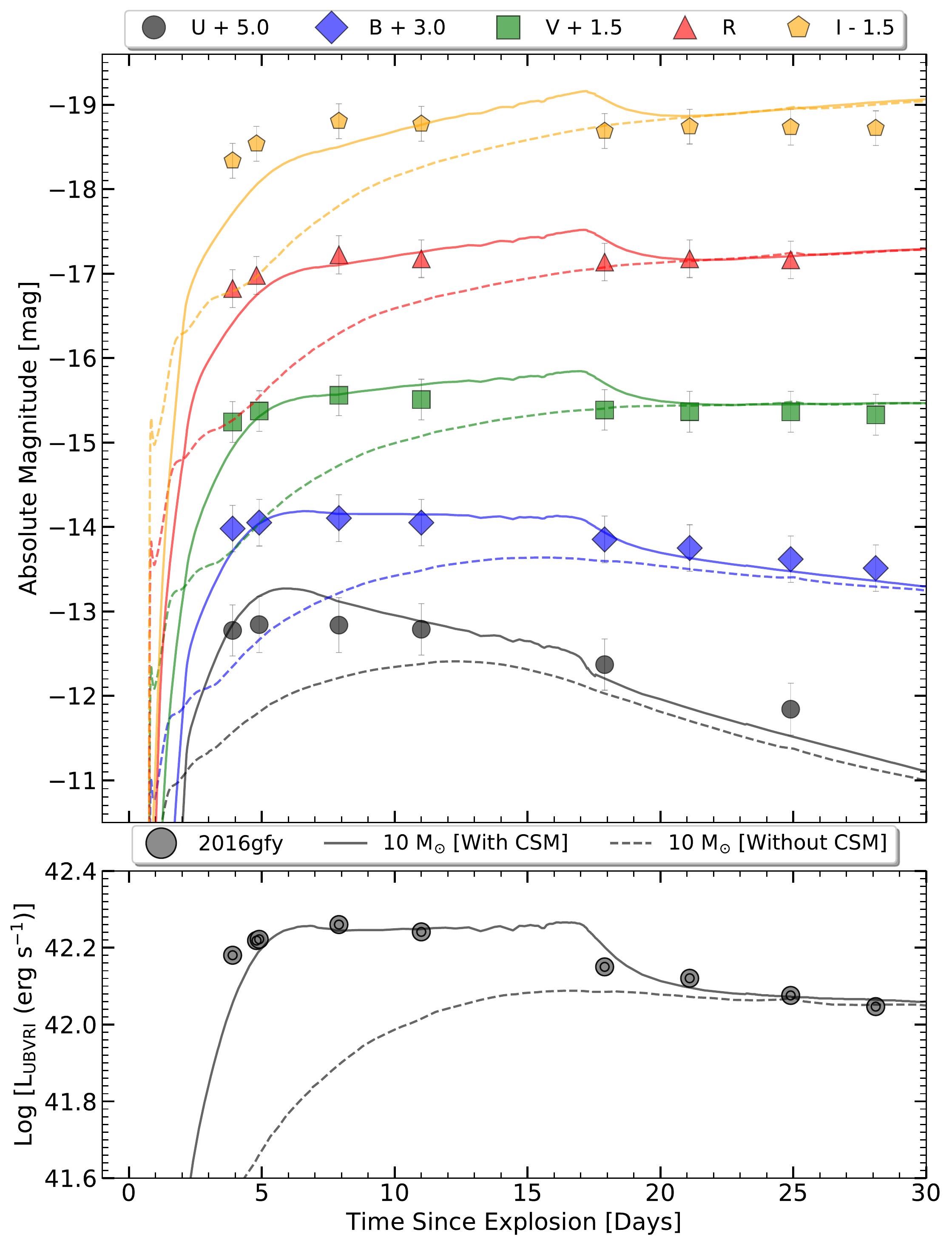}}
\caption{Comparison of \sniip\ with the 10 $\rm M_{\odot}$ progenitor model from \citet{2016sukhbold} with CSM ($solid$) and without CSM ($dashed$). $Top\ panel$ compares the broadband absolute magnitude LCs whereas the $bottom\ panel$ compares the bolometric LCs.}
\label{fig:modelearlylc}
\end{figure}

\section{Discussion}\label{sec:discussion}

\subsection{Early phase CSM-ejecta interaction}\label{sec:earlycsm}

The boxy emission profile of H\,$\rm \alpha$ seen in the spectra of \sniip\ during $\sim$\,11--21 d is an indication of interaction between the fast-moving SN shock and the slow-moving shell-shaped CSM \citep{1994chevalier,2017morozova}. Figure~\ref{fig:specboxyha} shows the evolution of the boxy features in \sniip\ in the top panel, with a comparison to other Type II SNe that show similar features in the middle panel and those that don't in the bottom panel. Narrow Balmer emission lines are also seen in case of an interaction with a massive CSM shell \citep{2018nakaoka}. However, the contamination from narrow features of the parent \ion{H}{2} region in the spectra of \sniip\ makes it difficult to isolate such signatures.

The boxy profile is not seen in the spectrum of $\sim$\,5 d and fades away past the spectrum of $\sim$\,25 d in the case of \sniip, giving an estimated length of interaction with the CSM as 17\,$\pm$\,3 d (epoch of interaction $\sim$\,8--25 d). Using an ejecta velocity of $\sim$\,13,000 \kms\ on day 8 (see Figure~\ref{fig:photvel}), the inner radius of the CSM is estimated as $\sim$\,60 AU. The duration of interaction coupled with the average H\,$\rm \alpha$ velocity during the period ($\sim$\,10,000 \kms) gives a thickness of 110 AU for the CSM shell. Assuming a wind velocity of 10 \kms\ \citep[for an RSG,]{2014smith}, the progenitor of \sniip\ experienced an episode of enhanced mass-loss 30-80 years preceding the explosion.

The interaction of the ejecta with the slowly moving CSM is seen in the spectra of \sniip\ beyond $\sim$\,25 d in the form of HV features of H\,$\rm \alpha$ and H\,$\rm \beta$ which evolve slowly throughout the spectra (9500\,--\,8500 \kms\ in a period of $\sim$\,80 d). This is similar to the case of SN~2013ej where weak CSM interaction was inferred in the early phase \citep{2015boseej,2017das}. The broad emission lines of H\,$\rm \alpha$ and [\ion{O}{1}] $\rm \lambda \lambda$6300,\,6364 seen in the late-phase optical spectra of Type II SN~1980K \citep{1994chevalier} and SN~2007od \citep{2011inserra} also signify CSM interaction. However, no such features are seen in the late-phase spectra of \sniip, possibly due to the absence of CSM at that distance and/or low signal-to-noise ratio (SNR) of the spectra.

As presented in previous studies, not only SN spectra but also early LCs are likely affected by the dense CSM
\citep[e.g.,][]{2017moriya,2018moriya,2017morozova,2018forster}. This interaction converts the kinetic energy of the ejecta upon collision with the nearby CSM into radiative energy and boosts up the early phase luminosity of \sniip. It was shown in the previous section that the early bolometric LC of \sniip\ has the \lq \lq shell\rq \rq\ component which likely originates from the CSM interaction (N16). To estimate the amount of the dense CSM required to explain the early LC bump, numerical LC modeling of the interaction between the SN ejecta and the dense CSM was performed. The method adopted is similar to \citet{2018moriya} and we refer the reader to their study for the complete details of the numerical modeling. 

Briefly, the radiation hydrodynamics code \texttt{STELLA} \citep{1998blinnikov,2000blinnikov,2006blinnikov} is used. The progenitor model of 10 $\rm M_\odot$ at ZAMS and solar metallicity from S16 is used (see Figure~\ref{fig:modelearlylc}). The mass cut is set at 1.4 $\rm M_\odot$ and the explosion is triggered by putting thermal energy just above the mass cut. The explosion energy is $10^{51}~\mathrm{erg}$ and the $^{56}$Ni mass is 0.055~$M_\odot$ in the given model. A dense CSM with a mass-loss rate of $10^{-3}~M_\odot~\mathrm{yr^{-1}}$ and the terminal wind velocity of $10\,\mathrm{km~s^{-1}}$ is put taking wind acceleration into account with the wind acceleration parameter, $\beta$\,=\,2.5 in determining the CSM density structure \citep{2017moriya}. High-mass loss rate here can be explained by wave-driven mass loss \citep{2012quataert}. The dense CSM is extended to $10^{15}~\mathrm{cm}$ ($\sim$\,70 AU) with a total mass of 0.15 $M_\odot$. These CSM parameters are often found in Type II SNe \citep{2018forster}.

A late-plateau bump (besides the early bump) is prominently seen in the $VRI$ LCs of \sniip, which could emerge from an extended interaction with the CSM. This interaction may result in the presence of narrow Balmer features, the signature of which is not seen in our spectral sequence during this phase. However, the narrow lines from the interaction can be enveloped by the SN photosphere as in the case of PTF11iqb \citep{2015smith} and iPTF14hls \citep{2018andrews}. This scenario cannot be ruled out for \sniip\ due to the lack of very late-phase data, in which the photosphere would have receded enough to reveal the hidden CSM-ejecta interaction region. Hence, CSM interaction is a plausible source of luminosity during this phase. 

\citet{2016nakar} explored the effect of $\rm ^{56}$Ni-mixing in the ejecta of Type II SNe and showed that such mixing can alter the plateau duration and/or the decline rate. We therefore, explore $\rm ^{56}Ni$-mixing as an alternate mechanism to explain the late-plateau bump.

\subsection{Case of Ni-Mixing in the Late-Plateau?}\label{sec:nimix}

Radioactivity does not extensively alter the plateau phase luminosity due to the long diffusion time in comparison with the recombination time (KW09). This is however untrue for Type II-P/L SNe that have a progenitor smaller (in radius) than an RSG (e.g. Blue Super Giant in case of SN~1987A) or synthesize large amount of $\rm ^{56}Ni$ ($>$0.1 $\rm M_{\odot}$). However, the plateau is lengthened in proportion to the $\rm ^{56}Ni$ synthesized as the energy from the radioactive decay keeps the ejecta gas ionised longer (S16). \sniip\ shows a bump in the late-plateau phase ($\sim$\,50--95 d, see Figure~\ref{fig:apparentlc}) which is not seen in a majority of bright Type II SNe ($<$\,--17.0 mag).

Light curves of Type II SNe past the photospheric phase show a significant drop to the radioactive tail. The contribution from the cooling envelope becomes negligible relative to the radioactive decay chain ($\rm ^{56}Ni \rightarrow\ ^{56}Co \rightarrow\ ^{56}Fe$) past the luminosity drop at the end of the transition phase, $t_{Ni}$. The fleeting deposition of energy into the ejecta as a result of the $\rm ^{56}Ni$ decay \citep{2016nakar} is given by:

\begin{equation}
\label{eqn:nickellum}
    \rm Q_{Ni}(t) = \frac{M_{Ni}}{M_{\odot}} (6.45e^{-t/8.8} + 1.45e^{-t/111.3}) \times 10^{43} erg s^{-1},
\end{equation}

where $t$ is the time since the explosion in days and $M_{Ni}$ is the mass of $\rm ^{56}Ni$ synthesized. To study the effect of $\rm ^{56}Ni$ on the early phase LC, \citet{2016nakar} defined the observable $\eta_{Ni}$ to disentangle the fraction of bolometric luminosity contributed by $Q_{Ni}(t)$ from the contribution due to the cooling envelope. The observable is defined as:

\begin{equation}
\label{eqn:nickelmix}
    \rm \eta_{Ni} = \frac{\int_{0}^{t_{Ni}} t\ Q_{Ni}(t) dt}{\int_{0}^{t_{Ni}} t\ (L_{bol}(t) - Q_{Ni}(t)) dt},
\end{equation}

where $L_{bol}(t)$ is the bolometric luminosity at time $t$. An $\rm \eta_{Ni}$ of 0.60 is obtained for \sniip\ which translates to a $\sim$\,38\% contribution by $\rm ^{56}Ni$ decay to the time-weighted bolometric luminosity during the plateau phase. The $\eta_{Ni}$ values inferred for the sample of Type II SNe in \citet{2016nakar} lie within the range 0.09\,--\,0.71 (except for SN~2009ib) and indicates a non-negligible contribution in the photospheric phase from the decay of $\rm ^{56}Ni$.

$\rm ^{56}Ni$ can either extend the plateau duration (without any change in the decline rate) and/or cause flattening of the plateau phase (lower the decline rate). A centrally concentrated $\rm ^{56}Ni$ is likely to lengthen the plateau because $\rm ^{56}Ni$ does not diffuse out until the end of the plateau phase \citep[see Figure 4 in][]{2019kozyreva}. If $\rm ^{56}Ni$ is uniformly mixed in the envelope, it increases the luminosity during the plateau phase (and flattens it) as $\rm ^{56}Ni$ diffuses out earlier in comparison with a centrally concentrated $\rm ^{56}Ni$. The phase during which $\rm ^{56}Ni$ starts affecting the LC is dependent on the degree of $\rm ^{56}Ni$ mixing in the envelope \citep{2019kozyreva}.

The effect of $\rm ^{56}Ni$ on the plateau phase is more pronounced in the case of higher $\rm ^{56}Ni$ mass and lower explosion energy \citep{2019kozyreva}. The case of an extremely long plateau in SN~2009ib \citep{2015takats} is partially due to the former reasons but could only be explained with complete mixing of $\rm ^{56}Ni$ in the envelope as it results in a smoother plateau evolution. No observable transition (due to the dominance of $\rm ^{56}Ni$) in the plateau phase is seen in such cases regardless of the value of $\rm \eta_{Ni}$.

At the intermediate value of $\eta_{Ni}$ (=0.60) inferred for \sniip, the emission from the cooling envelope and the $\rm ^{56}Ni$ decay becomes comparable during the late-plateau phase. Unlike the case of SN~2009ib, the slight bump noticed in \sniip\ could only be a result of centrally concentrated or partially mixed $\rm ^{56}Ni$. As the bump is evident only past $\sim$\,50 d, the theoretical light curves in \citet{2019kozyreva} point towards a $\rm ^{56}Ni$-mixing with one-third of the ejecta.

\citet{2016nakar} defined two other dimensionless variables to quantify the effect of $\rm ^{56}Ni$, given as:

\begin{eqnarray}
\label{sec:nickellambda}
\centering
    \Lambda \equiv \frac{L_{25} \cdot\ (80 d)^2}{\int_{0}^{t_{Ni}} t\ L_{bol}(t) dt},
    \Lambda_e \equiv \frac{L_{e,25} \cdot\ (80 d)^2}{\int_{0}^{t_{Ni}} t\ (L_{bol}(t) - Q_{Ni}(t)) dt},
\end{eqnarray}

where $\rm L_{25}$ and $\rm L_e$ are the observed and hypothetical bolometric luminosity (when no $\rm ^{56}Ni$ is synthesized in the explosion) on day 25, respectively. The quantities $\rm 2.5\ log_{10}\ \Lambda$ and $\rm 2.5\ log_{10}\ \Lambda_e$ are indicators of plateau decline rates in units of \magfifty, with and without the effect of $\rm ^{56}Ni$, respectively. A difference ($\rm \Lambda$\,--\,$\rm Lambda_e$) of $\sim$\,0.5 is estimated for \sniip\ which translates to a change in slope of $\sim$\,1 \maghundred\ due to the effect of $\rm ^{56}Ni$ during the plateau phase. This explains the bump in the late-plateau phase of the bolometric light curve wherein a decline is mostly seen.

This effect is similar to the transition from $s_1$ to $s_2$ seen in most Type II SNe \citep{2014anderson} but the degree of flattening varies across the sample. Brighter Type II SNe ($\rm Mv$\,$>$\,--17.0 mag) tend to have higher inherent luminosity (and explosion energy) and hence the effect of $\rm ^{56}Ni$ during the plateau is minimized, leading to steeper decline rates. However, the presence of this effect in \sniip\ ($\rm M_V$\,$\sim$\,--17.1 mag), signifies a lower explosion energy.

\subsection{Is \sniip\ a typical Type II SN?}

\begin{figure}
\centering
\resizebox{\hsize}{!}{\includegraphics{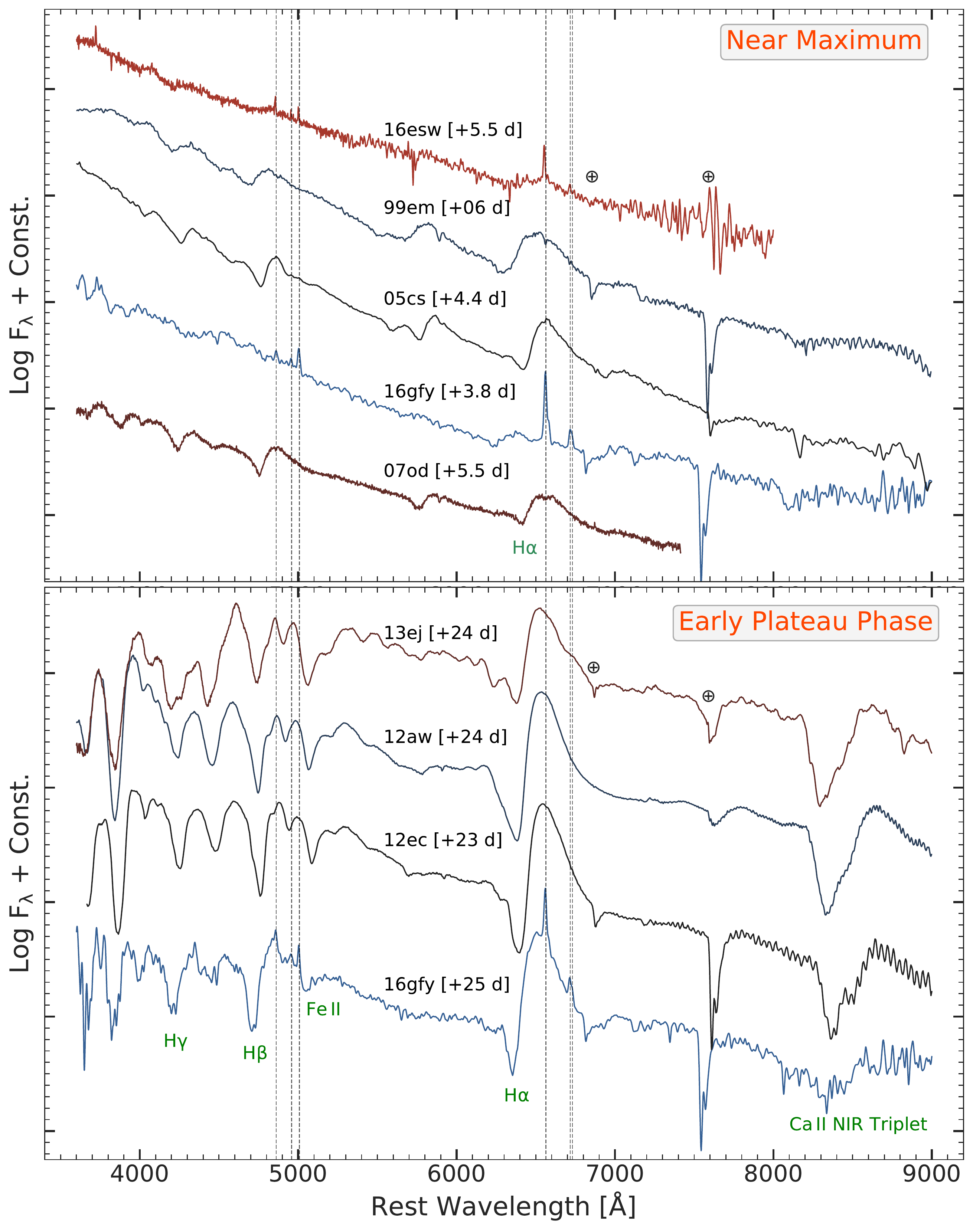}}
\caption{Comparison of the first week ($\sim$\,4 d) and early plateau phase ($\sim$\,25 d) spectrum of \sniip\ with other Type II SNe in the $top$ and $bottom$ panels, respectively. Host galactic lines are indicated with a vertical $dashed$ line. References: 1999em \citep{2002bleonard}; 2005cs \citep{2009pastorello}; 2007od \citep{2011inserra}; 2012aw \citep{2013bose}; 2012ec \citep{2015barbarino}; 2013ej \citep{2015boseej}; 2016esw \citep{2018bdejaeger}.}
\label{fig:compspec1}
\end{figure}

The first observed spectrum of \sniip\ is compared with the first week spectra of Type II SNe from the literature in the top panel of Figure~\ref{fig:compspec1}. \sniip\ shows a blue featureless continuum at this epoch, similar to the spectrum of SN~2016esw. However, this is in contrast to other Type II SNe that show P-Cygni Balmer features along with \ion{He}{2} $\rm \lambda$5876. The bottom panel in Figure~\ref{fig:compspec1} shows comparison during the steeper part of the plateau phase. Here, the overall spectrum of \sniip\ resembles the spectra of other Type II SNe, with noticeable differences only in the metal line strengths. This can be attributed to the lower metallicity of the progenitor (See Section~\ref{sec:progenitorprop}) in comparison to other Type II SNe.

\begin{figure}
\centering
\resizebox{\hsize}{!}{\includegraphics{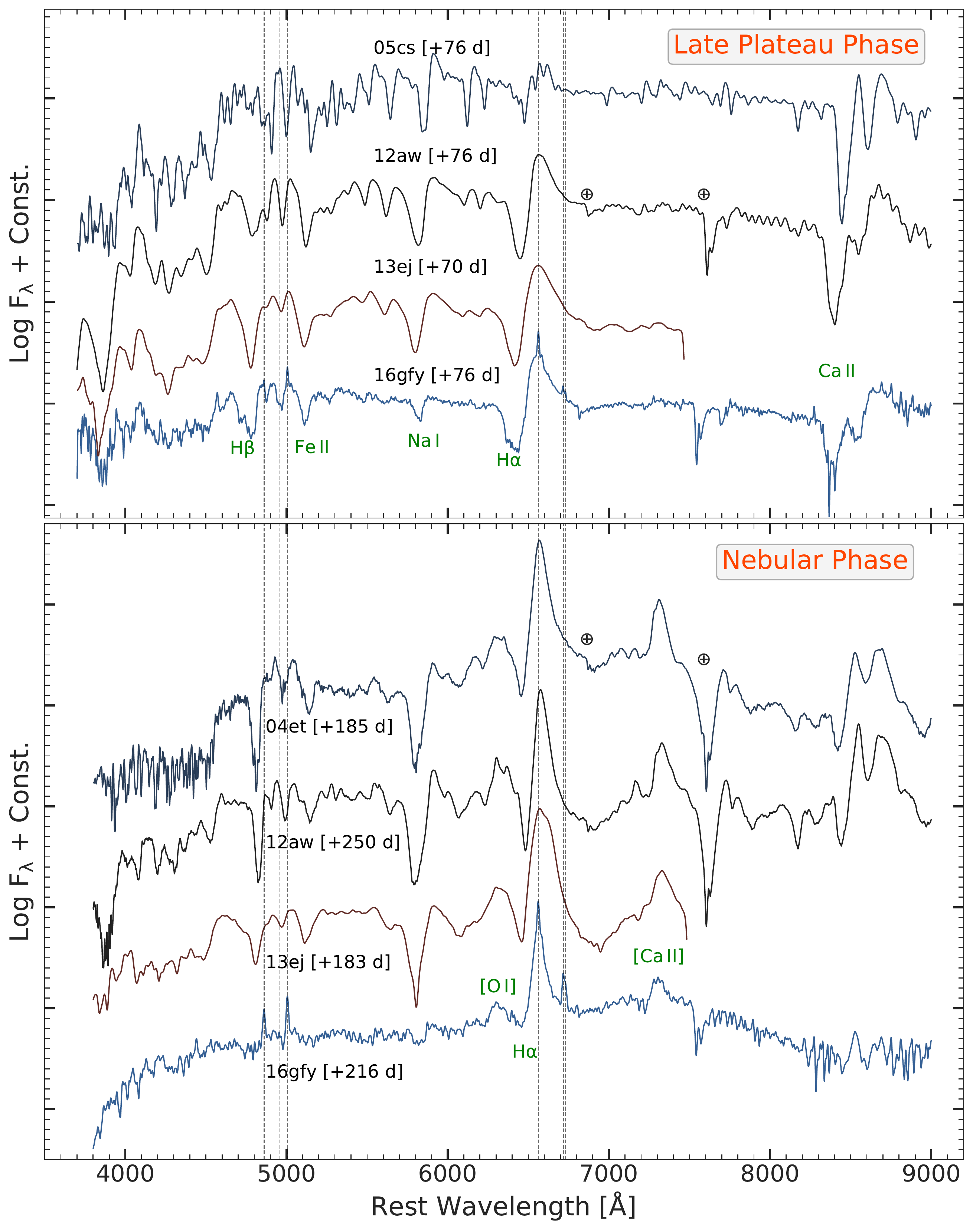}}
\caption{Comparison of the late-plateau ($\sim$\,76 d) and nebular phase spectrum of \sniip\ with other Type II SNe in the $top$ and $bottom$ panels, respectively. Host galactic lines are indicated with a vertical $dashed$ line. References: 2004et \citep{2006sahu}; 2005cs \citep{2009pastorello}; 2012aw \citep{2013bose}; 2013ej \citep{2015boseej}.}
\label{fig:compspec2}
\end{figure}

The top panel in Figure~\ref{fig:compspec2} shows the comparison during the late-plateau phase. The lack of richness in the metal features in the spectra of \sniip\ coupled with their weakness is clearly evident. Hence, the inference of metal-poor progenitor of \sniip\ is strengthened as the SN spectra traces the progenitor metallicity during the photospheric phase (A16). The comparison during the nebular phase is shown in the bottom panel of Figure~\ref{fig:compspec2}. The spectra of \sniip\ shows relatively weak signatures of \ion{Na}{1}D, [\ion{O}{1}] $\rm \lambda \lambda$6300, 6364 and the [\ion{Ca}{2}] $\rm \lambda \lambda$7291, 7324 and the \ion{Ca}{2} NIR triplet. Also, absorption associated with the H\,$\rm \alpha$ is almost negligible in comparison with other Type II SNe indicating that \sniip\ entered the nebular phase earlier, possibly due to a low-mass progenitor.

\begin{figure}
\centering
\resizebox{\hsize}{!}{\includegraphics{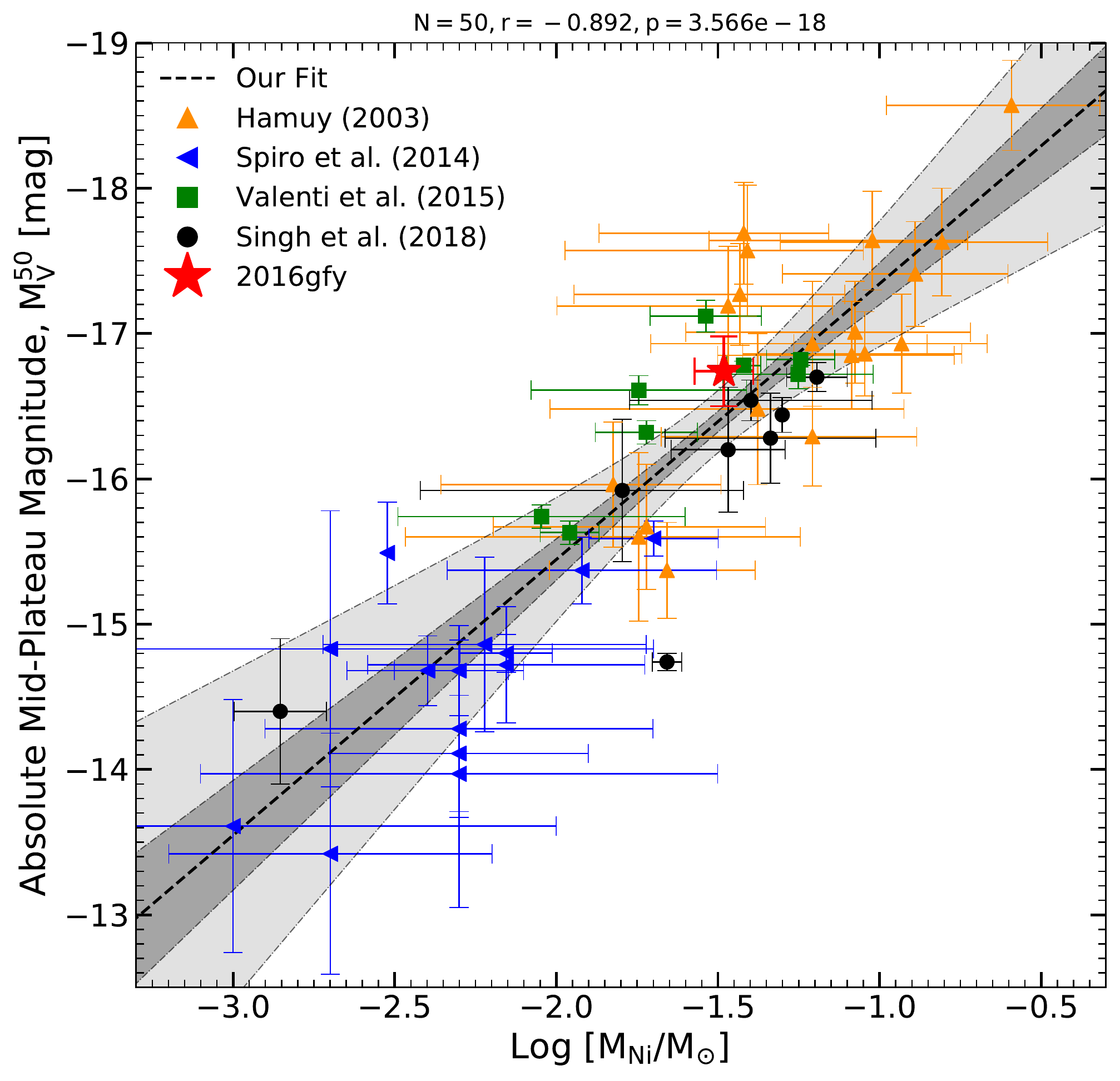}}
\caption{Plot of $\rm M_V^{50}$ vs $\rm Log M_{Ni}$, a correlation inferred by \citet{2003hamuy} for Type II SNe. The data is adopted from \citet{2003hamuy}, \citet{2014spiro}, \citet{2015valenti} and \citet{2018avinash}. The fits to the collective sample are shown with a $dotted$ line. The 1$\rm \sigma$ and 3$\rm \sigma$ confidence intervals of the fit are shaded in $dark-grey$ and $light-grey$, respectively.}
\label{fig:hamuycorr}
\end{figure}

\begin{figure*}
\centering
\resizebox{0.48\hsize}{!}{\includegraphics{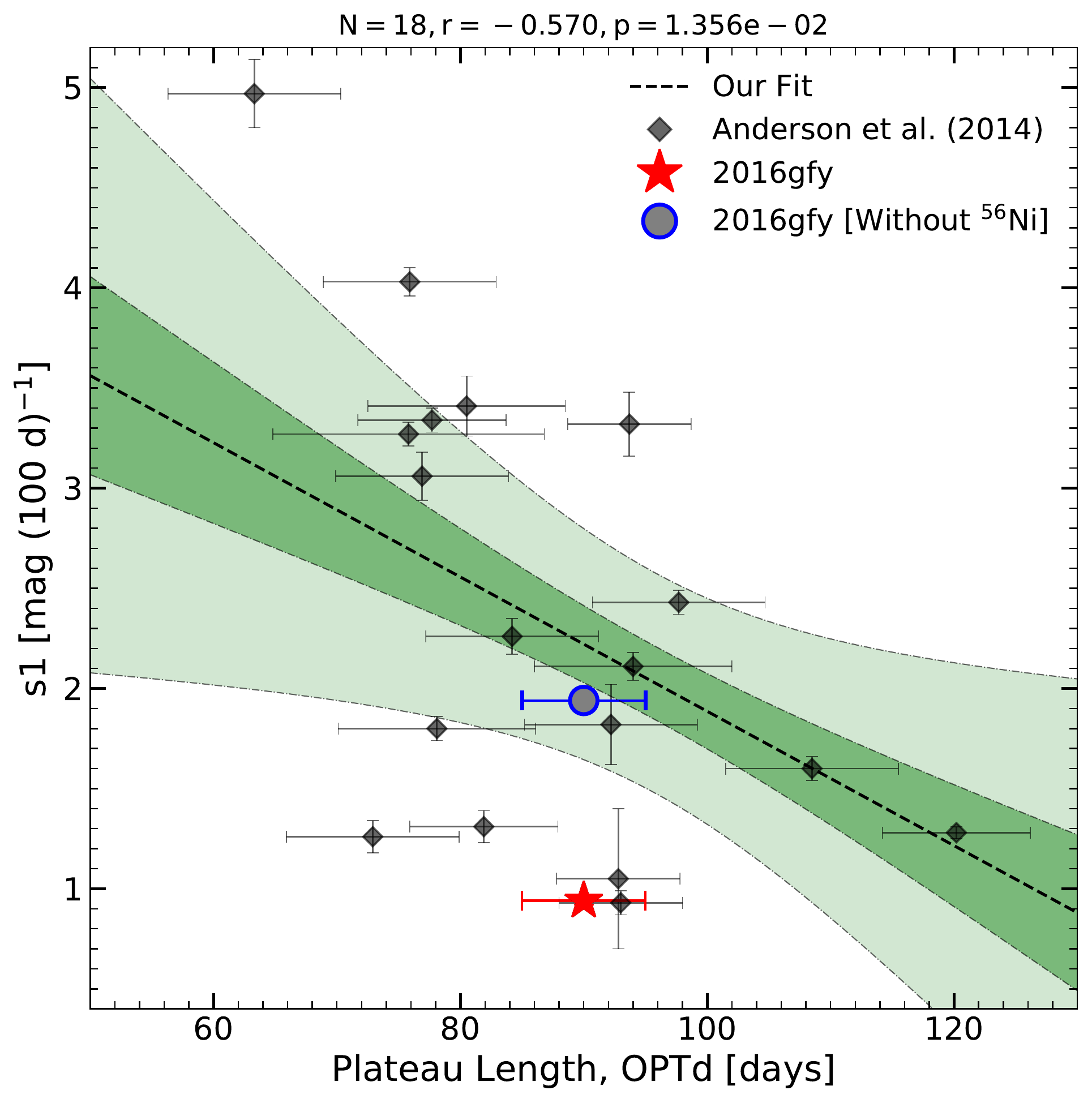}}
\resizebox{0.48\hsize}{!}{\includegraphics{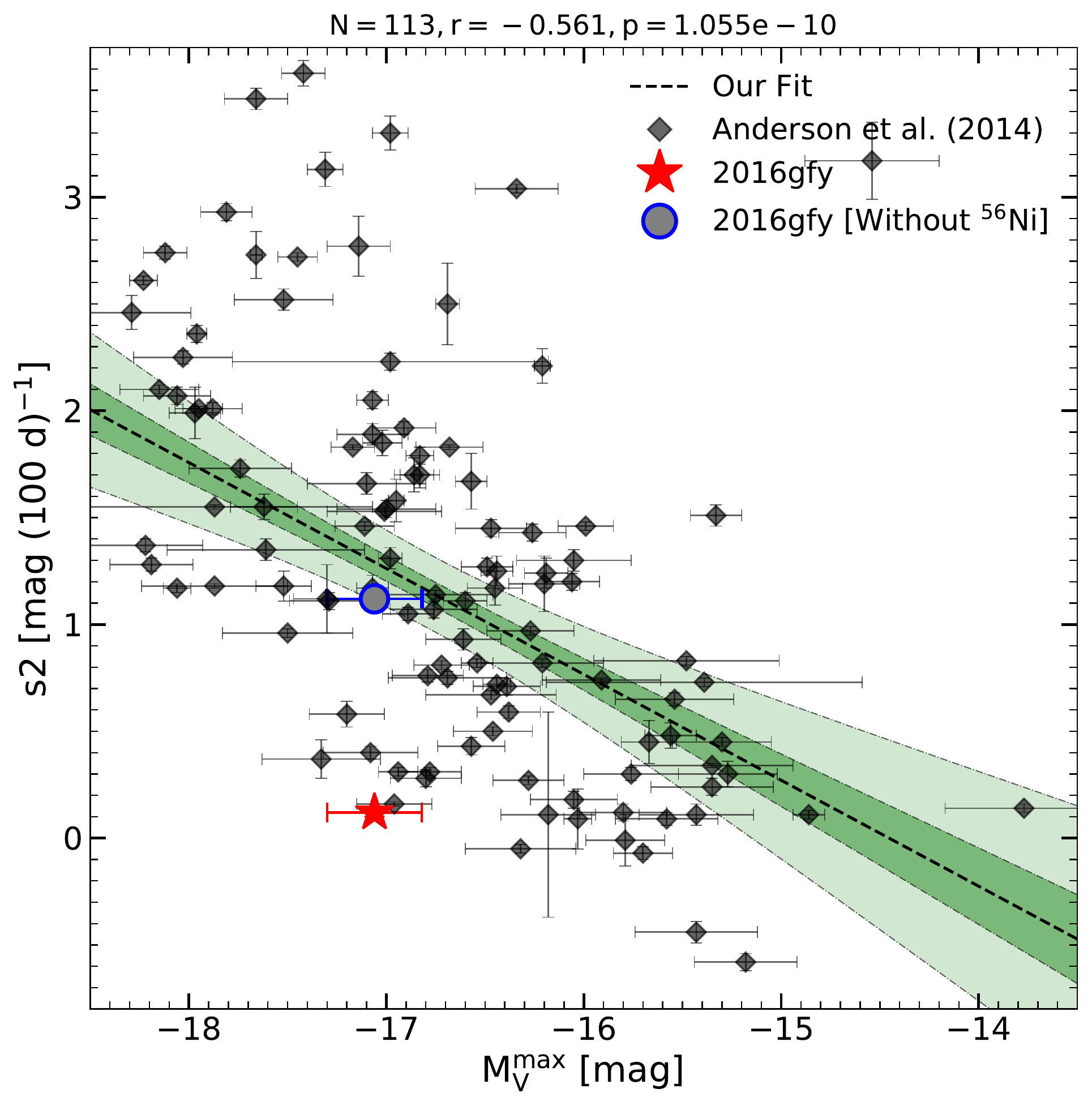}}
\caption{Decline rates $s_1$ and $s_2$ plotted against the plateau length (OPTd) and $\rm M_V^{Max}$, respectively. The sample for the comparison is adopted from \citet{2014anderson}. The $dashed$ line shows the fit to the sample with the 1$\rm \sigma$ and 3$\rm \sigma$ confidence intervals shown shaded in $dark$ and $light$, respectively.}
\label{fig:andercorr}
\end{figure*}

\sniip\ adds to the sample of Type II SNe in low-metallicity environments \citep{2016polshaw, 2018avinash,2018gutierrez,2018meza} which were earlier considered scarce. To picture \sniip\ in the parameter space of well-studied Type II SNe from the literature \citep{2003hamuy,2014spiro,2015valenti,2018avinash}, the absolute $V$-band magnitude during the mid-plateau is compared with the mass of $\rm ^{56}Ni$ synthesized in Figure~\ref{fig:hamuycorr}. \sniip\ lies within the 3$\rm \sigma$ dispersion of the fit shown and indicates no peculiarity w.r.t. these parameters. 

The plateau decline rates, $s_1$ and $s_2$ \citep{2014anderson} in $UBVRI$ were determined by a linear piece-wise fit to the LCs until the end of the plateau phase ($\sim$\,90 d). When the estimated decline rates of \sniip\ are compared versus the extensive sample of Type II SNe from \citet{2014anderson} in Figure~\ref{fig:andercorr}, \sniip\ clearly stands outside the 3$\sigma$ dispersion of the fit and shows extremely slow decline in comparison to SNe of similar luminosity. \sniip\ has decline rates of $s_1$\,=\,0.94 \maghundred and $s_2$\,=\,0.12 \maghundred in the plateau and are much lower than the mean decline rates for Type II SNe in \citet{2014anderson}, which have $s_1^{mean}$\,$\sim$\,2.65 \maghundred and $s_2^{mean}$\,$\sim$\,1.27 \maghundred. As discussed in Section~\ref{sec:nimix}, the effect of $\rm ^{56}Ni$ and its mixing on the bolometric LC of \sniip\ is significant and is clearly evident in the comparison.

However, if the change in slope of $\sim$\,1 \maghundred\ due to the effect of $\rm ^{56}Ni$ is taken into account, \sniip\ (shown as a $grey$ circle) lies directly on the expected correlation from the fits. It appears that the diversity of decline rates seen in Type II SNe is not only due to the range of envelope masses seen for progenitors of different masses but also due to varied amounts of $\rm^{56}Ni$ synthesized and its degree of mixing. The effects of $\rm ^{56}Ni$-mixing and the weak metal features in the spectra of \sniip\ indicate a metallicity at the lower end of the population of Type II SNe, making \sniip\ atypical and an interesting object to study.

\section{Summary}\label{sec:summary}
In this article, we presented the photometric and spectroscopic analysis of the slow-declining Type II \sniip. The properties of \sniip\ are outlined below:

\begin{itemize}
    \item \sniip\ is a luminous Type II SN with a peak $V$-band absolute magnitude of --17.06\,$\pm$\,0.24 mag.
    \item It is a slow-declining Type II SN ($s_1$\,=\,0.94 \maghundred and $s_2$\,=\,0.12 \maghundred) in comparison to the extensive sample of Type II SNe in \citet{2014anderson}.
    \item The host galaxy \hostglx\ is a starburst with an SFR $\sim$\,8.5 $\rm M_{\odot} yr^{-1}$. The spectrum of the parent \ion{H}{2} region yielded an oxygen abundance of 12 + log(O/H) = 8.50\,$\pm$\,0.11, indicating an average metallicity for its progenitor in comparison to the sample of Type II SNe \citep{2016anderson}.
    \item The progenitor of \sniip\ belongs to the class of RSGs with a radius in the range $\sim$\,350--700 $\rm R_{\odot}$. The progenitor has a mass in the range of 12--15 $\rm M_{\odot}$ and an explosion energy in the range of 0.9-1.4$\rm \times\ 10^{51}\ erg$.
    \item A boxy emission profile of $\rm H\alpha$ is seen in the spectra obtained during $\sim$\,11--21 d indicating a CSM-ejecta interaction. This CSM, in the immediate vicinity of the SN could be a result of the mass-loss episode 30-80 yrs before the explosion. Numerical modeling of \sniip\ suggests the presence of 0.15 $\rm M_{\odot}$ CSM spread to a radius of $\sim$\,70 AU around the progenitor.
    \item The late-plateau phase ($\sim$\,50--95 d) in \sniip\ shows a bump which is explained as a result of interaction with the CSM and/or partial mixing of $\rm ^{56}Ni$ in the SN ejecta.
    \item The spectral evolution of \sniip\ features metal-poor spectra compared to other Type II SNe and the theoretical models of \citet{2013bdessart}, signifying a low-metallicity of the progenitor and is consistent with the low-metallicity of the parent \ion{H}{2} region.
\end{itemize}

\section{Acknowledgements}

We thank the anonymous referee for their insightful suggestions. We thank Dr. Thomas de Jaeger and Nicolas Eduardo Meza for sharing data. We would also like to thank Dr. Sudhanshu Barway for discussions involving metallicity estimation of galaxies and galaxy morphology.

We thank the staff of IAO, Hanle and CREST, Hosakote, that made these observations possible. The facilities at IAO and CREST are operated by the Indian Institute of Astrophysics, Bangalore. Observations reported here were also obtained at the MMT Observatory, a joint facility of the University of Arizona and the Smithsonian Institution. We also thank the observers who helped us with the follow-up observations. BK acknowledges the Science and Engineering Research Board (SERB) under the Department of Science \& Technology (DST), Govt. of India, for financial assistance in the form of National Post-Doctoral Fellowship (Ref. no. PDF/2016/001563). BK, DKS and GCA acknowledge the BRICS grant, DST/IMRCD/BRICS/PilotCall1/MuMeSTU/2017(G), for the present work. DKS and GCA also acknowledge the DST/JSPS grant, DST/INT/JSPS/P/281/2018.

This research made use of \textsc{RedPipe}\footnote{\url{https://github.com/sPaMFouR/RedPipe}}, an assemblage of data reduction and analysis scripts written by AS. This work also made use of the NASA Astrophysics Data System and the NASA/IPAC Extragalactic Database (NED\footnote{\url{https://ned.ipac.caltech.edu}}) which is operated by the Jet Propulsion Laboratory, California Institute of Technology. We acknowledge Wiezmann Interactive Supernova data REPository\footnote{\url{https://wiserep.weizmann.ac.il}} (WISeREP) \citep{2012yaron}.

\software{SciPy (v1.3.0, \citealp{2007scipy}), Matplotlib (v3.1.0, \citealp{2007hunter}), Pandas (v0.24.2, \citealp{2010pandas}), PyRAF (v2.1.14, \citealp{2012pyraf}), Astropy (v3.1.2, \citealp{2018astropy}), Seaborn (v0.9.0, \citealp{2018seaborn})
}


\bibliographystyle{AASJournal}
\bibliography{_Reference}

\begin{thebibliography}{}
\expandafter\ifx\csname natexlab\endcsname\relax\def\natexlab#1{#1}\fi
\providecommand{\url}[1]{\href{#1}{#1}}

\bibitem[{{Anderson} {et~al.}(2014){Anderson}, {Gonz{\'a}lez-Gait{\'a}n},
  {Hamuy}, {Guti{\'e}rrez}, {Stritzinger}, {Olivares E.}, {Phillips},
  {Schulze}, {Antezana}, {Bolt}, {Campillay}, {Castell{\'o}n}, {Contreras}, {de
  Jaeger}, {Folatelli}, {F{\"o}rster}, {Freedman}, {Gonz{\'a}lez}, {Hsiao},
  {Krzemi{\'n}ski}, {Krisciunas}, {Maza}, {McCarthy}, {Morrell}, {Persson},
  {Roth}, {Salgado}, {Suntzeff}, \& {Thomas-Osip}}]{2014anderson}
{Anderson}, J.~P., {Gonz{\'a}lez-Gait{\'a}n}, S., {Hamuy}, M., {et~al.} 2014,
  \apj, 786, 67

\bibitem[{{Anderson} {et~al.}(2016){Anderson}, {Guti{\'e}rrez}, {Dessart},
  {Hamuy}, {Galbany}, {Morrell}, {Stritzinger}, {Phillips}, {Folatelli},
  {Boffin}, {de Jaeger}, {Kuncarayakti}, \& {Prieto}}]{2016anderson}
{Anderson}, J.~P., {Guti{\'e}rrez}, C.~P., {Dessart}, L., {et~al.} 2016, \aap,
  589, A110

\bibitem[{{Andrews} \& {Smith}(2018)}]{2018andrews}
{Andrews}, J.~E., \& {Smith}, N. 2018, \mnras, 477, 74

\bibitem[{{Andrews} {et~al.}(2010){Andrews}, {Gallagher}, {Clayton},
  {Sugerman}, {Chatelain}, {Clem}, {Welch}, {Barlow}, {Ercolano}, \&
  {Fabbri}}]{2010andrews}
{Andrews}, J.~E., {Gallagher}, J.~S., {Clayton}, G.~C., {et~al.} 2010, \apj,
  715, 541

\bibitem[{{Arcavi} {et~al.}(2012){Arcavi}, {Gal-Yam}, {Cenko}, {Fox},
  {Leonard}, {Moon}, {Sand}, {Soderberg}, {Kiewe}, {Yaron}, {Becker}, {Scheps},
  {Birenbaum}, {Chamudot}, \& {Zhou}}]{2012arcavi}
{Arcavi}, I., {Gal-Yam}, A., {Cenko}, S.~B., {et~al.} 2012, \apjl, 756, L30

\bibitem[{{Arcavi} {et~al.}(2017){Arcavi}, {Howell}, {Kasen}, {Bildsten},
  {Hosseinzadeh}, {McCully}, {Wong}, {Katz}, {Gal-Yam}, {Sollerman}, {Taddia},
  {Leloudas}, {Fremling}, {Nugent}, {Horesh}, {Mooley}, {Rumsey}, {Cenko},
  {Graham}, {Perley}, {Nakar}, {Shaviv}, {Bromberg}, {Shen}, {Ofek}, {Cao},
  {Wang}, {Huang}, {Rui}, {Zhang}, {Li}, {Li}, {Zhang}, {Valenti}, {Guevel},
  {Shappee}, {Kochanek}, {Holoien}, {Filippenko}, {Fender}, {Nyholm}, {Yaron},
  {Kasliwal}, {Sullivan}, {Blagorodnova}, {Walters}, {Lunnan}, {Khazov},
  {Andreoni}, {Laher}, {Konidaris}, {Wozniak}, \& {Bue}}]{2017arcavi}
{Arcavi}, I., {Howell}, D.~A., {Kasen}, D., {et~al.} 2017, \nat, 551, 210

\bibitem[{{Arnett}(1980)}]{1980arnett}
{Arnett}, W.~D. 1980, \apj, 237, 541

\bibitem[{{Arnett}(1982)}]{1982arnett}
---. 1982, \apj, 253, 785

\bibitem[{{Asplund} {et~al.}(2009){Asplund}, {Grevesse}, {Sauval}, \&
  {Scott}}]{2009asplund}
{Asplund}, M., {Grevesse}, N., {Sauval}, A.~J., \& {Scott}, P. 2009, \araa, 47,
  481

\bibitem[{{Astropy Collaboration} {et~al.}(2018){Astropy Collaboration},
  {Price-Whelan}, {Sip{\H o}cz}, {G{\"u}nther}, {Lim}, {Crawford}, {Conseil},
  {Shupe}, {Craig}, {Dencheva}, {Ginsburg}, {VanderPlas}, {Bradley},
  {P{\'e}rez-Su{\'a}rez}, {de Val-Borro}, {Aldcroft}, {Cruz}, {Robitaille},
  {Tollerud}, {Ardelean}, {Babej}, {Bach}, {Bachetti}, {Bakanov}, {Bamford},
  {Barentsen}, {Barmby}, {Baumbach}, {Berry}, {Biscani}, {Boquien}, {Bostroem},
  {Bouma}, {Brammer}, {Bray}, {Breytenbach}, {Buddelmeijer}, {Burke},
  {Calderone}, {Cano Rodr{\'{\i}}guez}, {Cara}, {Cardoso}, {Cheedella},
  {Copin}, {Corrales}, {Crichton}, {D'Avella}, {Deil}, {Depagne}, {Dietrich},
  {Donath}, {Droettboom}, {Earl}, {Erben}, {Fabbro}, {Ferreira}, {Finethy},
  {Fox}, {Garrison}, {Gibbons}, {Goldstein}, {Gommers}, {Greco}, {Greenfield},
  {Groener}, {Grollier}, {Hagen}, {Hirst}, {Homeier}, {Horton}, {Hosseinzadeh},
  {Hu}, {Hunkeler}, {Ivezi{\'c}}, {Jain}, {Jenness}, {Kanarek}, {Kendrew},
  {Kern}, {Kerzendorf}, {Khvalko}, {King}, {Kirkby}, {Kulkarni}, {Kumar},
  {Lee}, {Lenz}, {Littlefair}, {Ma}, {Macleod}, {Mastropietro}, {McCully},
  {Montagnac}, {Morris}, {Mueller}, {Mumford}, {Muna}, {Murphy}, {Nelson},
  {Nguyen}, {Ninan}, {N{\"o}the}, {Ogaz}, {Oh}, {Parejko}, {Parley}, {Pascual},
  {Patil}, {Patil}, {Plunkett}, {Prochaska}, {Rastogi}, {Reddy Janga},
  {Sabater}, {Sakurikar}, {Seifert}, {Sherbert}, {Sherwood-Taylor}, {Shih},
  {Sick}, {Silbiger}, {Singanamalla}, {Singer}, {Sladen}, {Sooley},
  {Sornarajah}, {Streicher}, {Teuben}, {Thomas}, {Tremblay}, {Turner},
  {Terr{\'o}n}, {van Kerkwijk}, {de la Vega}, {Watkins}, {Weaver}, {Whitmore},
  {Woillez}, {Zabalza}, \& {Astropy Contributors}}]{2018astropy}
{Astropy Collaboration}, {Price-Whelan}, A.~M., {Sip{\H o}cz}, B.~M., {et~al.}
  2018, \aj, 156, 123

\bibitem[{{Baldwin} {et~al.}(1981){Baldwin}, {Phillips}, \&
  {Terlevich}}]{1981baldwin}
{Baldwin}, J.~A., {Phillips}, M.~M., \& {Terlevich}, R. 1981, \pasp, 93, 5

\bibitem[{{Barbarino} {et~al.}(2015){Barbarino}, {Dall'Ora}, {Botticella},
  {Della Valle}, {Zampieri}, {Maund}, {Pumo}, {Jerkstrand}, {Benetti},
  {Elias-Rosa}, {Fraser}, {Gal-Yam}, {Hamuy}, {Inserra}, {Knapic}, {LaCluyze},
  {Molinaro}, {Ochner}, {Pastorello}, {Pignata}, {Reichart}, {Ries},
  {Riffeser}, {Schmidt}, {Schmidt}, {Smareglia}, {Smartt}, {Smith},
  {Sollerman}, {Sullivan}, {Tomasella}, {Turatto}, {Valenti}, {Yaron}, \&
  {Young}}]{2015barbarino}
{Barbarino}, C., {Dall'Ora}, M., {Botticella}, M.~T., {et~al.} 2015, \mnras,
  448, 2312

\bibitem[{{Barbon} {et~al.}(1979){Barbon}, {Ciatti}, \& {Rosino}}]{1979barbon}
{Barbon}, R., {Ciatti}, F., \& {Rosino}, L. 1979, \aap, 72, 287

\bibitem[{{Bersten} {et~al.}(2011){Bersten}, {Benvenuto}, \&
  {Hamuy}}]{2011bersten}
{Bersten}, M.~C., {Benvenuto}, O., \& {Hamuy}, M. 2011, \apj, 729, 61

\bibitem[{{Blinnikov} {et~al.}(2000){Blinnikov}, {Lundqvist}, {Bartunov},
  {Nomoto}, \& {Iwamoto}}]{2000blinnikov}
{Blinnikov}, S., {Lundqvist}, P., {Bartunov}, O., {Nomoto}, K., \& {Iwamoto},
  K. 2000, \apj, 532, 1132

\bibitem[{{Blinnikov} \& {Bartunov}(1993)}]{1993blinnikov}
{Blinnikov}, S.~I., \& {Bartunov}, O.~S. 1993, \aap, 273, 106

\bibitem[{{Blinnikov} {et~al.}(1998){Blinnikov}, {Eastman}, {Bartunov},
  {Popolitov}, \& {Woosley}}]{1998blinnikov}
{Blinnikov}, S.~I., {Eastman}, R., {Bartunov}, O.~S., {Popolitov}, V.~A., \&
  {Woosley}, S.~E. 1998, \apj, 496, 454

\bibitem[{{Blinnikov} {et~al.}(2006){Blinnikov}, {R{\"o}pke}, {Sorokina},
  {Gieseler}, {Reinecke}, {Travaglio}, {Hillebrandt}, \&
  {Stritzinger}}]{2006blinnikov}
{Blinnikov}, S.~I., {R{\"o}pke}, F.~K., {Sorokina}, E.~I., {et~al.} 2006, \aap,
  453, 229

\bibitem[{{Bose} \& {Kumar}(2014)}]{2014bose}
{Bose}, S., \& {Kumar}, B. 2014, \apj, 782, 98

\bibitem[{{Bose} {et~al.}(2013){Bose}, {Kumar}, {Sutaria}, {Kumar}, {Roy},
  {Bhatt}, {Pandey}, {Chandola}, {Sagar}, {Misra}, \& {Chakraborti}}]{2013bose}
{Bose}, S., {Kumar}, B., {Sutaria}, F., {et~al.} 2013, \mnras, 433, 1871

\bibitem[{{Bose} {et~al.}(2015{\natexlab{a}}){Bose}, {Valenti}, {Misra},
  {Pumo}, {Zampieri}, {Sand}, {Kumar}, {Pastorello}, {Sutaria}, {Maccarone},
  {Kumar}, {Graham}, {Howell}, {Ochner}, {Chandola}, \& {Pandey}}]{2015boseab}
{Bose}, S., {Valenti}, S., {Misra}, K., {et~al.} 2015{\natexlab{a}}, \mnras,
  450, 2373

\bibitem[{{Bose} {et~al.}(2015{\natexlab{b}}){Bose}, {Sutaria}, {Kumar},
  {Duggal}, {Misra}, {Brown}, {Singh}, {Dwarkadas}, {York}, {Chakraborti},
  {Chandola}, {Dahlstrom}, {Ray}, \& {Safonova}}]{2015boseej}
{Bose}, S., {Sutaria}, F., {Kumar}, B., {et~al.} 2015{\natexlab{b}}, \apj, 806,
  160

\bibitem[{{Bradley} {et~al.}(2017){Bradley}, Sipocz, Robitaille, Vinícius,
  Tollerud, Deil, Barbary, Günther, Cara, Busko, Droettboom, Bostroem, Bray,
  Bratholm, Pickering, Craig, Barentsen, Pascual, Conseil, adonath, Greco,
  Kerzendorf, de~Val-Borro, StuartLittlefair, Ogaz, Lim, Ferreira, D'Eugenio,
  \& Weaver}]{2017photutils}
{Bradley}, L., Sipocz, B., Robitaille, T., {et~al.} 2017, astropy/photutils:
  v0.4,  zenodo, doi:10.5281/zenodo.1039309

\bibitem[{{Breeveld} {et~al.}(2011){Breeveld}, {Landsman}, {Holland}, {Roming},
  {Kuin}, \& {Page}}]{2011breeveld}
{Breeveld}, A.~A., {Landsman}, W., {Holland}, S.~T., {et~al.} 2011, in American
  Institute of Physics Conference Series, Vol. 1358, GAMMA RAY BURSTS 2010. AIP
  Conference Proceedings, ed. J.~E. {McEnery}, J.~L. {Racusin}, \&
  N.~{Gehrels}, 373--376

\bibitem[{{Brown} {et~al.}(2014){Brown}, {Breeveld}, {Holland}, {Kuin}, \&
  {Pritchard}}]{2014brown}
{Brown}, P.~J., {Breeveld}, A.~A., {Holland}, S., {Kuin}, P., \& {Pritchard},
  T. 2014, A\&SS, 354, 89

\bibitem[{{Brown} {et~al.}(2007){Brown}, {Dessart}, {Holland}, {Immler},
  {Landsman}, {Blondin}, {Blustin}, {Breeveld}, {Dewangan}, {Gehrels},
  {Hutchins}, {Kirshner}, {Mason}, {Mazzali}, {Milne}, {Modjaz}, \&
  {Roming}}]{2007brown}
{Brown}, P.~J., {Dessart}, L., {Holland}, S.~T., {et~al.} 2007, \apj, 659, 1488

\bibitem[{{Cardelli} {et~al.}(1989){Cardelli}, {Clayton}, \&
  {Mathis}}]{1989cardelli}
{Cardelli}, J.~A., {Clayton}, G.~C., \& {Mathis}, J.~S. 1989, \apj, 345, 245

\bibitem[{{Chabrier}(2003)}]{2003chabrier}
{Chabrier}, G. 2003, \pasp, 115, 763

\bibitem[{{Chevalier} \& {Fransson}(1994)}]{1994chevalier}
{Chevalier}, R.~A., \& {Fransson}, C. 1994, \apj, 420, 268

\bibitem[{{Chugai}(1994)}]{1994chugai}
{Chugai}, N.~N. 1994, \apjl, 428, L17

\bibitem[{{Chugai} {et~al.}(2007){Chugai}, {Chevalier}, \&
  {Utrobin}}]{2007chugai}
{Chugai}, N.~N., {Chevalier}, R.~A., \& {Utrobin}, V.~P. 2007, \apj, 662, 1136

\bibitem[{{Colgate}(1974)}]{1974colgate}
{Colgate}, S.~A. 1974, \apj, 187, 333

\bibitem[{{Cowen} {et~al.}(2010){Cowen}, {Franckowiak}, \&
  {Kowalski}}]{2010cowen}
{Cowen}, D.~F., {Franckowiak}, A., \& {Kowalski}, M. 2010, Astroparticle
  Physics, 33, 19

\bibitem[{{Das} \& {Ray}(2017)}]{2017das}
{Das}, S., \& {Ray}, A. 2017, \apj, 851, 138

\bibitem[{{Davies} \& {Beasor}(2018)}]{2018davies}
{Davies}, B., \& {Beasor}, E.~R. 2018, \mnras, 474, 2116

\bibitem[{{Davis} {et~al.}(1997){Davis}, {Keel}, {Mulchaey}, \&
  {Henning}}]{1997davis}
{Davis}, D.~S., {Keel}, W.~C., {Mulchaey}, J.~S., \& {Henning}, P.~A. 1997,
  \aj, 114, 613

\bibitem[{{de Jaeger} {et~al.}(2015){de Jaeger}, {Gonz{\'a}lez-Gait{\'a}n},
  {Anderson}, {Galbany}, {Hamuy}, {Phillips}, {Stritzinger}, {Guti{\'e}rrez},
  {Bolt}, {Burns}, {Campillay}, {Castell{\'o}n}, {Contreras}, {Folatelli},
  {Freedman}, {Hsiao}, {Krisciunas}, {Krzeminski}, {Kuncarayakti}, {Morrell},
  {Olivares E.}, {Persson}, \& {Suntzeff}}]{2015dejaeger}
{de Jaeger}, T., {Gonz{\'a}lez-Gait{\'a}n}, S., {Anderson}, J.~P., {et~al.}
  2015, \apj, 815, 121

\bibitem[{{de Jaeger} {et~al.}(2018{\natexlab{a}}){de Jaeger}, {Anderson},
  {Galbany}, {Gonz{\'a}lez-Gait{\'a}n}, {Hamuy}, {Phillips}, {Stritzinger},
  {Contreras}, {Folatelli}, {Guti{\'e}rrez}, {Hsiao}, {Morrell}, {Suntzeff},
  {Dessart}, \& {Filippenko}}]{2018adejaeger}
{de Jaeger}, T., {Anderson}, J.~P., {Galbany}, L., {et~al.} 2018{\natexlab{a}},
  \mnras, 476, 4592

\bibitem[{{de Jaeger} {et~al.}(2018{\natexlab{b}}){de Jaeger}, {Galbany},
  {Guti{\'e}rrez}, {Filippenko}, {Zheng}, {Brink}, {Foley}, {S{\'a}nchez},
  {Channa}, {de Kouchkovsky}, {Halevi}, {Kilpatrick}, {Kumar}, {Molloy}, {Pan},
  {Ross}, {Shivvers}, {Siebert}, {Stahl}, {Stegman}, \&
  {Yunus}}]{2018bdejaeger}
{de Jaeger}, T., {Galbany}, L., {Guti{\'e}rrez}, C.~P., {et~al.}
  2018{\natexlab{b}}, \mnras, 478, 3776

\bibitem[{{de Vaucouleurs} {et~al.}(1991){de Vaucouleurs}, {de Vaucouleurs},
  {Corwin}, {Buta}, {Paturel}, \& {Fouqu{\'e}}}]{1991devac}
{de Vaucouleurs}, G., {de Vaucouleurs}, A., {Corwin}, Jr., H.~G., {et~al.}
  1991, {Third Reference Catalogue of Bright Galaxies. Volume I: Explanations
  and references. Volume II: Data for galaxies between 0$^{h}$ and 12$^{h}$.
  Volume III: Data for galaxies between 12$^{h}$ and 24$^{h}$.}
  (Springer-Verlag, New York)

\bibitem[{{Dessart} \& {Hillier}(2005)}]{2005adessart}
{Dessart}, L., \& {Hillier}, D.~J. 2005, \aap, 439, 671

\bibitem[{{Dessart} {et~al.}(2013){Dessart}, {Hillier}, {Waldman}, \&
  {Livne}}]{2013bdessart}
{Dessart}, L., {Hillier}, D.~J., {Waldman}, R., \& {Livne}, E. 2013, \mnras,
  433, 1745

\bibitem[{{Dessart} {et~al.}(2014){Dessart}, {Gutierrez}, {Hamuy}, {Hillier},
  {Lanz}, {Anderson}, {Folatelli}, {Freedman}, {Ley}, {Morrell}, {Persson},
  {Phillips}, {Stritzinger}, \& {Suntzeff}}]{2014dessart}
{Dessart}, L., {Gutierrez}, C.~P., {Hamuy}, M., {et~al.} 2014, \mnras, 440,
  1856

\bibitem[{{Dickinson} {et~al.}(2003){Dickinson}, {Papovich}, {Ferguson}, \&
  {Budav{\'a}ri}}]{2003dickinson}
{Dickinson}, M., {Papovich}, C., {Ferguson}, H.~C., \& {Budav{\'a}ri}, T. 2003,
  \apj, 587, 25

\bibitem[{{Dimai}(2016)}]{2016discovery}
{Dimai}, A. 2016, Transient Name Server Discovery Report, 673

\bibitem[{{Dimai} {et~al.}(2005){Dimai}, {Migliardi}, \& {Manzini}}]{2005dimai}
{Dimai}, A., {Migliardi}, M., \& {Manzini}, F. 2005, \iaucirc, 8588

\bibitem[{{Dom{\'{\i}}nguez} {et~al.}(2013){Dom{\'{\i}}nguez}, {Siana},
  {Henry}, {Scarlata}, {Bedregal}, {Malkan}, {Atek}, {Ross}, {Colbert},
  {Teplitz}, {Rafelski}, {McCarthy}, {Bunker}, {Hathi}, {Dressler}, {Martin},
  \& {Masters}}]{2013dominguez}
{Dom{\'{\i}}nguez}, A., {Siana}, B., {Henry}, A.~L., {et~al.} 2013, \apj, 763,
  145

\bibitem[{{Eldridge} {et~al.}(2017){Eldridge}, {Stanway}, {Xiao}, {McClelland},
  {Taylor}, {Ng}, {Greis}, \& {Bray}}]{2017eldridge}
{Eldridge}, J.~J., {Stanway}, E.~R., {Xiao}, L., {et~al.} 2017, \pasa, 34, e058

\bibitem[{{Elias-Rosa} {et~al.}(2011){Elias-Rosa}, {Van Dyk}, {Li},
  {Silverman}, {Foley}, {Ganeshalingam}, {Mauerhan}, {Kankare}, {Jha},
  {Filippenko}, {Beckman}, {Berger}, {Cuillandre}, \& {Smith}}]{2011elias}
{Elias-Rosa}, N., {Van Dyk}, S.~D., {Li}, W., {et~al.} 2011, \apj, 742, 6

\bibitem[{{Elmhamdi} {et~al.}(2003{\natexlab{a}}){Elmhamdi}, {Chugai}, \&
  {Danziger}}]{2003elmhamdi}
{Elmhamdi}, A., {Chugai}, N.~N., \& {Danziger}, I.~J. 2003{\natexlab{a}}, \aap,
  404, 1077

\bibitem[{{Elmhamdi} {et~al.}(2003{\natexlab{b}}){Elmhamdi}, {Danziger},
  {Chugai}, {Pastorello}, {Turatto}, {Cappellaro}, {Altavilla}, {Benetti},
  {Patat}, \& {Salvo}}]{2003aelmhamdi}
{Elmhamdi}, A., {Danziger}, I.~J., {Chugai}, N., {et~al.} 2003{\natexlab{b}},
  \mnras, 338, 939

\bibitem[{{Epinat} {et~al.}(2008){Epinat}, {Amram}, \& {Marcelin}}]{2008epinat}
{Epinat}, B., {Amram}, P., \& {Marcelin}, M. 2008, \mnras, 390, 466

\bibitem[{{Falk} \& {Arnett}(1977)}]{1977falk}
{Falk}, S.~W., \& {Arnett}, W.~D. 1977, \apjs, 33, 515

\bibitem[{{Faran} {et~al.}(2014{\natexlab{a}}){Faran}, {Poznanski},
  {Filippenko}, {Chornock}, {Foley}, {Ganeshalingam}, {Leonard}, {Li},
  {Modjaz}, {Serduke}, \& {Silverman}}]{2014bfaran}
{Faran}, T., {Poznanski}, D., {Filippenko}, A.~V., {et~al.} 2014{\natexlab{a}},
  \mnras, 445, 554

\bibitem[{{Faran} {et~al.}(2014{\natexlab{b}}){Faran}, {Poznanski},
  {Filippenko}, {Chornock}, {Foley}, {Ganeshalingam}, {Leonard}, {Li},
  {Modjaz}, {Nakar}, {Serduke}, \& {Silverman}}]{2014afaran}
---. 2014{\natexlab{b}}, \mnras, 442, 844

\bibitem[{{Filippenko}(1982)}]{1982filippenko}
{Filippenko}, A.~V. 1982, \pasp, 94, 715

\bibitem[{{Filippenko}(1997)}]{1997filippenko}
---. 1997, \araa, 35, 309

\bibitem[{{Fitzpatrick}(1999)}]{1999fitzpatrick}
{Fitzpatrick}, E.~L. 1999, \pasp, 111, 63

\bibitem[{{Forster} {et~al.}(2018){Forster}, {Moriya}, {Maureira}, {Anderson},
  {Blinnikov}, {Bufano}, {Cabrera-Vives}, {Clocchiatti}, {de Jaeger},
  {Estevez}, {Galbany}, {Gonzalez-Gaitan}, {Grafener}, {Hamuy}, {Hsiao},
  {Huentelemu}, {Huijse}, {Kuncarayakti}, {Martinez}, {Medina}, {Olivares},
  {Pignata}, {Razza}, {Reyes}, {San}, {Smith}, {Vera}, {Vivas}, {de Ugarte
  Postigo}, {Yoon}, {Ashall}, {Fraser}, {Gal-Yam}, {Kankare}, {Le Guillou},
  {Mazzali}, {Walton}, \& {Young}}]{2018forster}
{Forster}, F., {Moriya}, T.~J., {Maureira}, J.~C., {et~al.} 2018, Nature
  Astronomy, 2, 808

\bibitem[{{Fransson} \& {Chevalier}(1989)}]{1989fransson}
{Fransson}, C., \& {Chevalier}, R.~A. 1989, \apj, 343, 323

\bibitem[{{Gall} {et~al.}(2015){Gall}, {Polshaw}, {Kotak}, {Jerkstrand},
  {Leibundgut}, {Rabinowitz}, {Sollerman}, {Sullivan}, {Smartt}, {Anderson},
  {Benetti}, {Baltay}, {Feindt}, {Fraser}, {Gonz{\'a}lez-Gait{\'a}n},
  {Inserra}, {Maguire}, {McKinnon}, {Valenti}, \& {Young}}]{2015gall}
{Gall}, E.~E.~E., {Polshaw}, J., {Kotak}, R., {et~al.} 2015, \aap, 582, A3

\bibitem[{{Gehrels} {et~al.}(2004){Gehrels}, {Chincarini}, {Giommi}, {Mason},
  {Nousek}, {Wells}, {White}, {Barthelmy}, {Burrows}, {Cominsky}, {Hurley},
  {Marshall}, {M{\'e}sz{\'a}ros}, {Roming}, {Angelini}, {Barbier}, {Belloni},
  {Campana}, {Caraveo}, {Chester}, {Citterio}, {Cline}, {Cropper}, {Cummings},
  {Dean}, {Feigelson}, {Fenimore}, {Frail}, {Fruchter}, {Garmire}, {Gendreau},
  {Ghisellini}, {Greiner}, {Hill}, {Hunsberger}, {Krimm}, {Kulkarni}, {Kumar},
  {Lebrun}, {Lloyd-Ronning}, {Markwardt}, {Mattson}, {Mushotzky}, {Norris},
  {Osborne}, {Paczynski}, {Palmer}, {Park}, {Parsons}, {Paul}, {Rees},
  {Reynolds}, {Rhoads}, {Sasseen}, {Schaefer}, {Short}, {Smale}, {Smith},
  {Stella}, {Tagliaferri}, {Takahashi}, {Tashiro}, {Townsley}, {Tueller},
  {Turner}, {Vietri}, {Voges}, {Ward}, {Willingale}, {Zerbi}, \&
  {Zhang}}]{2004gehrels}
{Gehrels}, N., {Chincarini}, G., {Giommi}, P., {et~al.} 2004, \apj, 611, 1005

\bibitem[{{Gonz{\'a}lez-Gait{\'a}n} {et~al.}(2015){Gonz{\'a}lez-Gait{\'a}n},
  {Tominaga}, {Molina}, {Galbany}, {Bufano}, {Anderson}, {Gutierrez},
  {F{\"o}rster}, {Pignata}, {Bersten}, {Howell}, {Sullivan}, {Carlberg}, {de
  Jaeger}, {Hamuy}, {Baklanov}, \& {Blinnikov}}]{2015gonzalez}
{Gonz{\'a}lez-Gait{\'a}n}, S., {Tominaga}, N., {Molina}, J., {et~al.} 2015,
  \mnras, 451, 2212

\bibitem[{{Guti{\'e}rrez} {et~al.}(2017){Guti{\'e}rrez}, {Anderson}, {Hamuy},
  {Morrell}, {Gonz{\'a}lez-Gaitan}, {Stritzinger}, {Phillips}, {Galbany},
  {Folatelli}, {Dessart}, {Contreras}, {Della Valle}, {Freedman}, {Hsiao},
  {Krisciunas}, {Madore}, {Maza}, {Suntzeff}, {Prieto}, {Gonz{\'a}lez},
  {Cappellaro}, {Navarrete}, {Pizzella}, {Ruiz}, {Smith}, \&
  {Turatto}}]{2017gutierrez}
{Guti{\'e}rrez}, C.~P., {Anderson}, J.~P., {Hamuy}, M., {et~al.} 2017, \apj,
  850, 89

\bibitem[{{Guti{\'e}rrez} {et~al.}(2018){Guti{\'e}rrez}, {Anderson},
  {Sullivan}, {Dessart}, {Gonz{\'a}lez-Gaitan}, {Galbany}, {Dimitriadis},
  {Arcavi}, {Bufano}, {Chen}, {Dennefeld}, {Gromadzki}, {Haislip},
  {Hosseinzadeh}, {Howell}, {Inserra}, {Kankare}, {Leloudas}, {Maguire},
  {McCully}, {Morrell}, {E}, {Pignata}, {Reichart}, {Reynolds}, {Smartt},
  {Sollerman}, {Taddia}, {Tak{\'a}ts}, {Terreran}, {Valenti}, \&
  {Young}}]{2018gutierrez}
{Guti{\'e}rrez}, C.~P., {Anderson}, J.~P., {Sullivan}, M., {et~al.} 2018,
  \mnras, arXiv:1806.03855

\bibitem[{{Hamuy}(2003)}]{2003hamuy}
{Hamuy}, M. 2003, \apj, 582, 905

\bibitem[{{Hamuy}(2004)}]{2004hamuy}
---. 2004, Measuring and Modeling the Universe, 2

\bibitem[{{Hamuy} \& {Pinto}(2002)}]{2002hamuy}
{Hamuy}, M., \& {Pinto}, P.~A. 2002, \apjl, 566, L63

\bibitem[{{Hamuy} \& {Suntzeff}(1990)}]{1990hamuy}
{Hamuy}, M., \& {Suntzeff}, N.~B. 1990, \aj, 99, 1146

\bibitem[{{Hamuy} {et~al.}(2001){Hamuy}, {Pinto}, {Maza}, {Suntzeff},
  {Phillips}, {Eastman}, {Smith}, {Corbally}, {Burstein}, {Li}, {Ivanov},
  {Moro-Martin}, {Strolger}, {de Souza}, {dos Anjos}, {Green}, {Pickering},
  {Gonz{\'a}lez}, {Antezana}, {Wischnjewsky}, {Galaz}, {Roth}, {Persson}, \&
  {Schommer}}]{2001hamuy}
{Hamuy}, M., {Pinto}, P.~A., {Maza}, J., {et~al.} 2001, \apj, 558, 615

\bibitem[{{Heger} {et~al.}(2003){Heger}, {Fryer}, {Woosley}, {Langer}, \&
  {Hartmann}}]{2003heger}
{Heger}, A., {Fryer}, C.~L., {Woosley}, S.~E., {Langer}, N., \& {Hartmann},
  D.~H. 2003, \apj, 591, 288

\bibitem[{{Henry} \& {Worthey}(1999)}]{1999henry}
{Henry}, R.~B.~C., \& {Worthey}, G. 1999, \pasp, 111, 919

\bibitem[{{Horiuchi} {et~al.}(2014){Horiuchi}, {Nakamura}, {Takiwaki},
  {Kotake}, \& {Tanaka}}]{2014horiuchi}
{Horiuchi}, S., {Nakamura}, K., {Takiwaki}, T., {Kotake}, K., \& {Tanaka}, M.
  2014, \mnras, 445, L99

\bibitem[{{Huang} {et~al.}(2018){Huang}, {Wang}, {Hosseinzadeh}, {Brown}, {Mo},
  {Zhang}, {Zhang}, {Zhang}, {Howell}, {Arcavi}, {McCully}, {Valenti}, {Rui},
  {Song}, {Xiang}, {Li}, {Lin}, \& {Wang}}]{2018huang}
{Huang}, F., {Wang}, X.-F., {Hosseinzadeh}, G., {et~al.} 2018, \mnras, 475,
  3959

\bibitem[{Hunter(2007)}]{2007hunter}
Hunter, J.~D. 2007, Computing In Science \& Engineering, 9, 90

\bibitem[{{Inserra} {et~al.}(2011){Inserra}, {Turatto}, {Pastorello},
  {Benetti}, {Cappellaro}, {Pumo}, {Zampieri}, {Agnoletto}, {Bufano},
  {Botticella}, {Della Valle}, {Elias Rosa}, {Iijima}, {Spiro}, \&
  {Valenti}}]{2011inserra}
{Inserra}, C., {Turatto}, M., {Pastorello}, A., {et~al.} 2011, \mnras, 417, 261

\bibitem[{{Iskudarian}(1968)}]{1968iskudarian}
{Iskudarian}, S.~G. 1968, Astronomicheskij Tsirkulyar, 480, 1

\bibitem[{{Iskudaryan} \& {Shakhbazyan}(1967)}]{1967iskudarian}
{Iskudaryan}, S.~G., \& {Shakhbazyan}, R.~K. 1967, Astrophysics, 3, 67

\bibitem[{{Jerkstrand} {et~al.}(2012){Jerkstrand}, {Fransson}, {Maguire},
  {Smartt}, {Ergon}, \& {Spyromilio}}]{2012jerkstrand}
{Jerkstrand}, A., {Fransson}, C., {Maguire}, K., {et~al.} 2012, \aap, 546, A28

\bibitem[{{Jerkstrand} {et~al.}(2014){Jerkstrand}, {Smartt}, {Fraser},
  {Fransson}, {Sollerman}, {Taddia}, \& {Kotak}}]{2014jerkstrand}
{Jerkstrand}, A., {Smartt}, S.~J., {Fraser}, M., {et~al.} 2014, \mnras, 439,
  3694

\bibitem[{{Jordi} {et~al.}(2006){Jordi}, {Grebel}, \& {Ammon}}]{2006jordi}
{Jordi}, K., {Grebel}, E.~K., \& {Ammon}, K. 2006, \aap, 460, 339

\bibitem[{{Karachentsev} \& {Kaisina}(2013)}]{2013karachentsev}
{Karachentsev}, I.~D., \& {Kaisina}, E.~I. 2013, \aj, 146, 46

\bibitem[{{Kasen} \& {Woosley}(2009)}]{2009kasen}
{Kasen}, D., \& {Woosley}, S.~E. 2009, \apj, 703, 2205

\bibitem[{{Kauffmann} {et~al.}(2003){Kauffmann}, {Heckman}, {Tremonti},
  {Brinchmann}, {Charlot}, {White}, {Ridgway}, {Brinkmann}, {Fukugita}, {Hall},
  {Ivezi{\'c}}, {Richards}, \& {Schneider}}]{2003kauffmann}
{Kauffmann}, G., {Heckman}, T.~M., {Tremonti}, C., {et~al.} 2003, \mnras, 346,
  1055

\bibitem[{{Kennicutt}(1984)}]{1984kennicutt}
{Kennicutt}, Jr., R.~C. 1984, \apj, 277, 361

\bibitem[{{Kennicutt}(1998)}]{1998kennicutt}
---. 1998, \araa, 36, 189

\bibitem[{{Khazov} {et~al.}(2016){Khazov}, {Yaron}, {Gal-Yam}, {Manulis},
  {Rubin}, {Kulkarni}, {Arcavi}, {Kasliwal}, {Ofek}, {Cao}, {Perley},
  {Sollerman}, {Horesh}, {Sullivan}, {Filippenko}, {Nugent}, {Howell}, {Cenko},
  {Silverman}, {Ebeling}, {Taddia}, {Johansson}, {Laher}, {Surace},
  {Rebbapragada}, {Wozniak}, \& {Matheson}}]{2016khazov}
{Khazov}, D., {Yaron}, O., {Gal-Yam}, A., {et~al.} 2016, \apj, 818, 3

\bibitem[{{Kirshner} \& {Kwan}(1974)}]{1974kirshner}
{Kirshner}, R.~P., \& {Kwan}, J. 1974, \apj, 193, 27

\bibitem[{{Kochanek} {et~al.}(2012){Kochanek}, {Khan}, \& {Dai}}]{2012kochanek}
{Kochanek}, C.~S., {Khan}, R., \& {Dai}, X. 2012, \apj, 759, 20

\bibitem[{{Kozyreva} {et~al.}(2019){Kozyreva}, {Nakar}, \&
  {Waldman}}]{2019kozyreva}
{Kozyreva}, A., {Nakar}, E., \& {Waldman}, R. 2019, \mnras, 483, 1211

\bibitem[{{Kumar} {et~al.}(2018){Kumar}, {Singh}, {Srivastav}, {Sahu}, \&
  {Anupama}}]{2018brajesh}
{Kumar}, B., {Singh}, A., {Srivastav}, S., {Sahu}, D.~K., \& {Anupama}, G.~C.
  2018, \mnras, 473, 3776

\bibitem[{{Kuncarayakti} {et~al.}(2013{\natexlab{a}}){Kuncarayakti}, {Doi},
  {Aldering}, {Arimoto}, {Maeda}, {Morokuma}, {Pereira}, {Usuda}, \&
  {Hashiba}}]{2013akuncarayakti}
{Kuncarayakti}, H., {Doi}, M., {Aldering}, G., {et~al.} 2013{\natexlab{a}},
  \aj, 146, 30

\bibitem[{{Kuncarayakti} {et~al.}(2013{\natexlab{b}}){Kuncarayakti}, {Doi},
  {Aldering}, {Arimoto}, {Maeda}, {Morokuma}, {Pereira}, {Usuda}, \&
  {Hashiba}}]{2013bkuncarayakti}
---. 2013{\natexlab{b}}, \aj, 146, 31

\bibitem[{{Kuncarayakti} {et~al.}(2016){Kuncarayakti}, {Reynolds}, {Mattila},
  {Elias-Rosa}, {Lundqvist}, {Stritzinger}, {Harmanen}, {Pastorello},
  {Benetti}, {Cappellaro}, {Blagorodnova}, {Davis}, {Dong}, {Fraser}, {Gall},
  {Harrison}, {Hodgkin}, {Hsiao}, {Jonker}, {Kangas}, {Kankare},
  {Kostrzewa-Rutkowska}, {Nielsen}, {Ochner}, {Prieto}, {Romero-Canizales},
  {Taddia}, {Tartaglia}, {Terreran}, {Tomasella}, \&
  {Wyrzykowski}}]{2016classify}
{Kuncarayakti}, H., {Reynolds}, T., {Mattila}, S., {et~al.} 2016, The
  Astronomer's Telegram, 9498

\bibitem[{{Landolt}(1992)}]{1992landolt}
{Landolt}, A.~U. 1992, \aj, 104, 340

\bibitem[{{Leonard} {et~al.}(2003){Leonard}, {Kanbur}, {Ngeow}, \&
  {Tanvir}}]{2003leonard}
{Leonard}, D.~C., {Kanbur}, S.~M., {Ngeow}, C.~C., \& {Tanvir}, N.~R. 2003,
  \apj, 594, 247

\bibitem[{{Leonard} {et~al.}(2002{\natexlab{a}}){Leonard}, {Filippenko},
  {Gates}, {Li}, {Eastman}, {Barth}, {Bus}, {Chornock}, {Coil}, {Frink},
  {Grady}, {Harris}, {Malkan}, {Matheson}, {Quirrenbach}, \&
  {Treffers}}]{2002bleonard}
{Leonard}, D.~C., {Filippenko}, A.~V., {Gates}, E.~L., {et~al.}
  2002{\natexlab{a}}, \pasp, 114, 35

\bibitem[{{Leonard} {et~al.}(2002{\natexlab{b}}){Leonard}, {Filippenko}, {Li},
  {Matheson}, {Kirshner}, {Chornock}, {Van Dyk}, {Berlind}, {Calkins},
  {Challis}, {Garnavich}, {Jha}, \& {Mahdavi}}]{2002aleonard}
{Leonard}, D.~C., {Filippenko}, A.~V., {Li}, W., {et~al.} 2002{\natexlab{b}},
  \aj, 124, 2490

\bibitem[{{Li} {et~al.}(2011){Li}, {Leaman}, {Chornock}, {Filippenko},
  {Poznanski}, {Ganeshalingam}, {Wang}, {Modjaz}, {Jha}, {Foley}, \&
  {Smith}}]{2011li}
{Li}, W., {Leaman}, J., {Chornock}, R., {et~al.} 2011, \mnras, 412, 1441

\bibitem[{{Litvinova} \& {Nadezhin}(1985)}]{1985litvinova}
{Litvinova}, I.~Y., \& {Nadezhin}, D.~K. 1985, Soviet Astronomy Letters, 11,
  145

\bibitem[{{Lovegrove} \& {Woosley}(2013)}]{2013lovegrove}
{Lovegrove}, E., \& {Woosley}, S.~E. 2013, \apj, 769, 109

\bibitem[{{Maguire} {et~al.}(2010){Maguire}, {Di Carlo}, {Smartt},
  {Pastorello}, {Tsvetkov}, {Benetti}, {Spiro}, {Arkharov}, {Beccari},
  {Botticella}, {Cappellaro}, {Cristallo}, {Dolci}, {Elias-Rosa}, {Fiaschi},
  {Gorshanov}, {Harutyunyan}, {Larionov}, {Navasardyan}, {Pietrinferni},
  {Raimondo}, {di Rico}, {Valenti}, {Valentini}, \& {Zampieri}}]{2010maguire}
{Maguire}, K., {Di Carlo}, E., {Smartt}, S.~J., {et~al.} 2010, \mnras, 404, 981

\bibitem[{{McCall}(2004)}]{2004mccall}
{McCall}, M.~L. 2004, \aj, 128, 2144

\bibitem[{McKinney(2010)}]{2010pandas}
McKinney, W. 2010, in Proceedings of the 9th Python in Science Conference, ed.
  S.~van~der Walt \& J.~Millman, 51 -- 56

\bibitem[{{Meza} {et~al.}(2018){Meza}, {Prieto}, {Clocchiatti}, {Galbany},
  {Anderson}, {Falco}, {Kochanek}, {Kuncarayakti}, {Brimacombe}, {Holoien},
  {Shappee}, {Stanek}, \& {Thompson}}]{2018meza}
{Meza}, N., {Prieto}, J.~L., {Clocchiatti}, A., {et~al.} 2018, arXiv e-prints,
  arXiv:1811.11771

\bibitem[{{Minkowski}(1941)}]{1941minkowski}
{Minkowski}, R. 1941, \pasp, 53, 224

\bibitem[{{Moriya} {et~al.}(2018){Moriya}, {F{\"o}rster}, {Yoon},
  {Gr{\"a}fener}, \& {Blinnikov}}]{2018moriya}
{Moriya}, T.~J., {F{\"o}rster}, F., {Yoon}, S.-C., {Gr{\"a}fener}, G., \&
  {Blinnikov}, S.~I. 2018, \mnras, 476, 2840

\bibitem[{{Moriya} {et~al.}(2017){Moriya}, {Yoon}, {Gr{\"a}fener}, \&
  {Blinnikov}}]{2017moriya}
{Moriya}, T.~J., {Yoon}, S.-C., {Gr{\"a}fener}, G., \& {Blinnikov}, S.~I. 2017,
  \mnras, 469, L108

\bibitem[{{Morozova} {et~al.}(2016){Morozova}, {Piro}, {Renzo}, \&
  {Ott}}]{2016morozova}
{Morozova}, V., {Piro}, A.~L., {Renzo}, M., \& {Ott}, C.~D. 2016, \apj, 829,
  109

\bibitem[{{Morozova} {et~al.}(2015){Morozova}, {Piro}, {Renzo}, {Ott},
  {Clausen}, {Couch}, {Ellis}, \& {Roberts}}]{2015morozova}
{Morozova}, V., {Piro}, A.~L., {Renzo}, M., {et~al.} 2015, \apj, 814, 63

\bibitem[{{Morozova} {et~al.}(2017){Morozova}, {Piro}, \&
  {Valenti}}]{2017morozova}
{Morozova}, V., {Piro}, A.~L., \& {Valenti}, S. 2017, \apj, 838, 28

\bibitem[{{Morrissey} {et~al.}(2007){Morrissey}, {Conrow}, {Barlow}, {Small},
  {Seibert}, {Wyder}, {Budav{\'a}ri}, {Arnouts}, {Friedman}, {Forster},
  {Martin}, {Neff}, {Schiminovich}, {Bianchi}, {Donas}, {Heckman}, {Lee},
  {Madore}, {Milliard}, {Rich}, {Szalay}, {Welsh}, \& {Yi}}]{2007morrissey}
{Morrissey}, P., {Conrow}, T., {Barlow}, T.~A., {et~al.} 2007, \apjs, 173, 682

\bibitem[{{Mould} {et~al.}(2000){Mould}, {Huchra}, {Freedman}, {Kennicutt},
  {Ferrarese}, {Ford}, {Gibson}, {Graham}, {Hughes}, {Illingworth}, {Kelson},
  {Macri}, {Madore}, {Sakai}, {Sebo}, {Silbermann}, \& {Stetson}}]{2000mould}
{Mould}, J.~R., {Huchra}, J.~P., {Freedman}, W.~L., {et~al.} 2000, \apj, 529,
  786

\bibitem[{{Nagy} {et~al.}(2014){Nagy}, {Ordasi}, {Vink{\'o}}, \&
  {Wheeler}}]{2014nagy}
{Nagy}, A.~P., {Ordasi}, A., {Vink{\'o}}, J., \& {Wheeler}, J.~C. 2014, \aap,
  571, A77

\bibitem[{{Nagy} \& {Vink{\'o}}(2016)}]{2016nagy}
{Nagy}, A.~P., \& {Vink{\'o}}, J. 2016, \aap, 589, A53

\bibitem[{{Nakaoka} {et~al.}(2018){Nakaoka}, {Kawabata}, {Maeda}, {Tanaka},
  {Yamanaka}, {Moriya}, {Tominaga}, {Morokuma}, {Takaki}, {Kawabata},
  {Kawahara}, {Itoh}, {Shiki}, {Mori}, {Hirochi}, {Abe}, {Uemura}, {Yoshida},
  {Akitaya}, {Moritani}, {Ueno}, {Urano}, {Isogai}, {Hanayama}, \&
  {Nagayama}}]{2018nakaoka}
{Nakaoka}, T., {Kawabata}, K.~S., {Maeda}, K., {et~al.} 2018, \apj, 859, 78

\bibitem[{{Nakar} {et~al.}(2016){Nakar}, {Poznanski}, \& {Katz}}]{2016nakar}
{Nakar}, E., {Poznanski}, D., \& {Katz}, B. 2016, \apj, 823, 127

\bibitem[{{Nugent} {et~al.}(2006){Nugent}, {Sullivan}, {Ellis}, {Gal-Yam},
  {Leonard}, {Howell}, {Astier}, {Carlberg}, {Conley}, {Fabbro}, {Fouchez},
  {Neill}, {Pain}, {Perrett}, {Pritchet}, \& {Regnault}}]{2006nugent}
{Nugent}, P., {Sullivan}, M., {Ellis}, R., {et~al.} 2006, \apj, 645, 841

\bibitem[{{Oke} \& {Gunn}(1983)}]{1983oke}
{Oke}, J.~B., \& {Gunn}, J.~E. 1983, \apj, 266, 713

\bibitem[{Oliphant(2007)}]{2007scipy}
Oliphant, T.~E. 2007, Computing in Science \& Engineering, 9, 10.
\newblock \url{https://aip.scitation.org/doi/abs/10.1109/MCSE.2007.58}

\bibitem[{Olivares {et~al.}(2010)Olivares, Hamuy, Pignata, Maza, Bersten,
  Phillips, Suntzeff, Filippenko, Morrel, Kirshner, \& Matheson}]{2010olivares}
Olivares, F., Hamuy, M., Pignata, G., {et~al.} 2010, The Astrophysical Journal,
  715, 833.
\newblock \url{http://stacks.iop.org/0004-637X/715/i=2/a=833}

\bibitem[{{Osterbrock}(1989)}]{1989osterbrock}
{Osterbrock}, D.~E. 1989, {Astrophysics of gaseous nebulae and active galactic
  nuclei} (University Science Books, Mill Valley, California)

\bibitem[{{Pastorello} {et~al.}(2009){Pastorello}, {Valenti}, {Zampieri},
  {Navasardyan}, {Taubenberger}, {Smartt}, {Arkharov}, {B{\"a}rnbantner},
  {Barwig}, {Benetti}, {Birtwhistle}, {Botticella}, {Cappellaro}, {Del
  Principe}, {di Mille}, {di Rico}, {Dolci}, {Elias-Rosa}, {Efimova},
  {Fiedler}, {Harutyunyan}, {H{\"o}flich}, {Kloehr}, {Larionov}, {Lorenzi},
  {Maund}, {Napoleone}, {Ragni}, {Richmond}, {Ries}, {Spiro}, {Temporin},
  {Turatto}, \& {Wheeler}}]{2009pastorello}
{Pastorello}, A., {Valenti}, S., {Zampieri}, L., {et~al.} 2009, \mnras, 394,
  2266

\bibitem[{{Patat} {et~al.}(1994){Patat}, {Barbon}, {Cappellaro}, \&
  {Turatto}}]{1994patat}
{Patat}, F., {Barbon}, R., {Cappellaro}, E., \& {Turatto}, M. 1994, \aap, 282,
  731

\bibitem[{{Pettini} \& {Pagel}(2004)}]{2004pettini}
{Pettini}, M., \& {Pagel}, B.~E.~J. 2004, \mnras, 348, L59

\bibitem[{{Polshaw} {et~al.}(2016){Polshaw}, {Kotak}, {Dessart}, {Fraser},
  {Gal-Yam}, {Inserra}, {Sim}, {Smartt}, {Sollerman}, {Baltay}, {Rabinowitz},
  {Benetti}, {Botticella}, {Campbell}, {Chen}, {Galbany}, {McKinnon},
  {Nicholl}, {Smith}, {Sullivan}, {Tak{\'a}ts}, {Valenti}, \&
  {Young}}]{2016polshaw}
{Polshaw}, J., {Kotak}, R., {Dessart}, L., {et~al.} 2016, \aap, 588,
  doi:10.1051/0004-6361/201527682

\bibitem[{{Poznanski} {et~al.}(2010){Poznanski}, {Nugent}, \&
  {Filippenko}}]{2010poznanski}
{Poznanski}, D., {Nugent}, P.~E., \& {Filippenko}, A.~V. 2010, \apj, 721, 956

\bibitem[{{Poznanski} {et~al.}(2012){Poznanski}, {Prochaska}, \&
  {Bloom}}]{2012poznanski}
{Poznanski}, D., {Prochaska}, J.~X., \& {Bloom}, J.~S. 2012, \mnras, 426, 1465

\bibitem[{{Poznanski} {et~al.}(2009){Poznanski}, {Butler}, {Filippenko},
  {Ganeshalingam}, {Li}, {Bloom}, {Chornock}, {Foley}, {Nugent}, {Silverman},
  {Cenko}, {Gates}, {Leonard}, {Miller}, {Modjaz}, {Serduke}, {Smith}, {Swift},
  \& {Wong}}]{2009poznanski}
{Poznanski}, D., {Butler}, N., {Filippenko}, A.~V., {et~al.} 2009, \apj, 694,
  1067

\bibitem[{{Prieto} {et~al.}(2008){Prieto}, {Stanek}, \& {Beacom}}]{2008prieto}
{Prieto}, J.~L., {Stanek}, K.~Z., \& {Beacom}, J.~F. 2008, \apj, 673, 999

\bibitem[{{Pritchard} {et~al.}(2014){Pritchard}, {Roming}, {Brown}, {Bayless},
  \& {Frey}}]{2014pritchard}
{Pritchard}, T.~A., {Roming}, P.~W.~A., {Brown}, P.~J., {Bayless}, A.~J., \&
  {Frey}, L.~H. 2014, \apj, 787, 157

\bibitem[{{Quataert} \& {Shiode}(2012)}]{2012quataert}
{Quataert}, E., \& {Shiode}, J. 2012, \mnras, 423, L92

\bibitem[{{Rabinak} \& {Waxman}(2011)}]{2011rabinak}
{Rabinak}, I., \& {Waxman}, E. 2011, \apj, 728, 63

\bibitem[{{Rasmussen} {et~al.}(2006){Rasmussen}, {Ponman}, \&
  {Mulchaey}}]{2006rasmussen}
{Rasmussen}, J., {Ponman}, T.~J., \& {Mulchaey}, J.~S. 2006, \mnras, 370, 453

\bibitem[{{Riess} {et~al.}(2018){Riess}, {Casertano}, {Yuan}, {Macri},
  {Bucciarelli}, {Lattanzi}, {MacKenty}, {Bowers}, {Zheng}, {Filippenko},
  {Huang}, \& {Anderson}}]{2018riess}
{Riess}, A.~G., {Casertano}, S., {Yuan}, W., {et~al.} 2018, \apj, 861, 126

\bibitem[{{Rodr{\'{\i}}guez} {et~al.}(2014){Rodr{\'{\i}}guez}, {Clocchiatti},
  \& {Hamuy}}]{2014rodriguez}
{Rodr{\'{\i}}guez}, {\'O}., {Clocchiatti}, A., \& {Hamuy}, M. 2014, \aj, 148,
  107

\bibitem[{{Roming} {et~al.}(2005){Roming}, {Kennedy}, {Mason}, {Nousek}, {Ahr},
  {Bingham}, {Broos}, {Carter}, {Hancock}, {Huckle}, {Hunsberger}, {Kawakami},
  {Killough}, {Koch}, {McLelland}, {Smith}, {Smith}, {Soto}, {Boyd},
  {Breeveld}, {Holland}, {Ivanushkina}, {Pryzby}, {Still}, \&
  {Stock}}]{2005roming}
{Roming}, P.~W.~A., {Kennedy}, T.~E., {Mason}, K.~O., {et~al.} 2005, Space
  Science Reviews, 120, 95

\bibitem[{{Rubin} {et~al.}(2013){Rubin}, {Knop}, {Rykoff}, {Aldering},
  {Amanullah}, {Barbary}, {Burns}, {Conley}, {Connolly}, {Deustua}, {Fadeyev},
  {Fakhouri}, {Fruchter}, {Gibbons}, {Goldhaber}, {Goobar}, {Hsiao}, {Huang},
  {Kowalski}, {Lidman}, {Meyers}, {Nordin}, {Perlmutter}, {Saunders},
  {Spadafora}, {Stanishev}, {Suzuki}, {Wang}, \& {Supernova Cosmology
  Project}}]{2013rubin}
{Rubin}, D., {Knop}, R.~A., {Rykoff}, E., {et~al.} 2013, \apj, 763, 35

\bibitem[{{Sahu} {et~al.}(2013){Sahu}, {Anupama}, \& {Chakradhari}}]{2013sahu}
{Sahu}, D.~K., {Anupama}, G.~C., \& {Chakradhari}, N.~K. 2013, \mnras, 433, 2

\bibitem[{{Sahu} {et~al.}(2018){Sahu}, {Anupama}, {Chakradhari}, {Srivastav},
  {Tanaka}, {Maeda}, \& {Nomoto}}]{2018sahu}
{Sahu}, D.~K., {Anupama}, G.~C., {Chakradhari}, N.~K., {et~al.} 2018, \mnras,
  475, 2591

\bibitem[{{Sahu} {et~al.}(2006){Sahu}, {Anupama}, {Srividya}, \&
  {Muneer}}]{2006sahu}
{Sahu}, D.~K., {Anupama}, G.~C., {Srividya}, S., \& {Muneer}, S. 2006, \mnras,
  372, 1315

\bibitem[{{Sanders} {et~al.}(2015){Sanders}, {Soderberg}, {Gezari},
  {Betancourt}, {Chornock}, {Berger}, {Foley}, {Challis}, {Drout}, {Kirshner},
  {Lunnan}, {Marion}, {Margutti}, {McKinnon}, {Milisavljevic}, {Narayan},
  {Rest}, {Kankare}, {Mattila}, {Smartt}, {Huber}, {Burgett}, {Draper},
  {Hodapp}, {Kaiser}, {Kudritzki}, {Magnier}, {Metcalfe}, {Morgan}, {Price},
  {Tonry}, {Wainscoat}, \& {Waters}}]{2015sanders}
{Sanders}, N.~E., {Soderberg}, A.~M., {Gezari}, S., {et~al.} 2015, \apj, 799,
  208

\bibitem[{{Schlafly} \& {Finkbeiner}(2011)}]{2011schlafly}
{Schlafly}, E.~F., \& {Finkbeiner}, D.~P. 2011, \apj, 737, 103

\bibitem[{{Schmidt} {et~al.}(1989){Schmidt}, {Weymann}, \&
  {Foltz}}]{1989schmidt}
{Schmidt}, G.~D., {Weymann}, R.~J., \& {Foltz}, C.~B. 1989, \pasp, 101, 713

\bibitem[{{Schoniger} \& {Sofue}(1994)}]{1994schoniger}
{Schoniger}, F., \& {Sofue}, Y. 1994, \aap, 283, 21

\bibitem[{{Science Software Branch at STScI}(2012)}]{2012pyraf}
{Science Software Branch at STScI}. 2012, {PyRAF: Python alternative for IRAF},
  Astrophysics Source Code Library,  Science Software Branch, STScI,
  ascl:1207.011

\bibitem[{{Shakhbazyan}(1968)}]{1968shakhbazyan}
{Shakhbazyan}, R.~K. 1968, Astrophysics, 4, 123

\bibitem[{{Singh} {et~al.}(2018){Singh}, {Srivastav}, {Kumar}, {Anupama}, \&
  {Sahu}}]{2018avinash}
{Singh}, A., {Srivastav}, S., {Kumar}, B., {Anupama}, G.~C., \& {Sahu}, D.~K.
  2018, \mnras, 480, 2475

\bibitem[{{Smartt}(2009)}]{2009smartt}
{Smartt}, S.~J. 2009, \araa, 47, 63

\bibitem[{{Smith}(2014)}]{2014smith}
{Smith}, N. 2014, \araa, 52, 487

\bibitem[{{Smith} {et~al.}(2015){Smith}, {Mauerhan}, {Cenko}, {Kasliwal},
  {Silverman}, {Filippenko}, {Gal-Yam}, {Clubb}, {Graham}, {Leonard}, {Horst},
  {Williams}, {Andrews}, {Kulkarni}, {Nugent}, {Sullivan}, {Maguire}, {Xu}, \&
  {Ben-Ami}}]{2015smith}
{Smith}, N., {Mauerhan}, J.~C., {Cenko}, S.~B., {et~al.} 2015, \mnras, 449,
  1876

\bibitem[{{Spiro} {et~al.}(2014){Spiro}, {Pastorello}, {Pumo}, {Zampieri},
  {Turatto}, {Smartt}, {Benetti}, {Cappellaro}, {Valenti}, {Agnoletto},
  {Altavilla}, {Aoki}, {Brocato}, {Corsini}, {Di Cianno}, {Elias-Rosa},
  {Hamuy}, {Enya}, {Fiaschi}, {Folatelli}, {Desidera}, {Harutyunyan}, {Howell},
  {Kawka}, {Kobayashi}, {Leibundgut}, {Minezaki}, {Navasardyan}, {Nomoto},
  {Mattila}, {Pietrinferni}, {Pignata}, {Raimondo}, {Salvo}, {Schmidt},
  {Sollerman}, {Spyromilio}, {Taubenberger}, {Valentini}, {Vennes}, \&
  {Yoshii}}]{2014spiro}
{Spiro}, S., {Pastorello}, A., {Pumo}, M.~L., {et~al.} 2014, \mnras, 439, 2873

\bibitem[{{Stanway} \& {Eldridge}(2018)}]{2018stanway}
{Stanway}, E.~R., \& {Eldridge}, J.~J. 2018, \mnras, 479, 75

\bibitem[{{Sukhbold} {et~al.}(2016){Sukhbold}, {Ertl}, {Woosley}, {Brown}, \&
  {Janka}}]{2016sukhbold}
{Sukhbold}, T., {Ertl}, T., {Woosley}, S.~E., {Brown}, J.~M., \& {Janka}, H.-T.
  2016, \apj, 821, 38

\bibitem[{{Swartz} {et~al.}(1991){Swartz}, {Wheeler}, \&
  {Harkness}}]{1991swartz}
{Swartz}, D.~A., {Wheeler}, J.~C., \& {Harkness}, R.~P. 1991, \apj, 374, 266

\bibitem[{{Taddia} {et~al.}(2015){Taddia}, {Sollerman}, {Fremling},
  {Pastorello}, {Leloudas}, {Fransson}, {Nyholm}, {Stritzinger}, {Ergon},
  {Roy}, \& {Migotto}}]{2015taddia}
{Taddia}, F., {Sollerman}, J., {Fremling}, C., {et~al.} 2015, \aap, 580, A131

\bibitem[{{Tak{\'a}ts} \& {Vink{\'o}}(2012)}]{2012takats}
{Tak{\'a}ts}, K., \& {Vink{\'o}}, J. 2012, \mnras, 419, 2783

\bibitem[{{Tak{\'a}ts} {et~al.}(2015){Tak{\'a}ts}, {Pignata}, {Pumo},
  {Paillas}, {Zampieri}, {Elias-Rosa}, {Benetti}, {Bufano}, {Cappellaro},
  {Ergon}, {Fraser}, {Hamuy}, {Inserra}, {Kankare}, {Smartt}, {Stritzinger},
  {Van Dyk}, {Haislip}, {LaCluyze}, {Moore}, \& {Reichart}}]{2015takats}
{Tak{\'a}ts}, K., {Pignata}, G., {Pumo}, M.~L., {et~al.} 2015, \mnras, 450,
  3137

\bibitem[{{Terreran} {et~al.}(2016){Terreran}, {Jerkstrand}, {Benetti},
  {Smartt}, {Ochner}, {Tomasella}, {Howell}, {Morales-Garoffolo},
  {Harutyunyan}, {Kankare}, {Arcavi}, {Cappellaro}, {Elias-Rosa},
  {Hosseinzadeh}, {Kangas}, {Pastorello}, {Tartaglia}, {Turatto}, {Valenti},
  {Wiggins}, \& {Yuan}}]{2016terreran}
{Terreran}, G., {Jerkstrand}, A., {Benetti}, S., {et~al.} 2016, \mnras, 462,
  137

\bibitem[{{Tomi{\v c}i{\'c}} {et~al.}(2018){Tomi{\v c}i{\'c}}, {Hughes},
  {Kreckel}, {Renaud}, {Pety}, {Schinnerer}, {Saito}, {Querejeta}, {Faesi}, \&
  {Garcia-Burillo}}]{2018tomi}
{Tomi{\v c}i{\'c}}, N., {Hughes}, A., {Kreckel}, K., {et~al.} 2018, \apjl, 869,
  L38

\bibitem[{{Treffers} {et~al.}(1993){Treffers}, {Filippenko}, {Leibundgut},
  {Paik}, {Lee}, \& {Richmond}}]{1993treffers}
{Treffers}, R.~R., {Filippenko}, A.~V., {Leibundgut}, B., {et~al.} 1993,
  \iaucirc, 5850

\bibitem[{{Turatto} {et~al.}(2003){Turatto}, {Benetti}, \&
  {Cappellaro}}]{2003turatto}
{Turatto}, M., {Benetti}, S., \& {Cappellaro}, E. 2003, in From Twilight to
  Highlight: The Physics of Supernovae, ed. W.~{Hillebrandt} \&
  B.~{Leibundgut}, 200

\bibitem[{{Turatto} {et~al.}(1998){Turatto}, {Mazzali}, {Young}, {Nomoto},
  {Iwamoto}, {Benetti}, {Cappellaro}, {Danziger}, {de Mello}, {Phillips},
  {Suntzeff}, {Clocchiatti}, {Piemonte}, {Leibundgut}, {Covarrubias}, {Maza},
  \& {Sollerman}}]{1998turatto}
{Turatto}, M., {Mazzali}, P.~A., {Young}, T.~R., {et~al.} 1998, \apjl, 498,
  L129

\bibitem[{{Valenti} {et~al.}(2015){Valenti}, {Sand}, {Stritzinger}, {Howell},
  {Arcavi}, {McCully}, {Childress}, {Hsiao}, {Contreras}, {Morrell},
  {Phillips}, {Gromadzki}, {Kirshner}, \& {Marion}}]{2015valenti}
{Valenti}, S., {Sand}, D., {Stritzinger}, M., {et~al.} 2015, \mnras, 448, 2608

\bibitem[{{Valenti} {et~al.}(2016){Valenti}, {Howell}, {Stritzinger}, {Graham},
  {Hosseinzadeh}, {Arcavi}, {Bildsten}, {Jerkstrand}, {McCully}, {Pastorello},
  {Piro}, {Sand}, {Smartt}, {Terreran}, {Baltay}, {Benetti}, {Brown},
  {Filippenko}, {Fraser}, {Rabinowitz}, {Sullivan}, \& {Yuan}}]{2016valenti}
{Valenti}, S., {Howell}, D.~A., {Stritzinger}, M.~D., {et~al.} 2016, \mnras,
  459, 3939

\bibitem[{{Van Dyk} {et~al.}(2019){Van Dyk}, {Zheng}, {Maund}, {Brink},
  {Srinivasan}, {Andrews}, {Smith}, {Leonard}, {Morozova}, {Filippenko},
  {Conner}, {Milisavljevic}, {de Jaeger}, {Long}, {Isaacson}, {Crossfield},
  {Kosiarek}, {Howard}, {Fox}, {Kelly}, {Piro}, {Littlefair}, {Dhillon},
  {Wilson}, {Butterley}, {Yunus}, {Channa}, {Jeffers}, {Falcon}, {Ross},
  {Hestenes}, {Stegman}, {Zhang}, \& {Kumar}}]{2019vandyk}
{Van Dyk}, S.~D., {Zheng}, W., {Maund}, J.~R., {et~al.} 2019, \apj, 875, 136

\bibitem[{{Walmswell} \& {Eldridge}(2012)}]{2012walmswell}
{Walmswell}, J.~J., \& {Eldridge}, J.~J. 2012, \mnras, 419, 2054

\bibitem[{Waskom {et~al.}(2018)Waskom, Botvinnik, O'Kane, Hobson, Ostblom,
  Lukauskas, Gemperline, Augspurger, Halchenko, Cole, Warmenhoven, de~Ruiter,
  Pye, Hoyer, Vanderplas, Villalba, Kunter, Quintero, Bachant, Martin, Meyer,
  Miles, Ram, Brunner, Yarkoni, Williams, Evans, Fitzgerald, Brian, \&
  Qalieh}]{2018seaborn}
Waskom, M., Botvinnik, O., O'Kane, D., {et~al.} 2018, mwaskom/seaborn: v0.9.0
  (July 2018),  zenodo, doi:10.5281/zenodo.1313201

\bibitem[{{Waxman} {et~al.}(2007){Waxman}, {M{\'e}sz{\'a}ros}, \&
  {Campana}}]{2007waxman}
{Waxman}, E., {M{\'e}sz{\'a}ros}, P., \& {Campana}, S. 2007, \apj, 667, 351

\bibitem[{{Wolter} {et~al.}(2015){Wolter}, {Esposito}, {Mapelli}, {Pizzolato},
  \& {Ripamonti}}]{2015wolter}
{Wolter}, A., {Esposito}, P., {Mapelli}, M., {Pizzolato}, F., \& {Ripamonti},
  E. 2015, \mnras, 448, 781

\bibitem[{{Woosley} \& {Heger}(2007)}]{2007woosley}
{Woosley}, S.~E., \& {Heger}, A. 2007, \physrep, 442, 269

\bibitem[{{Woosley} \& {Heger}(2012)}]{2012woosley}
---. 2012, \apj, 752, 32

\bibitem[{{Woosley} \& {Weaver}(1995)}]{1995woosley}
{Woosley}, S.~E., \& {Weaver}, T.~A. 1995, \apjs, 101, 181

\bibitem[{{Yaron} \& {Gal-Yam}(2012)}]{2012yaron}
{Yaron}, O., \& {Gal-Yam}, A. 2012, \pasp, 124, 668

\bibitem[{{Yuan} {et~al.}(2016){Yuan}, {Jerkstrand}, {Valenti}, {Sollerman},
  {Seitenzahl}, {Pastorello}, {Schulze}, {Chen}, {Childress}, {Fraser},
  {Fremling}, {Kotak}, {Ruiter}, {Schmidt}, {Smartt}, {Taddia}, {Terreran},
  {Tucker}, {Barbarino}, {Benetti}, {Elias-Rosa}, {Gal-Yam}, {Howell},
  {Inserra}, {Kankare}, {Lee}, {Li}, {Maguire}, {Margheim}, {Mehner}, {Ochner},
  {Sullivan}, {Tomasella}, \& {Young}}]{2016yuan}
{Yuan}, F., {Jerkstrand}, A., {Valenti}, S., {et~al.} 2016, \mnras, 461, 2003

\end{thebibliography}

\appendix

\section{Template Subtraction}\label{sec:tempsub}

\begin{figure*}
\label{fig:tempsub}
\centering
\resizebox{\hsize}{!}{\includegraphics[width=\linewidth]{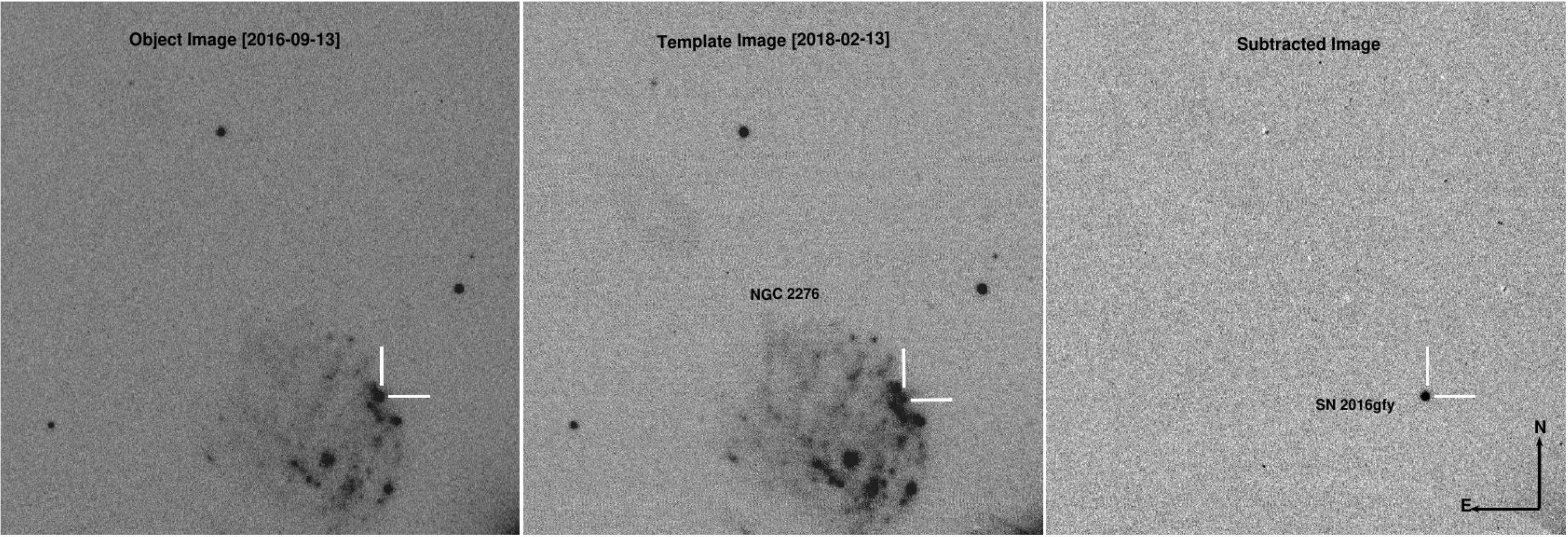}}
\caption{$Left\ panel$: $U$-band image of \sniip\ obtained with the HCT on 2016 Sep 13; $Middle\ panel$: Template image observed with HCT on 2018 Feb 13; $Right\ panel$: Subtracted image procured after PSF-matching of the background subtracted images in the first two panels.}
\end{figure*}
 
Since \sniip\ exploded in the bright spiral arm of the host galaxy, there is a significant contribution from the host environment in its optical photometry. Template images of \hostglx\ were obtained in a good seeing condition ($\leq 2^{\prime\prime}$) from the 2\,m HCT (Himalayan Chandra Telescope) on 2018 February 13, almost 1.5 years from the date of explosion when the SN had diminished enough to allow for the the imaging of the bright galaxy background. The templates were aligned to the object frame, PSF-matched, background-subtracted and scaled in order to subtract the host galaxy contribution in the photometric frames of the \sniip. 

\section{Distance Estimation using SCM}\label{sec:scm}

Type Ia SNe have been studied up to a redshift of $\sim$\,1.7 \citep{2013rubin}. To allow the study beyond the above redshift, wherein Type II SNe are in abundance due to their shorter lifetimes and hence becomes an entrancing choice even though they are fainter. The \lq \lq Standard Candle Method\rq \rq\ \citep[SCM,][]{2002hamuy}, helps estimate distance using the correlation of bolometric luminosity with the expansion velocity of the ejecta during the plateau phase. The latest value of the Hubble constant determined by SNe Ia, i.e. $\rm H_0$ = 73.52\,$\pm$\,1.62 $\rm km\ s^{-1}\ Mpc^{-1}$\citep{2018riess} is used here to compute the distances. The implementation of these techniques is discussed in the following subsections.

\subsection{Apparent light curve fits}

The apparent $BVRI$ light curves of \sniip\ were fit using an analytic function \citep{2010olivares} comprising of three components (see Equation~\ref{eqn:fitolivares}). Nelder-Mead optimization was employed to minimize the ${\chi}^{2}$ of the fit. The interpolated magnitudes from the fit were extracted at a step size of 1 d until 250 d and the 3$\sigma$ deviation of the fit were adopted as errors. The quantities inferred from the fit are compiled in Table~\ref{tab:fitolivares}. The parameter $t_{PT}$ derived here can be used as an alternative to the date of explosion to define the epoch for calculating observables for the SCM.

\begin{equation}
\label{eqn:fitolivares}
    f(t) = \underbrace{\frac{-a_0}{1 + e^{(t-t_{PT})/ w_0}}}_\text{Fermi-Dirac term} + \underbrace{p_0 (t-t_{PT}) + m_0}_\text{Linear term} - \underbrace{P e^{-(\frac{t-Q}{R})^2}}_\text{Gaussian term}
\end{equation}

\subsection{Expansion velocities}

Photospheric velocities measured from the minimum of the of \ion{Fe}{2} $\rm \lambda$5169 absorption are accurate up to of 5$\%$-10$\%$ \citep{2005adessart}. An alternative is to use the H\,$\rm \beta$ feature which has a higher SNR and which correlates as $v_{Fe\,II}$ = (0.82\,$\pm$\,0.05) $v_{H\beta}$ during the plateau phase \citep{2010poznanski,2012takats}. A power law ($v(t) = \alpha\ \times\ (t - A) ^{\beta}$) is fit to the \ion{Fe}{2} $\rm \lambda$5169 velocity curve during the plateau phase (until the time of inflection, $\sim$\,110 d) where $\alpha$, $A$, $\beta$ have no direct physical interpretations. The variance-weighted least squares minimization (WLS) fit is shown in an inset in Figure~\ref{fig:photvel} and helped extract expansion velocities from 20 - 110 d at an interval of 1 d without the need for extrapolation. The errors computed are the 3$\sigma$ deviations of the fit.

\subsection{Colour evolution fits}

The Galactic extinction corrected colours $U-B$, $B-V$, $V-R$ and $V-I$ were fitted with a Legendre polynomial until 150 d from the date of the explosion, some of which are shown in Figure~\ref{fig:colour} with a $red-dashed$ line. The colour values were obtained in a continuous grid with a spacing of 1 d until the transition phase ($\sim$\,110 d).

\subsection{Methodology}

The SCM technique was further inspected by \citet{2004hamuy,2006nugent,2009poznanski,2010olivares}(hereafter H04, N06, P09 and O10, respectively) using different samples of Type II SNe and using distinct epochs as reference for computing the correlated quantities.

\begin{itemize}
\item \citet{2004hamuy}:
H04 investigated the SCM technique with a sample of 24 Type II SNe and found that SCM has a precision of 15\%. H04 used a reference epoch of 50 days from the date of explosion to measure the SN observables required for SCM and estimated the distances using the Equation~\ref{eqn:scmhamuy}.

\item \citet{2006nugent}:
N06 utilized the $(V-I)$ colour during the mid-plateau phase ($\sim$\,50 d) to perform reddening correction using the extinction law from \citet{1989cardelli}. They adopted an un-reddened $V-I$ colour of 0.53 mag for Type II SNe and used an $\rm H_0$\,=\,70 $\rm km\ s^{-1}\ Mpc^{-1}$ for computing the distances using Equation~\ref{eqn:scmnugent}.

\item \citet{2009poznanski}:
P09 remodeled the relation from N06 with a sample of 34 Type II SNe with the most tangible assumption that not all the SNe must follow the same extinction law as \citet{1989cardelli}. The value of $\beta$ used here differs from the value in P09 because of the different value of $\rm H_0$ adopted here instead of $\rm H_0$\,=\,70 $\rm km\ s^{-1}\ Mpc^{-1}$ in P09. P09 computed the distances using the Equation~\ref{eqn:scmpoznanski}.

\item \citet{2010olivares}:
The refined SCM method by O10 makes use of the expansion velocities, magnitudes and colour terms estimated 30 d before the middle of the transition phase and was calibrated using a sample of 37 Type II SNe. The calibrated relation for $BVI$ bands is given in Equation~\ref{eqn:scmolivares}.
\end{itemize}

\begin{align}
    5\,\log[H_0 D_{\lambda}] = m_{\lambda} - A_{\lambda} + \alpha \times \log[v_{Fe\,II} / 5000] + \beta \label{eqn:scmhamuy}\\
    5\,\log[D_{\lambda}] - 5 = m_{\lambda} + \alpha \times \log[v_{Fe\,II} / 5000] + \gamma[(V-I) - 0.53] - \beta \label{eqn:scmnugent}\\
    5\,\log[H_0 D_{\lambda}] =  m_{\lambda} + \alpha \times \log[v_{Fe\,II} / 5000] - \gamma[(V-I) - 0.53] - \beta \label{eqn:scmpoznanski}\\
    5\,\log[H_0 D_{\lambda}] = m_{\lambda} + \alpha \times \log[v_{Fe\,II} / 5000] - \gamma(V-I) - \beta \label{eqn:scmolivares}
\end{align}

where $\rm m_{\lambda}$ is the apparent magnitude, $\rm A_{\lambda}$ is the extinction, $v_{Fe\,II}$ is in $\rm km\ s^{-1}$, D$_{\lambda}$ is in Mpc, and $\alpha$, $\beta$ and $\gamma$ are dimensionless constants mentioned in Table~\ref{tab:resultsscm}.\\

\begin{table*}
\centering
\caption{Parameters extracted from the analytic fit \citep{2010olivares} $BVRI$ LC of \sniip.}
\label{tab:fitolivares}
\begin{tabular}{|c| c c c c c c c c c c|}
\toprule
Filter  &  $a0$           &   $t_{PT}$  &     $w_0$           &  $p_0$     &  $m_0$  & $P$    &   $Q$   &   $R$  & $s_{1}^{*}$   &   $s_{2}^{*}$  \\
        &  (mag)          &   (d)$^{a}$ &      (d)            &  (mag / d) &  (mag)  & (mag)  &   (d)   &   (d)  & (mag / 100 d) &  (mag / 100 d) \\
\midrule
$B$     &  2.19\,$\pm$\,0.04  &  107.4\,$\pm$\,0.8  &  8.79\,$\pm$\,1.39  &  0.0059    &  19.79  &  0.92  &   4.24  & 27.37  &   3.10        &   1.15         \\
$V$     &  1.16\,$\pm$\,0.01  &  113.2\,$\pm$\,0.5  &  2.65\,$\pm$\,0.34  &  0.0107    &  18.27  &  0.50  &  82.37  & 31.39  &   0.94        &   0.12         \\
$R$     &  1.42\,$\pm$\,0.01  &  112.9\,$\pm$\,0.3  &  3.67\,$\pm$\,0.31  &  0.0086    &  17.35  & -0.59  &  11.25  & 52.77  &   0.30        &  -0.02         \\
$I$     &  1.25\,$\pm$\,0.01  &  114.0\,$\pm$\,0.5  &  3.54\,$\pm$\,0.42  &  0.0109    &  16.84  & -0.98  &   7.65  & 53.64  &   0.11        &  -0.27         \\
\bottomrule
\multicolumn{4}{c}{$^a$\footnotesize{Time since explosion epoch (JD 2457641.40)}}\\
\multicolumn{5}{c}{$^*$\footnotesize{Extracted from a linear piece-wise fit to the light curves.}}
\end{tabular}
\end{table*}
\begin{table*}
\centering
\setlength{\tabcolsep}{7pt}
\caption{Comparison sample of Type II SNe.}
\label{tab:compsample}
\begin{tabular}{|c| c c c c c|}
\toprule
SN           &  Explosion Epoch &   Distance        &   $\rm M^V_{50}$  &   $\rm M_{Ni}$       &   Reference   \\
(Name)       &  (JD)            &   (Mpc)           &   (mag)           &   ($\rm M_{\odot}$)  &               \\
\midrule
1987A        &  2446859.82      &    0.04$\pm$0.003 &       ---         &   0.075$\pm$0.005    &   1, 16       \\
1999em       &  2451475.60      &   11.70$\pm$0.99  &  -15.90$\pm$0.20  &   0.042$^{0.027}_{0.019}$  &   2, 3, 17 \\
2004et       &  2453270.25      &    5.60$\pm$0.10  &  -17.14$\pm$0.10  &   0.060$\pm$0.020    &   4           \\
2005cs       &  2453549.00      &    8.90$\pm$0.50  &  -14.83$\pm$0.10  &   0.006$\pm$0.003    &   5, 18       \\
2007od       &  2454404.00      &   25.70$\pm$0.80  &  -17.64$\pm$0.22  &   0.003              &   6           \\
2009ib       &  2455041.30      &   19.80$\pm$2.80  &   --              &   0.046$\pm$0.015    &   7           \\
2012aw       &  2456002.59      &    9.90$\pm$0.10  &  -16.67$\pm$0.04  &   0.056$\pm$0.013    &   8, 13       \\
2012ec       &  2456143.00      &   17.30$\pm$0.96  &  -16.54$\pm$0.14  &   0.040$\pm$0.015    &   9           \\
2013ab       &  2456340.00      &   24.30$\pm$1.00  &  -16.70$\pm$0.10  &   0.064$\pm$0.006    &   10          \\
2013ej       &  2456497.30      &    9.57$\pm$0.70  &  -16.60$\pm$0.10  &   0.018$\pm$0.006    &   11, 13      \\
ASASSN-14dq  &  2456841.50      &   44.80$\pm$3.10  &  -16.90$\pm$0.20  &   0.029$\pm$0.005    &   12          \\
2014cx       &  2456901.89      &   22.28$\pm$1.60  &  -17.20$\pm$0.20  &   0.056$\pm$0.008    &   13          \\
2016X        &  2457405.92      &   15.20$\pm$3.30  &  -16.20$\pm$0.43  &   0.034$\pm$0.006    &   14          \\
2016esw      &  2457608.33      &  123.60           &  -17.35$\pm$0.11  &        ---           &   15          \\
\bottomrule
\end{tabular}
\newline \newline
References:\
(1) \citet{1990hamuy};
(2) \citet{2002bleonard};
(3) \citet{2003leonard};
(4) \citet{2006sahu};
(5) \citet{2009pastorello};
(6) \citet{2011inserra};
(7) \citet{2015takats};
(8) \citet{2013bose};
(9) \citet{2015barbarino};
(10) \citet{2015boseab};
(11) \citet{2015boseej};
(12) \citet{2018avinash};
(13) \citet{2015valenti};
(14) \citet{2018huang};
(15) \citet{2018bdejaeger};
(16) \citet{2003turatto};
(17) \citet{2001hamuy};
(18) \citet{2014spiro};
\end{table*}
\begin{table*}
\centering
\renewcommand{\arraystretch}{1.6}
\caption{Distances to the host galaxy \hostglx}
\label{tab:dist_host}

\begin{tabular}{l c c c}
\toprule
Distance Method					                & Distance (in Mpc)	& Distance Modulus (in mag)	&   Reference 	\\
\midrule
Hubble Flow Distance (Virgo + GA + Shapley)     & 37.1$\pm$2.6 		& 32.85$\pm$0.15		    &	2	        \\
CO-Line Tully-Fisher relation           	    & 22.6  	        & 31.77         		    &	3	        \\
\midrule
Luminosity Distance  				            & 33.1   	        & 32.60		                &	1           \\
\midrule
Mean SCM            				            & 29.64$\pm$2.65  	& 32.36$\pm$0.18		    &	1           \\
\bottomrule
\end{tabular}
\newline
(1) This paper;
(2) \citet{2000mould};
(3) \citet{1994schoniger}
\end{table*}

\begin{table*}
\centering
\renewcommand{\arraystretch}{1.1}
\caption{$UBVRI$ magnitudes of secondary standards in the field of SN~2016gfy. The magnitudes reported are in Vega system.}
\label{tab:secstd}
\begin{tabular}{c c c c c c c c}
\toprule
ID  & $\alpha$  &     $\delta$               &        U         &       B           &         V         &         R         &        I          \\
    & (h:m:s)   & ($^{\circ}: \arcmin: \arcmin\arcmin$) &      (mag)       &     (mag)         &       (mag)       &       (mag)       &       (mag)       \\
\midrule
  1 &   07:28:02.81     & +85:48:20.89       &  15.90$\pm$0.05  &  15.75$\pm$0.02   &  15.03$\pm$0.01   &  14.60$\pm$0.01   &  14.15$\pm$0.03   \\
  2 &   07:26:02.70     & +85:46:48.94       &  15.50$\pm$0.05  &  15.53$\pm$0.02   &  14.94$\pm$0.01   &  14.57$\pm$0.01   &  14.16$\pm$0.03   \\
  3 &   07:29:35.46     & +85:45:41.42       &  17.33$\pm$0.05  &  16.25$\pm$0.02   &  15.19$\pm$0.01   &  14.58$\pm$0.01   &  14.01$\pm$0.03   \\
  4 &   07:27:47.80     & +85:43:46.73       &  15.02$\pm$0.05  &  15.06$\pm$0.02   &  14.53$\pm$0.01   &  14.19$\pm$0.01   &  13.81$\pm$0.03   \\
  5 &   07:27:22.28     & +85:42:26.58       &  15.96$\pm$0.05  &  14.94$\pm$0.02   &  13.91$\pm$0.01   &  13.32$\pm$0.01   &  12.73$\pm$0.03   \\
  6 &   07:23:53.97     & +85:42:41.15       &  13.16$\pm$0.05  &  13.28$\pm$0.02   &  12.80$\pm$0.01   &  12.49$\pm$0.01   &  12.11$\pm$0.03   \\
  7 &   07:23:27.40     & +85:45:01.93       &  15.76$\pm$0.05  &  15.68$\pm$0.02   &  15.06$\pm$0.01   &  14.65$\pm$0.01   &  14.23$\pm$0.03   \\
  8 &   07:28:11.34     & +85:42:23.56       &  16.46$\pm$0.05  &  16.03$\pm$0.02   &  15.15$\pm$0.01   &  14.64$\pm$0.01   &  14.11$\pm$0.03   \\
  9 &   07:30:20.55     & +85:43:15.26       &  15.78$\pm$0.05  &  15.83$\pm$0.02   &  15.24$\pm$0.01   &  14.89$\pm$0.01   &  14.60$\pm$0.03   \\
 10 &   07:30:01.53     & +85:41:53.41       &  15.88$\pm$0.05  &  15.52$\pm$0.02   &  14.71$\pm$0.01   &  14.26$\pm$0.01   &  13.78$\pm$0.03   \\
 11 &   07:24:16.53     & +85:41:53.73       &  17.67$\pm$0.05  &  16.40$\pm$0.02   &  15.14$\pm$0.01   &  14.36$\pm$0.01   &  13.58$\pm$0.03   \\
 12 &   07:30:16.42     & +85:47:52.66       &  17.47$\pm$0.05  &  17.30$\pm$0.02   &  16.60$\pm$0.01   &  16.17$\pm$0.02   &  15.76$\pm$0.03   \\
 13 &   07:29:28.41     & +85:44:09.11       &  15.84$\pm$0.05  &  15.87$\pm$0.02   &  15.27$\pm$0.01   &  14.88$\pm$0.01   &  14.47$\pm$0.03   \\
\bottomrule
\end{tabular}
\end{table*}
\begin{table*}
\centering
\caption{Photometric observations of SN~2016gfy from SWIFT-UVOT. The magnitudes reported are in Vega system.}
\label{tab:photswift}
\begin{tabular}{c c c c c c c c c}
\toprule
      Date      &    JD         & Phase$^*$ &   $uvw2$          &       $uvm2$      &   $uvw1$        &     $uvu$       &   $uvb$           &   $uvv$           \\
(yyyy/mm/dd)    & (245 7600+)   &   (d)     &   (mag)           &      (mag)        &    (mag)        &      (mag)      &   (mag)           &   (mag)           \\
\midrule                                                                                                                                                      
 2016-09-15     &  46.6         &     +5.2  &  15.38$\pm$0.09   &  15.29$\pm$0.07   &  15.24$\pm$0.08 &  15.08$\pm$0.08 &  16.318$\pm$0.085  &  16.02$\pm$0.10   \\
 2016-09-16     &  47.9         &     +6.5  &  15.88$\pm$0.10   &  15.64$\pm$0.08   &  15.36$\pm$0.08 &  15.12$\pm$0.08 &  16.255$\pm$0.081  &  16.01$\pm$0.10   \\
 2016-09-19     &  51.3         &     +9.9  &  16.54$\pm$0.14   &  16.25$\pm$0.10   &  15.69$\pm$0.09 &  15.24$\pm$0.08 &  16.255$\pm$0.082  &  16.12$\pm$0.11   \\
 2016-09-21     &  53.2         &    +11.8  &  16.75$\pm$0.15   &             ---   &  15.93$\pm$0.10 &  15.47$\pm$0.08 &  16.339$\pm$0.082  &             ---   \\
 2016-09-24     &  55.7         &    +14.3  &  17.26$\pm$0.23   &  17.36$\pm$0.21   &  16.42$\pm$0.14 &  15.59$\pm$0.10 &  16.403$\pm$0.096  &  16.17$\pm$0.12   \\
 2016-10-01     &  62.6         &    +21.2  &             ---   &             ---   &  17.47$\pm$0.26 &  16.30$\pm$0.12 &  16.745$\pm$0.104  &  16.23$\pm$0.12   \\
 2016-10-02     &  64.3         &    +22.9  &             ---   &             ---   &             --- &  16.58$\pm$0.14 &  16.719$\pm$0.099  &  16.08$\pm$0.10   \\
 2016-10-04     &  66.0         &    +24.6  &             ---   &             ---   &             --- &  16.74$\pm$0.16 &  16.842$\pm$0.109  &  16.19$\pm$0.12   \\
 2016-10-06     &  68.4         &    +27.0  &             ---   &             ---   &             --- &  16.87$\pm$0.17 &  16.851$\pm$0.105  &  16.18$\pm$0.11   \\
\bottomrule
\multicolumn{3}{l}{$^a$\footnotesize{Time since explosion epoch (JD 2457641.4)}}
\end{tabular}
\end{table*}

\begin{table*}
\centering
\caption{Photometric observations of SN~2016gfy from HCT-HFOSC. The magnitudes reported are in Vega system.}
\label{tab:phothct}
\begin{tabular}{c c c c c c c c}
\toprule
       Date &      JD 	    & Phase$^a$ &       $U$ &             $B$ &             $V$ &             $R$ &             $I$ \\
(yyyy-mm-dd)& (245 7600+)   &   (d)     &   (mag)   &      (mag)      &    (mag)        &      (mag)      &  (mag)          \\
\midrule
 2016-09-13 &   45.30 &     +3.90 &  15.57$\pm$0.05 &  16.23$\pm$0.02 &  16.25$\pm$0.01 &  16.04$\pm$0.02 &  15.88$\pm$0.02 \\
 2016-09-14 &   46.25 &     +4.85 &  15.51$\pm$0.14 &  16.16$\pm$0.03 &  16.13$\pm$0.01 &  15.89$\pm$0.02 &  15.68$\pm$0.02 \\
 2016-09-17 &   49.31 &     +7.92 &  15.51$\pm$0.13 &  16.11$\pm$0.03 &  15.94$\pm$0.02 &  15.64$\pm$0.04 &  15.41$\pm$0.02 \\
 2016-09-20 &   52.40 &    +11.00 &  15.56$\pm$0.05 &  16.16$\pm$0.02 &  15.99$\pm$0.02 &  15.69$\pm$0.02 &  15.45$\pm$0.02 \\
 2016-09-27 &   59.32 &    +17.92 &  15.98$\pm$0.04 &  16.36$\pm$0.02 &  16.11$\pm$0.01 &  15.73$\pm$0.02 &  15.53$\pm$0.02 \\
 2016-09-30 &   62.48 &    +21.08 &             --- &  16.46$\pm$0.02 &  16.13$\pm$0.01 &  15.69$\pm$0.01 &  15.48$\pm$0.02 \\
 2016-10-04 &   66.34 &    +24.94 &  16.51$\pm$0.07 &  16.59$\pm$0.02 &  16.14$\pm$0.01 &  15.70$\pm$0.02 &  15.48$\pm$0.06 \\
 2016-10-07 &   69.48 &    +28.08 &             --- &  16.70$\pm$0.02 &  16.17$\pm$0.02 &  15.71$\pm$0.03 &  15.50$\pm$0.02 \\
 2016-10-12 &   74.45 &    +33.05 &  17.05$\pm$0.09 &             --- &  16.20$\pm$0.01 &  15.74$\pm$0.02 &  15.46$\pm$0.02 \\
 2016-10-15 &   77.39 &    +35.99 &  17.16$\pm$0.10 &  16.93$\pm$0.02 &  16.21$\pm$0.02 &  15.72$\pm$0.02 &  15.46$\pm$0.03 \\
 2016-10-17 &   79.20 &    +37.80 &  17.33$\pm$0.11 &  17.00$\pm$0.03 &  16.27$\pm$0.02 &  15.76$\pm$0.01 &  15.49$\pm$0.02 \\
 2016-10-18 &   80.19 &    +38.79 &  17.42$\pm$0.06 &  17.01$\pm$0.02 &  16.24$\pm$0.01 &  15.78$\pm$0.02 &  15.48$\pm$0.02 \\
 2016-10-24 &   86.42 &    +45.02 &  17.69$\pm$0.03 &  17.15$\pm$0.01 &  16.27$\pm$0.01 &  15.78$\pm$0.01 &  15.48$\pm$0.02 \\
 2016-11-04 &   97.44 &    +56.04 &  17.95$\pm$0.04 &  17.27$\pm$0.01 &  16.25$\pm$0.01 &  15.74$\pm$0.01 &  15.43$\pm$0.02 \\
 2016-11-16 &  109.43 &    +68.03 &  18.12$\pm$0.06 &  17.36$\pm$0.02 &  16.21$\pm$0.01 &  15.72$\pm$0.02 &  15.35$\pm$0.03 \\
 2016-11-26 &  119.15 &    +77.75 &             --- &  17.50$\pm$0.02 &  16.29$\pm$0.01 &  15.74$\pm$0.02 &  15.37$\pm$0.02 \\
 2016-12-02 &  125.40 &    +84.00 &  18.77$\pm$0.09 &  17.62$\pm$0.02 &  16.31$\pm$0.01 &  15.76$\pm$0.01 &  15.36$\pm$0.02 \\
 2016-12-05 &  128.42 &    +87.03 &  18.95$\pm$0.05 &             --- &  16.33$\pm$0.01 &  15.77$\pm$0.01 &  15.39$\pm$0.02 \\
 2016-12-07 &  130.50 &    +89.10 &             --- &  17.77$\pm$0.02 &  16.37$\pm$0.01 &  15.79$\pm$0.02 &  15.40$\pm$0.02 \\
 2016-12-11 &  134.25 &    +92.85 &  18.89$\pm$0.27 &  17.99$\pm$0.10 &  16.41$\pm$0.04 &  15.85$\pm$0.02 &  15.47$\pm$0.05 \\
 2016-12-12 &  135.33 &    +93.93 &  19.22$\pm$0.16 &  17.93$\pm$0.03 &  16.47$\pm$0.02 &  15.87$\pm$0.01 &  15.51$\pm$0.02 \\
 2016-12-14 &  137.45 &    +96.05 &  19.22$\pm$0.18 &  17.95$\pm$0.03 &  16.50$\pm$0.02 &  15.86$\pm$0.01 &  15.48$\pm$0.03 \\
 2016-12-15 &  138.34 &    +96.94 &  19.36$\pm$0.22 &  18.01$\pm$0.04 &  16.61$\pm$0.06 &  15.89$\pm$0.03 &  15.46$\pm$0.10 \\
 2016-12-18 &  141.45 &   +100.05 &             --- &  18.15$\pm$0.06 &  16.63$\pm$0.04 &  15.95$\pm$0.05 &  15.54$\pm$0.09 \\
 2016-12-26 &  149.40 &   +108.00 &             --- &             --- &  16.96$\pm$0.01 &  16.22$\pm$0.02 &  15.78$\pm$0.03 \\
 2016-12-29 &  152.23 &   +110.83 &  20.45$\pm$0.17 &  18.92$\pm$0.02 &  17.19$\pm$0.01 &  16.43$\pm$0.01 &  15.94$\pm$0.02 \\
 2017-01-05 &  159.43 &   +118.03 &             --- &             --- &  18.01$\pm$0.02 &  17.11$\pm$0.03 &  16.56$\pm$0.05 \\
 2017-01-10 &  164.29 &   +122.89 &             --- &             --- &  18.34$\pm$0.03 &  17.36$\pm$0.02 &  16.75$\pm$0.04 \\
 2017-01-11 &  165.14 &   +123.74 &             --- &  19.94$\pm$0.10 &  18.30$\pm$0.03 &  17.39$\pm$0.02 &  16.94$\pm$0.03 \\
 2017-01-13 &  167.39 &   +125.99 &             --- &  19.88$\pm$0.09 &  18.30$\pm$0.03 &  17.44$\pm$0.01 &  16.98$\pm$0.03 \\
 2017-01-28 &  182.42 &   +141.02 &             --- &  20.02$\pm$0.06 &  18.49$\pm$0.02 &             --- &  17.16$\pm$0.03 \\
 2017-02-09 &  194.25 &   +152.85 &             --- &  19.90$\pm$0.15 &  18.67$\pm$0.05 &  17.57$\pm$0.06 &  17.25$\pm$0.07 \\
 2017-02-12 &  197.30 &   +155.90 &             --- &  20.08$\pm$0.09 &  18.66$\pm$0.02 &  17.68$\pm$0.02 &  17.28$\pm$0.02 \\
 2017-02-18 &  203.13 &   +161.73 &             --- &  20.00$\pm$0.06 &  18.82$\pm$0.06 &  17.79$\pm$0.04 &  17.39$\pm$0.09 \\
 2017-02-24 &  209.25 &   +167.85 &             --- &  20.03$\pm$0.04 &  18.84$\pm$0.02 &  17.83$\pm$0.02 &  17.35$\pm$0.03 \\
 2017-03-04 &  217.17 &   +175.77 &             --- &  20.37$\pm$0.05 &  18.90$\pm$0.02 &  17.88$\pm$0.01 &  17.54$\pm$0.02 \\
 2017-03-27 &  240.08 &   +198.68 &             --- &             --- &  19.31$\pm$0.02 &  18.07$\pm$0.02 &  17.80$\pm$0.02 \\
 2017-04-10 &  254.24 &   +212.84 &             --- &             --- &             --- &  18.15$\pm$0.02 &  17.90$\pm$0.04 \\
 2017-05-05 &  279.16 &   +237.76 &             --- &             --- &  19.51$\pm$0.03 &  18.41$\pm$0.02 &  18.19$\pm$0.04 \\
 2017-05-08 &  282.19 &   +240.79 &             --- &             --- &  19.58$\pm$0.05 &  18.51$\pm$0.02 &  18.25$\pm$0.04 \\
 2017-08-11 &  377.17 &   +335.77 &             --- &             --- &             --- &  19.70$\pm$0.08 &             --- \\
 2017-10-01 &  428.42 &   +387.02 &             --- &             --- &  21.19$\pm$0.16 &  20.34$\pm$0.06 &  20.07$\pm$0.12 \\
\bottomrule
\multicolumn{3}{l}{$^a$\footnotesize{Time since explosion epoch (JD 2457641.4)}}
\end{tabular}
\end{table*}
\begin{table}
\centering
\setlength\tabcolsep{2.8pt}
\caption{Spectroscopic observations of SN~2016gfy from HCT-HFOSC}
\label{tab:speclog}
\resizebox{0.5\linewidth}{!}{\begin{tabular}{c c c c}
\toprule
Date         & JD           & Phase$^a$ & Range                \\
(yyyy-mm-dd) & (245 7600+)  & (d)       & (\AA)                \\
\midrule
2016-09-13   &  45.24       &   +3.84   & 3500-7800; 5200-9250 \\
2016-09-14   &  46.27       &   +4.87   & 3500-7800; 5200-9250 \\
2016-09-20   &  52.42       &  +11.02   & 3500-7800; 5200-9250 \\
2016-09-27   &  59.37       &  +17.97   & 3500-7800; 5200-9250 \\
2016-09-30   &  62.49       &  +21.29   & 3500-7800; 5200-9250 \\
2016-10-04   &  66.41       &  +25.01   & 3500-7800; 5200-9250 \\
2016-10-12   &  74.46       &  +33.06   & 3500-7800            \\
2016-10-14   &  76.43       &  +35.03   & 3500-7800; 5200-9250 \\
2016-10-15   &  77.31       &  +35.91   & 3500-7800; 5200-9250 \\
2016-10-18   &  80.46       &  +39.06   & 3500-7800            \\
2016-10-29   &  91.46       &  +50.06   & 3500-7800; 5200-9250 \\
2016-11-04   &  97.38       &  +55.98   & 3500-7800; 5200-9250 \\
2016-11-16   & 109.44       &  +68.04   & 3500-7800; 5200-9250 \\
2016-11-24   & 117.46       &  +76.06   & 3500-7800; 5200-9250 \\
2016-12-05   & 128.45       &  +87.05   & 3500-7800; 5200-9250 \\
2016-12-11   & 134.26       &  +92.86   & 3500-7800; 5200-9250 \\
2016-12-12   & 135.39       &  +93.99   & 3500-7800; 5200-9250 \\
2016-12-14   & 137.46       &  +96.06   & 3500-7800; 5200-9250 \\
2016-12-26   & 149.41       & +108.01   & 3500-7800; 5200-9250 \\
2016-12-29   & 152.25       & +110.85   & 3500-7800; 5200-9250 \\
2017-01-05   & 159.35       & +117.95   & 3500-7800; 5200-9250 \\
2017-01-11   & 165.29       & +123.89   & 3500-7800; 5200-9250 \\
2017-01-28   & 182.43       & +141.03   & 3500-7800; 5200-9250 \\
2017-01-29   & 183.35       & +141.95   & 3500-7800; 5200-9250 \\
2017-02-09   & 194.27       & +152.87   & 3500-7800; 5200-9250 \\
2017-02-11   & 196.27       & +154.87   & 3500-7800; 5200-9250 \\
2017-02-12   & 197.32       & +155.92   & 3500-7800; 5200-9250 \\
2017-02-18   & 203.15       & +161.75   & 3500-7800; 5200-9250 \\
2017-02-24   & 209.27       & +167.87   & 3500-7800; 5200-9250 \\
2017-03-04   & 217.19       & +175.79   & 3500-7800; 5200-9250 \\
2017-03-27   & 240.11       & +198.71   & 3500-7800; 5200-9250 \\
2017-04-10   & 254.13       & +212.73   & 3500-7800; 5200-9250 \\
2017-04-13   & 257.22       & +215.82   & 3500-7800; 5200-9250 \\     
\bottomrule
\multicolumn{3}{l}{$^a$\footnotesize{Time since explosion epoch (JD 2457641.4)}}
\end{tabular}}
\end{table}

\begin{table}
\centering
\setlength\tabcolsep{2.8pt}
\caption{Spectroscopic observations of SN~2016gfy from MMT-BC}
\label{tab:speclogmmt}
\begin{tabular}{c c c c}
\toprule
Date         & JD           & Phase$^a$ & Grating              \\
(yyyy-mm-dd) & (245 7600+)  & (d)       & (lines/mm)           \\
\midrule
2017-01-05   &	160.7 	    & 119.3	& 1200  \\
2017-03-03   &  216.9 	    & 175.5	& 1200  \\
2017-05-21   &  294.7 	    & 253.3	& 1200  \\    
\bottomrule
\multicolumn{3}{l}{$^a$\footnotesize{Time since explosion epoch (JD 2457641.4)}}
\end{tabular}
\end{table}

\end{document}